\documentclass[journal]{IEEEtran}
\usepackage{cite}
\usepackage{graphicx}
\usepackage{float}
\usepackage{placeins}
\usepackage{subfigure}
\usepackage{caption}
\usepackage{pgfplots}
\pgfplotsset{compat=newest}
\usepackage{todonotes}
\usepackage{tabularx}
\usepackage{enumitem}
\usepackage{amsmath,amssymb}
\usepackage{bbm}
\usepackage{array, booktabs, ragged2e}
\tikzstyle{arrow} = [thick,->,>=latex]
\usetikzlibrary{positioning,arrows,patterns,fit,backgrounds,calc,svg.path,3d,decorations.text,shapes.arrows,spy}
\usepgfplotslibrary{groupplots}

\usepackage{mathtools}

\DeclarePairedDelimiter\floor{\lfloor}{\rfloor}

\DeclarePairedDelimiter\abs{\lvert}{\rvert}%

\newsavebox{\measurebox}

\renewcommand\figurename{Figure}

\ifCLASSINFOpdf
\else
\fi
\pgfdeclarelayer{background}
\pgfdeclarelayer{front}
\pgfsetlayers{background,main,front}

\makeatletter
\pgfkeys{%
	/tikz/on layer/.code={
		\pgfonlayer{#1}\begingroup
		\aftergroup\endpgfonlayer
		\aftergroup\endgroup
	},
	/tikz/node on layer/.code={
		\gdef\node@@on@layer{%
			\setbox\tikz@tempbox=\hbox\bgroup\pgfonlayer{#1}\unhbox\tikz@tempbox\endpgfonlayer\egroup}
		\aftergroup\node@on@layer
	},
	/tikz/end node on layer/.code={
		\endpgfonlayer\endgroup\endgroup
	}
}

\def\node@on@layer{\aftergroup\node@@on@layer}

\begin{document}
%
\title{Resource-efficient Deep Neural Networks for Automotive Radar Interference Mitigation}
%
%
%

\author{Johanna~Rock,
        Wolfgang~Roth,
        Mate~Toth,
        Paul~Meissner,
        and~Franz~Pernkopf%
\thanks{Johanna Rock, Wolfgang Roth and Franz Pernkopf  are with the Signal Processing and Speech Communication Laboratory, 
Graz University of Technology, Austria, e-mail: \{johanna.rock, roth, pernkopf\}@tugraz.at.}
\thanks{Mate Toth and Paul Meissner are with Infineon Technologies Austria AG, Graz, e-mail: \{mate.toth, paul.meissner\}@infineon.com.}
}

%
%

\markboth{}%
{}
%



\maketitle

\begin{abstract}
Radar sensors are crucial for environment perception of driver assistance systems as well as autonomous vehicles. With a rising number of radar sensors and the so far unregulated automotive radar frequency band, mutual interference is inevitable and must be dealt with.
Algorithms and models operating on radar data are required to run the early processing steps on specialized radar sensor hardware. This specialized hardware typically has strict resource-constraints, i.e. a low memory capacity and low computational power.
Convolutional Neural Network (CNN)-based approaches for denoising and interference mitigation yield promising results for radar processing in terms of performance. Regarding resource-constraints, however, CNNs typically exceed the hardware's capacities by far.

In this paper we investigate quantization techniques for CNN-based denoising and interference mitigation of radar signals. 
We analyze the quantization of (i) weights and (ii) activations of different CNN-based model architectures. This quantization results in reduced memory requirements for model storage and during inference.
We compare models with fixed and learned bit-widths and contrast two different methodologies for training quantized CNNs, i.e. the straight-through gradient estimator and training distributions over discrete weights.
We illustrate the importance of structurally small real-valued base models for quantization and show that learned bit-widths yield the smallest models.
We achieve a memory reduction of around 80\% compared to the real-valued baseline.
Due to practical reasons, however, we recommend the use of 8 bits for weights and activations, which results in models that require only 0.2 megabytes of memory.
\end{abstract}

\begin{IEEEkeywords}
Quantization aware training, resource-efficiency, binarized convolutional neural networks, straight-through estimator, discrete weight distributions, uncertainty maps, interference mitigation, automotive radar.
\end{IEEEkeywords}

%
\IEEEpeerreviewmaketitle

\definecolor{jstspgreen}{rgb}{0.172549019607843,0.627450980392157,0.172549019607843}
\definecolor{jstspred}{rgb}{0.83921568627451,0.152941176470588,0.156862745098039}
\definecolor{jstsppurple}{HTML}{9673A6} 
\definecolor{jstspblue}{rgb}{0.12156862745098,0.466666666666667,0.705882352941177}
\definecolor{jstsporange}{rgb}{1,0.498039215686275,0.0549019607843137}
\definecolor{interferenceyellow}{rgb}{0.92900,0.69400,0.12500}%

\section{Introduction}

\begin{figure}
	\centering
	\includegraphics[width=0.95\columnwidth]{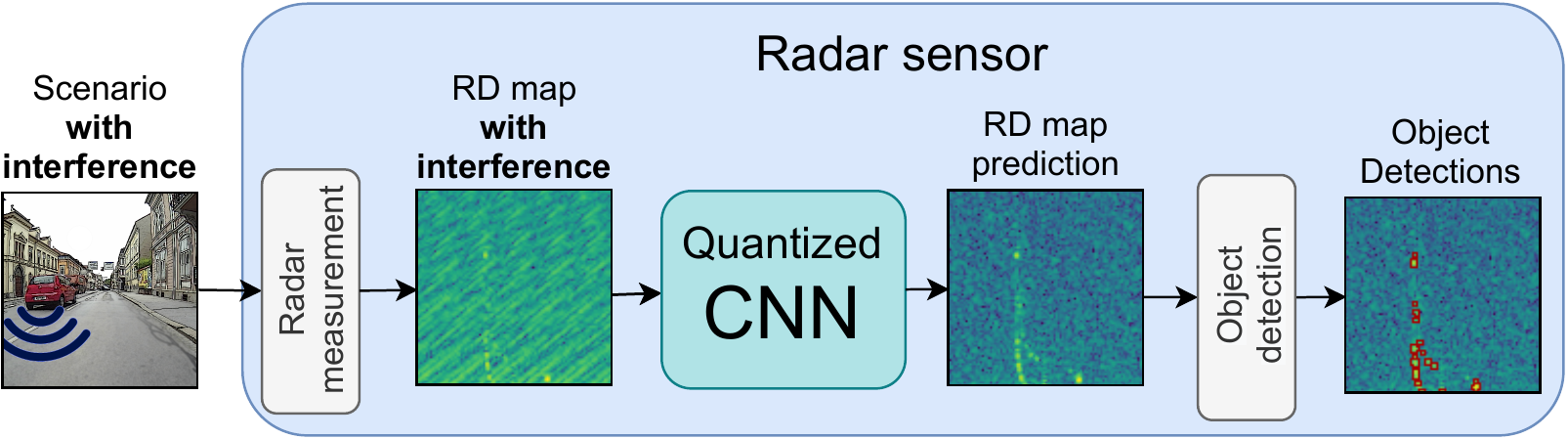}
	\captionof{figure}{Interference mitigation of radar signals using a quantized CNN to remove interference patterns, retain object signals, and provide a high detection sensitivity. To use this approach in practice, a resource-efficient model is essential.}
	\label{fig:vis_abstract}
\end{figure}

\begin{figure*}[t]
	\centering
	\footnotesize
	\resizebox{\textwidth}{!}{
		\tikzstyle{block}=[draw, fill=black!10, text width=5em, text centered, minimum height=6em]
\tikzstyle{blockfocus}=[block, draw=red, thick, dashed,rounded corners]
\tikzstyle{blockinactive}=[block, opacity=.3]

\tikzstyle{arrow}=[very thick, ->]
\tikzstyle{arrowinactive}=[arrow, opacity=.2]

\begin{tikzpicture}[scale=0.7, transform shape]
\begin{scope}[node distance=1cm and 1.9cm]

\node[block] (rs) at (0,0) {Radar Sensor};
\node[block, right = of rs] (tdp) {Time Domain Denoising (TDD)};
\node[block, right = of tdp] (dft1) {DFT over $n$ \\ for each $m$};

\node[block, right = of dft1] (rpd) {Range-Profile Denoising (RPD)};
\node[block, right = of rpd] (dft2) {DFT over $m$ \\ for each $n$};
\node[block, right = of dft2] (rdd) {Range-Doppler Denoising (RDD)};

\node[block, right = of rdd] (od) {Object Detection};
\node[block, right = of od] (ae) {Angle Estimation};
\node[block, right = of ae] (fp) {Further Processing};

\draw[arrow] (rs) -- (tdp) node[midway, above] () {$s_{\mathrm{IF}}[n,m]$} node [midway,below] () {${(N{\times}M)}$};
\draw[arrow] (tdp) -- (dft1) node[midway, above] () {$\tilde{s}_{\mathrm{IF}}[n,m]$};
\draw[arrow] (dft1) -- (rpd) node[midway, above] () {$S_{\mathrm{R}}[n,m]$};
\draw[arrow] (rpd) -- (dft2) node[midway, above] () {$\tilde{S}_{\mathrm{R}}[n,m]$};
\draw[arrow] (dft2) -- (rdd) node[midway, above] () {$S_{\mathrm{RD}}[n,m]$};
\draw[arrow] (rdd) -- (od) node[midway, above] () {$\tilde{S}_{\mathrm{RD}}[n,m]$};
\draw[arrow] (od) -- (ae) node[midway, above] () {object peaks};
\draw[arrow] (ae) -- (fp) node[midway, above] () {objects};

\draw[red,thick,dashed,rounded corners] ($(rpd.north west)+(-0.3,0.3)$)  rectangle ($(rpd.south east)+(0.3,-0.3)$);
\draw[red,thick,dashed,rounded corners] ($(rdd.north west)+(-0.2,0.2)$)  rectangle ($(rdd.south east)+(0.2,-0.2)$);

\draw[blue,thick,dashed,rounded corners] ($(tdp.north west)+(-0.2,0.2)$)  rectangle ($(tdp.south east)+(0.2,-0.2)$);
\draw[blue,thick,dashed,rounded corners] ($(rpd.north west)+(-0.2,0.2)$)  rectangle ($(rpd.south east)+(0.2,-0.2)$);

\end{scope}
\end{tikzpicture}
	}
	\caption{Block diagram of a basic FMCW/CS radar processing chain. Dashed boxes indicate the locations of optional interference mitigation steps, including CNN-based approaches ({\color{red}red}) and classical methods ({\color{blue}blue}).}
	\label{fig:spchain_classical}
	\vspace{-5mm}
\end{figure*}
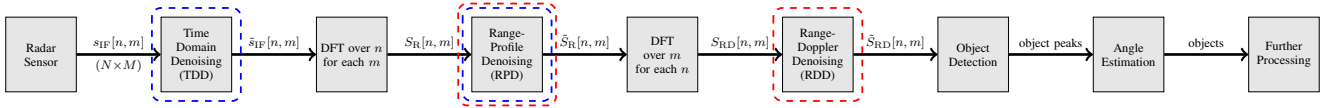

\IEEEPARstart{A}{dvanced} Driver Assistance Systems (ADAS) and Autonomous Vehicles (AV) heavily rely on a multitude of heterogeneous  sensors for environment perception. Among them are radar sensors that are used for object detection, classification and to directly measure relative object velocities. Advantages of radar sensors are a high range resolution and their robustness concerning difficult weather and lighting conditions.

Typically \emph{frequency modulated continuous wave (FMCW)/chirp sequence (CS)} radars are used in the automotive context. They transmit sequences of linearly modulated radio frequency (RF) chirps in a shared and non-regulated band. This may lead to mutual interference of multiple radar sensors; it becomes increasingly likely with a higher number of deployed radar-enhanced vehicles and larger chirp bandwidths of individual sensors used for better range-resolution.

For a non-regulated spectrum, the most common form of mutual interference is non-coherent, where the transmitters send with non-identical parameters. This results in burst-like interferences in time domain and a decreased detection sensitivity in the range-Doppler (RD) map~\cite{TOT18,Kim2018}. Thus, the detection and mitigation of interference is crucial in a safety context and must be addressed.

Several conventional signal processing algorithms for interference mitigation of mutual interference have been proposed. The most simplistic method is to substitute all interference-affected samples with zero \cite{Fischer}, followed by an optional smoothing of the boundaries. More advanced methods use nonlinear filtering in slow-time~\cite{WAG18}, iterative reconstruction using Fourier transforms and thresholding~\cite{MAR12}, estimation and subtraction of the interference component~\cite{BEC17}, an adaptive noise canceller~\cite{8651538}, or beamforming~\cite{Bechter2016}.

Recently, the use of deep learning has emerged for radar spectra denoising and interference mitigation. For this task deep neural networks (DNNs) are applied in time domain or in frequency domain, typically in a supervised manner. For interference mitigation in time domain, recurrent neural networks (RNNs) are used in \cite{8690848,9053013}. For interference mitigation in frequency domain, CNN-based models \cite{9114627,Ristea2020}, Convolutional Autoencoders \cite{9114719} and U-Net inspired CNNs \cite{9114641} are used.

While the results are impressive on simulated and real-world measurement data, the problem of high memory and computational requirements of DNN models has not been addressed in sufficient detail so far. In order to use the aforementioned methods for interference mitigation in practice, they have to comply with memory, computational as well as real-time constraints of specialized hardware, i.e. the radar sensor.

Typically, DNNs have thousands or even millions of parameters and require hundreds of megabytes memory to be stored and during computation. Note that memory is often the limiting factor also in terms of energy efficiency and execution time, because loading data dominates over arithmetic operations and loading from off-chip DRAM is magnitudes more costly than accessing data from on-chip SRAM \cite{han2015learning}. Thus, memory efficiency is particularly important for specialized embedded hardware such as radar sensors.

There are several, partly orthogonal, options to reduce memory and computational requirements. The initial network architecture contributes substantially to the resource requirements, thus a small model with few parameters and a small number of activations is preferable. \emph{Neural architecture search (NAS)} can be applied with resource-oriented objectives in order to find efficient models automatically \cite{cai2018proxylessnas}. Other approaches are network pruning techniques, weight sharing, knowledge distillation, special matrix structures and quantization \cite{roth2020resourceefficient}. In a quantized DNN, weights and activations are discretized and thus their bit-width is reduced. Typically, research on DNN quantization considers standard image classification data sets (e.g. MNIST, CIFAR-10 or ImageNet) rather than real-world data or regression tasks.

The aim of this paper (visually depicted in Figure~\ref{fig:vis_abstract}) is to build upon the approach from \cite{9114627}, and find small models with decent resource requirements that retain high interference mitigation performance. We compare two quantization techniques for CNN-based models from \cite{9114627} to reduce the total memory requirements on radar sensors. The first technique, known as quantization aware training, is based on the \emph{straight-through gradient estimator (STE)} \cite{NIPS20166573}. The second technique is based on \emph{training distributions over discrete weights} \cite{roth2020discrete}.

In our experiments we use real-world FMCW/CS radar measurements with simulated interference. The main contributions of this paper are:
\begin{itemize}
	\item We analyze the quantization capabilities according to different model architectures, sizes and quantization strategies, i.e. quantized weights, activations or both.
	\item We illustrate the importance of resource-efficient real-valued initial models w.r.t. quantization and the resulting memory requirements.
	\item We present results for quantizing exceptionally small models without significant performance degradation using fixed and learned bit-widths.
	\item We demonstrate how distributions over discrete weights can be used, in addition to denoising and interference mitigation, to obtain uncertainty estimates of the RD maps.
\end{itemize}

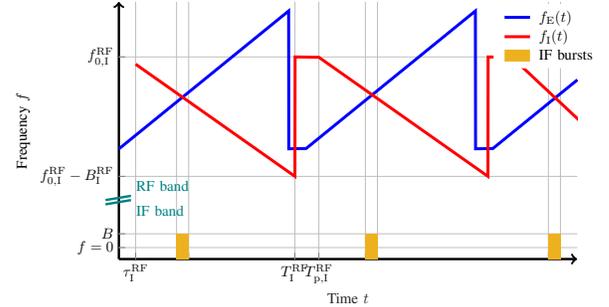
\begin{figure}[t]
	\centering
	\resizebox{0.9\columnwidth}{!} {

%
%
%
\begin{tikzpicture}

\definecolor{mycolor1}{rgb}{0.00000,0.44700,0.74100}%
\definecolor{mycolor2}{rgb}{0.85000,0.32500,0.09800}%
\definecolor{mycolor3}{rgb}{0.92900,0.69400,0.12500}%

\begin{axis}[%
scale=0.9,
width=4.725in,
height=2.654in,
at={(0.793in,0.358in)},
scale only axis,
unbounded coords=jump,
xmin=100,
xmax=2800,
xtick={\empty},
extra x ticks={1278,199,1137,440,511,1552,1623,2630,2701},	
extra x tick labels={$T^{\mathrm{RF}}_{\mathrm{p,I}}$,$\tau^{\mathrm{RF}}_{\mathrm{I}}$,$T^{\mathrm{RF}}_{\mathrm{I}}$,\empty},
extra x tick style={grid=major},
xlabel style={font=\color{white!15!black}},
xlabel={Time $t$},
ymin=-2,
ymax=26,
ytick={\empty},
extra y ticks={-0.75, 7, 20, 0.75},
extra y tick labels={$f=0$,$f^{\mathrm{RF}}_{0,\mathrm{I}} - B^{\mathrm{RF}}_{\mathrm{I}}$,$f^{\mathrm{RF}}_{0,\mathrm{I}}$,$B$},
extra y tick style={grid=major},
ylabel={Frequency $f$},
axis background/.style={fill=white},
axis x line*=bottom,
axis y line*=left,
axis line style = {ultra thick, ->},
legend style={at={(0.818,0.99)}, anchor=north west, legend cell align=left, align=left, draw=white}
]
\addplot [color=blue, line width=2.0pt]
  table[row sep=crcr]{%
99	10\\
103	10.03003003003\\
1100	25\\
1101	10\\
1203	10.03003003003\\
2200	25\\
2201	10\\
2303	10.03003003003\\
2801	17.5075075075074\\
};
\addlegendentry{$f_{\mathrm{E}}(t)$}

%

\addplot [color=red, line width=2.0pt]
  table[row sep=crcr]{%
199	19.2062062062064\\
1137	7\\
1138	20\\
1278	19.9609609609611\\
2274	7\\
2275	20\\
2415	19.9609609609611\\
2931	11.034034034034\\
};
\addlegendentry{$f_{\mathrm{I}}(t)$}


\draw[fill=mycolor3, draw = none] (440,-2) rectangle (511, 0.75);  
\draw[fill=mycolor3, draw = none] (1552,-2) rectangle (1623, 0.75);
\draw[fill=mycolor3, draw = none] (2630,-2) rectangle (2701, 0.75);

\addlegendimage{line legend, line width=8.0pt, color=mycolor3}
\addlegendentry{IF bursts}

\end{axis}

\draw[ultra thick, color = teal] (1.7,2.3) -- (2.3,2.4);
\draw[ultra thick, color = teal] (1.7,2.2) -- (2.3,2.3);
\node[color = teal, anchor = south west] at (2.3,2.4) {RF band};
\node[color = teal, anchor = north west] at (2.3,2.3) {IF band};

\end{tikzpicture}%

	}
	\caption{Non-coherent mutual FMCW/CS radar interference on the time-frequency plane. The ego radar $f_{\mathrm{E}}(t)$ and interferer $f_{\mathrm{I}}(t)$ frequency courses are depicted in blue and red, respectively. When both ramps cross, a time limited interference burst appears in the IF signal ({\color{interferenceyellow}yellow}).}
	\label{fig:fmcwcsint}
\end{figure}

\begin{figure}[t]
	\centering
	\resizebox{0.85\columnwidth}{!} {
		\input{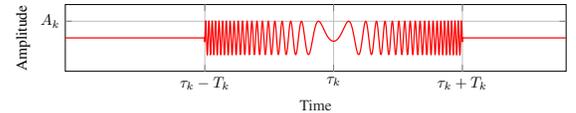}
	}
	\caption{Single, simulated and ideal interference burst in time-domain.}
	\label{fig:ideal_burst_ex}
\end{figure}

\section{Signal model} \label{sec:signal_model}

The RD processing chain of a common FMCW/CS radar is depicted in Figure~\ref{fig:spchain_classical}. The radar sensor transmits a set of linearly modulated RF chirps, also termed ramps. Object reflections are perceived by the receive antennas and mixed with the transmit signal resulting in the \emph{Intermediate Frequency (IF) Signal}. The objects' distances and velocities are contained in the sinusoidals' frequencies and their linear phase change over successive ramps~\cite{STO92,WIN07}, respectively.
The signal is processed as a $N \times M$ data matrix $s_{\mathrm{IF}}[n,m]$, containing $N$ \emph{fast time} samples for each of $M$ ramps. Discrete Fourier transforms (DFTs) are computed over both dimensions, yielding a two-dimensional spectrum, the RD map $S_{\mathrm{RD}}[n,m]$, on which peaks can be found at positions corresponding to the objects' distances and velocities. After peak detection, further processing can include angular estimation, tracking, and classification.

The IF signal $s_{\mathrm{IF}}[n,m]$ contains object reflections, noise, and may also include interference signals. It is modeled as

\begin{equation}
s_{\mathrm{IF}}[n,m]=\sum_{o=1}^{N_{\mathrm{O}}} s_{\mathrm{O},o}[n,m] + \sum_{i=1}^{N_{\mathrm{I}}} s_{\mathrm{I},i}[n,m] + \upsilon[n,m] \, ,
\label{eq:signal-model}
\end{equation}
where $s_{\mathrm{O},o}[n,m]$ are object reflections from $N_{\mathrm{O}}$ objects, $s_{\mathrm{I},i}[n,m]$ are interference signals from $N_I$ interfering radars and $\upsilon[n,m]$ models the noise.

The interference principle is illustrated in Figure~\ref{fig:fmcwcsint} on the time-frequency plane for two FMCW/CS radars. Depending on the RF transmit parameters of interferer and interfered (i.e. ego) radar, as well as the relative timing of the ramp sequences $\tau^{\mathrm{RF}}_{\mathrm{I}}$, interference bursts appear in the IF signal where the two ramp sequences cross. The characteristic 'chirp burst' form of the IF interference is due to the mixing and filtering process at the receiver. In discrete-time, after sampling with a frequency of $\frac{1}{\mathrm{T_s}}$, the resulting $K$ bursts caused by a single interferer can be parametrically modeled as
\begin{equation}\label{eq:fast_time_model}
	\footnotesize
	\begin{split}
		s_{\mathrm{I},1}[n,m] = \sum^{K}_{k=1} h[n] \ast A_{k} \cos{(-2 \pi \frac{\tau_{k}}{2T_k} n + \pi \frac{\mathrm{T_s}}{2T_k} n^2 + \varphi_{k})} \mathbbm{1}_{k}[m]
	\end{split}\,,
\end{equation}
where $h[n]$ is the combined impulse response of the radio channel and all filters in the receiver, such as a low-pass filter for anti-aliasing. $A_k, \tau_k, T_k$ and $\varphi_k$ are the amplitude, time delay, half-duration and initial phase of the $k$-th chirp burst, respectively. The indicator function $\mathbbm{1}_{k}[m]$ has a value of 1 for $m = m_k$ being the slow-time index of the k-th burst, and 0 else. More details on the chirp burst model can be found in~\cite{TOT18}; equivalent models and how they are related to RF transmit parameters are discussed in \cite{Kim2018,Fischer,WAG18,BEC17}. One such burst is illustrated in Figure~\ref{fig:ideal_burst_ex}, where an ideal $h[n] = \delta[n]$ is assumed with $\delta[n]$ being a Kronecker delta.

Classical interference mitigation methods are mostly signal processing algorithms that are applied either on the time domain signal $s_{\mathrm{IF}}[n,m]$ or on the frequency domain signal $S_{\mathrm{R}}[n,m]$ after the first DFT \cite{TOT18}. The CNN-based method used in this paper, also termed \emph{Range-Doppler Denoising (RDD)\label{rd-denoising}}, is applied on the RD map after the second DFT.

\begin{figure*}[t]
	\centering
	\begin{minipage}{.55\textwidth}
		\subfigure[CNN architecture]{
			\includegraphics[width=0.9\textwidth]{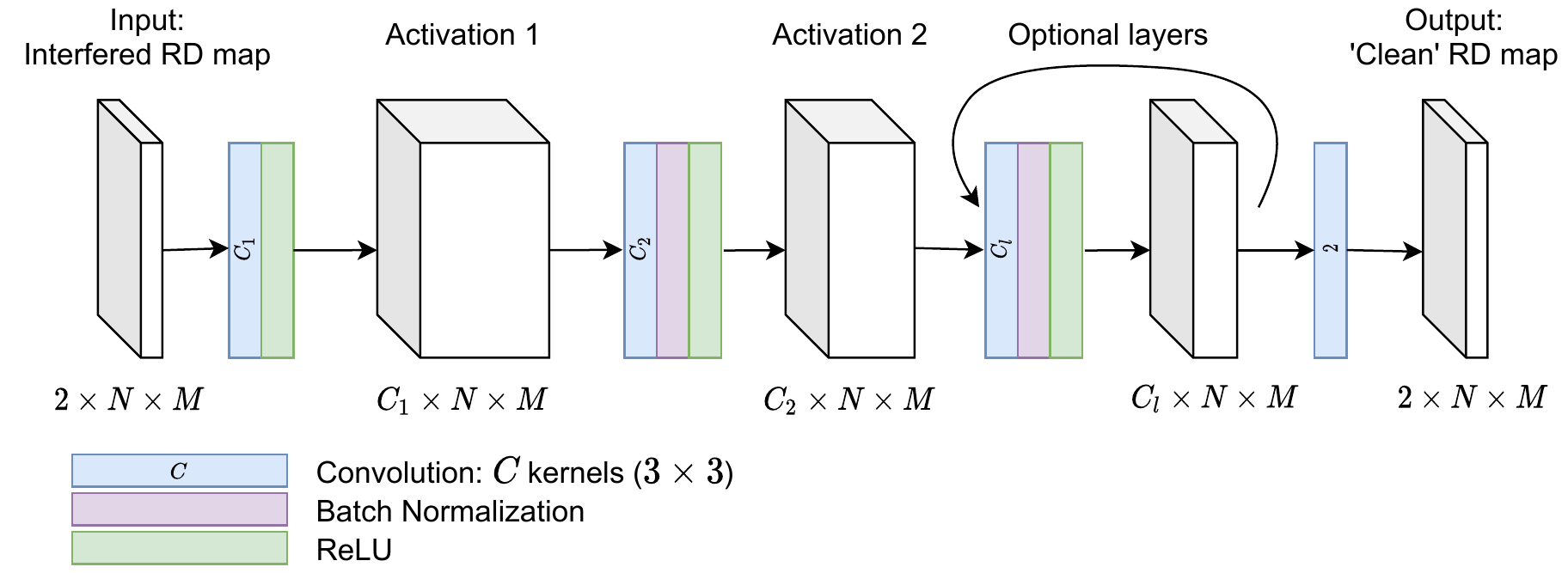}
			\label{fig:plots}
		}
	\end{minipage}%
	\begin{minipage}{0.45\textwidth}
		\centering
		\footnotesize
		\subfigure[Architecture A: same number of channels $C$ in all layers]{
			\includegraphics[width=0.85\textwidth]{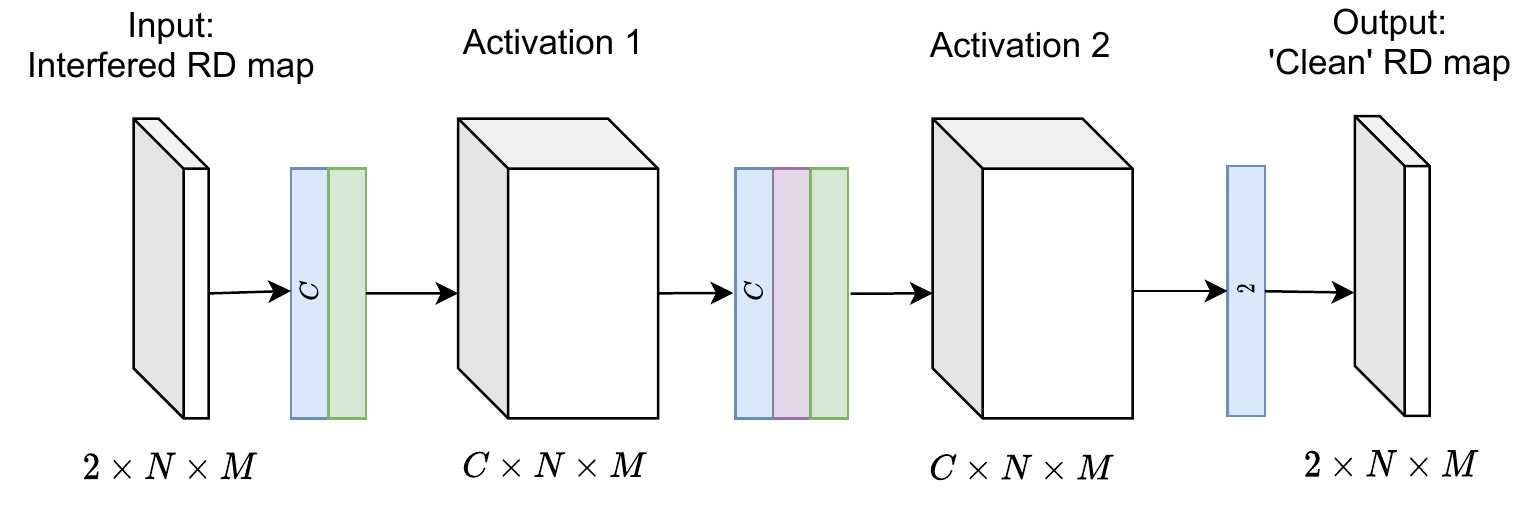}
			\label{fig:cnn_archA}
		}
		\subfigure[Architecture B: bottle-neck based architecture of channels, i.e. the number of channels $C$ is halved for each layer]{
			\includegraphics[width=0.85\textwidth]{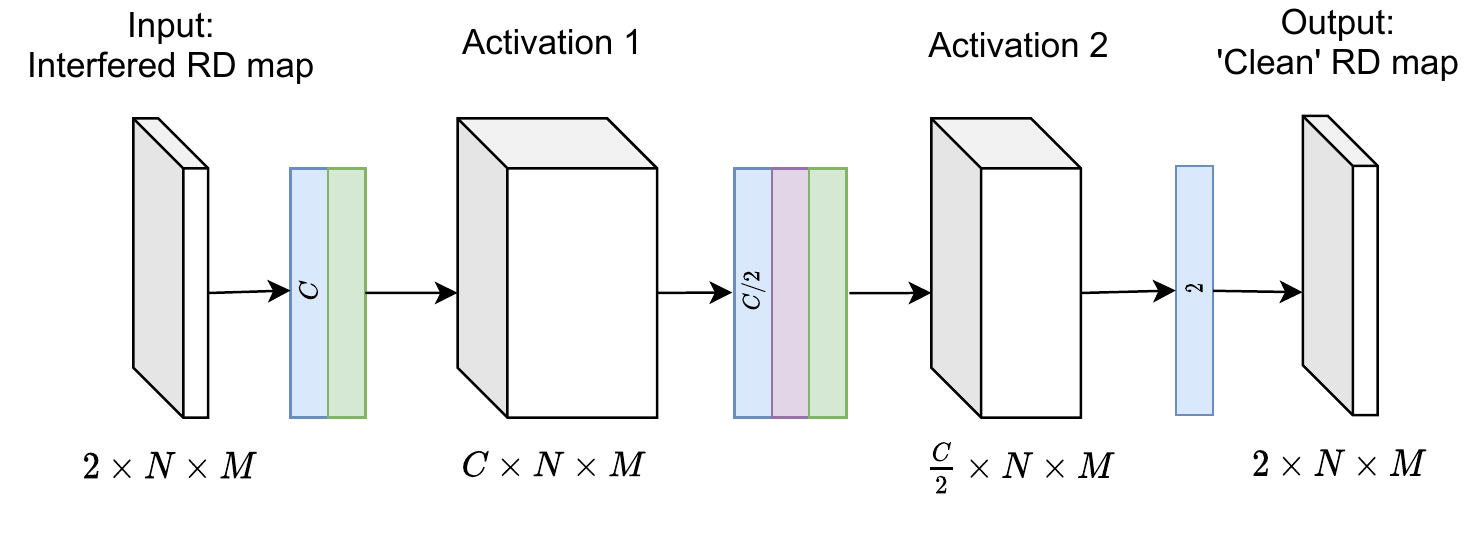}
			\label{fig:cnn_archB}
		}
	\end{minipage}
	\caption{CNN architecture for radar signal denoising and interference mitigation. It uses ReLU, \emph{Batch Normalization (BN)} and the convolution operation $\textrm{Conv}(i, o, (s_1\times s_2))$, for $i$ input channels, $o$ output channels, and a kernel size of $s_{1} \times s_{2}$. Figure (a) illustrates the general model architecture, while Figures (b) and (c) show concrete model variants with $L=3$ layers that are used in this paper.}
	\label{fig:cnn_arch_all}
\end{figure*}

\section{CNN model}
The CNN model architecture is illustrated in Figure~\ref{fig:cnn_arch_all} \cite{9114627}. The network contains $L$ layers, each being a composite function of operations including the convolution operation (Conv), ReLU activation function \cite{journals/jmlr/GlorotBB11} and \emph{Batch Normalization (BN)}. The last layer uses a linear activation function and two convolution channels\footnote[1]{Note that channels in the context of CNNs refer to the third dimension of activations in a convolution layer rather than physical or virtual signal channels as used in radar processing.} corresponding to the real and imaginary values of the complex-valued network output. From a signal processing perspective, the CNN model filters the RD map using learnable filter kernels.

The model is applied to radar snapshots for one antenna after the second DFT (RD maps), hence the input samples are complex valued patches of size $N \times M$. We use two input channels in order to represent the real and imaginary parts of the complex valued input. The network inputs are RD maps with interference and their targets are the corresponding 'clean' RD maps without interference. For the convolution, we employ square kernels and zero-padding, such that the inputs and outputs for each layer have the same spatial dimension. For training the network we use the \emph{mean squared error (MSE)} loss function and  the \emph{Adam} algorithm \cite{DBLP:journals/corr/KingmaB14}.
In this paper, we report results for two different variants of the CNN model:

\begin{description}
	\item[Architecture A] \label{sec:modelarchA} consists of the same number of channels in every layer, except for the output layer which has a fixed number of two channels. Hence, a model denoted as L3\_C32\_A has three layers with C = [32, 32, 2] channels. See Figure~\ref{fig:cnn_archA}.
	\item[Architecture B] \label{sec:modelarchB} has a bottleneck-based structure of channels where the number of channels is halved for each layer and the last layer always consists of two channels. A model denoted as L3\_C32\_B has three layers with C = [32, 16, 2] channels. See Figure~\ref{fig:cnn_archB}.
\end{description}

\section{Quantization}
The training of real-valued DNNs is typically performed using gradient-based algorithms that update the network weights according to some loss function. However, quantizers and piecewise constant activation functions are non-differentiable components, whose gradient is zero almost everywhere, such that conventional gradient-based optimization is not possible. Besides quantization of pre-trained real-valued DNNs as a post-processing step, quantization can also be incorporated in the training process by:
\begin{enumerate}
	\item \label{enum:q2} Quantization aware training using real-valued auxiliary weights and the \emph{straight-through gradient estimator (STE)} during the backward pass of quantization functions \cite{NIPS20166573}.
	\item \label{enum:q3} Inspired by methods from Bayesian inference, we can train \emph{weight distributions over discrete weights}. The most probable weights of the optimized weight distributions can be selected as discrete-valued, i.e. quantized, DNN \cite{shayer2018learning,roth2020discrete}.
\end{enumerate}
In this paper, we consider approaches \ref{enum:q2}) and \ref{enum:q3}) for trained quantization. We denote $\mathbb{D}^{D} = \{0,\pm w_1,\ldots,\pm w_{\floor*{D/2}}\}$ to be a set of symmetric discrete weights with $0 < w_1 < \ldots < w_{\floor*{D/2}}$ and a uniform step size $\delta_w = w_{d+1} - w_d$.
The quantizer $Q$ maps a real-valued number $x \in \mathbb{R}$ to one of the $D = 2^k - 1$ quantized weights $w_q \in \mathbb{D}^{D}$, assuming $k \ge 2$ bits are used to encode the weights $w_q$. The quantizer $Q$, and thus the set of discrete weights $\mathbb{D}^{D}$, are defined through three parameters: the bit-width $k \in \mathbb{N}$, the step size $\delta_w \in \mathbb{R}$ and the dynamic range $\alpha \in \mathbb{R}$. These parameters depend on each other according to $\alpha = (2^{k-1}-1) \delta_w$. Note that this implies $\alpha = w_{\floor*{D/2}}$. For binary quantization with $k = 1$ bit, we consider $\mathbb{D}^{2} = \{-w_1,+w_1\}$.

\subsection{Straight-through gradient estimator (STE)}
\label{subsec:ste}

The gradients required for gradient-based learning are typically computed using the backpropagation algorithm in a computation graph that specifies a loss function $\mathcal{L}$.
After evaluating the loss $\mathcal{L}$, backpropagation computes the gradients by repeated application of the chain rule.
It is important that all components involved during backpropagation exhibit non-zero derivatives, preventing the use of many interesting components such as piecewise constant quantizers.

The STE is a simple method to approximate the zero gradient of such components by a non-zero value.
More specifically, let $f(w)$ be some function within the computation graph with $df/dw=0$.
The STE approximates the gradient $df/dw$ during backpropagation by the non-zero derivative of a different function $\tilde{f}(w)$ with similar functional shape as $f(w)$, i.e.,
\begin{equation}
	\frac{\partial \mathcal{L}}{\partial w} = \frac{\partial \mathcal{L}}{\partial f} \frac{\partial f}{\partial w} \approx \frac{\partial \mathcal{L}}{\partial f} \frac{\partial \tilde{f}}{\partial w}.
\end{equation}
Note that for the commonly used identity function $\tilde{f}(w) = w$, the gradient is simply passed 'straight-through' to higher components in the computation graph.
Figure~\ref{fig:ste} illustrates the computation graph of the STE on a simplified convolutional layer with sign activation function. In the forward pass, the piecewise constant quantization and activation functions are applied, while their zero gradients are avoided during the backward pass. The gradient updates are then applied to the real-valued auxiliary weights.

{
\definecolor{colorredborder}{HTML}{9673A6}
\definecolor{colorgreenborder}{HTML}{82B366}
\begin{figure}
	\centering
	\footnotesize
	\includegraphics[width=0.86\columnwidth]{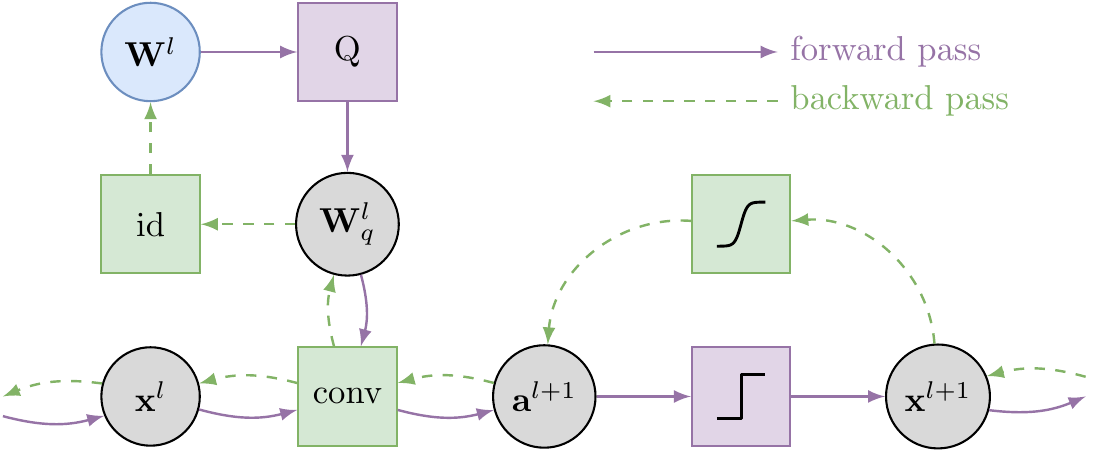}
	\caption{Computation of forward {\color{colorredborder}(purple)} and backward {\color{colorgreenborder}(green)} pass through a simplified DNN building block using the STE. The building block consists of a convolution with quantized weights followed by a sign activation function. In the forward pass of layer $l$ the convolution of the inputs $\textbf{x}^l$ and the quantized weights $\textbf{W}^l_q$ is performed. The quantization function $Q$ is applied to the real-valued auxiliary weights $\textbf{W}^l$ to retrieve the quantized weights $\textbf{W}^l_q$. The sign activation function is applied to the activations $\textbf{a}^{l+1}$ resulting in the next layer's inputs $\textbf{x}^{l+1}$. During backpropagation the green dashed line is followed, where the zero-gradient components {\color{colorredborder}(purple)} are avoided and substituted by the gradients of the tanh and identity, respectively. The gradient updates are then applied to the real-valued auxiliary weights $\textbf{W}^l$ according to the gradient based learning algorithm.}
	\label{fig:ste}
\end{figure}
}

\subsection{Quantization aware training using the STE}
\label{sec:quantization_fun}
Quantization aware training \cite{NIPS20166573} uses auxiliary real-valued weights and the STE to approximate the gradient of zero-gradient DNN components during the backward pass. Piecewise constant quantization functions are used to discretize real-valued weights or activations. While some quantization functions (e.g. sign) map to a fixed finite set of values and therefore result in a specific bit-width of discretized values, other quantization functions (e.g. rounding) can be applied with a variable bit-width $k \ge 2$.
Let $\alpha \in \mathbb{R}$ be the dynamic range. Then we can define the following quantization schemes:

\begin{enumerate}[label=(\roman*)]
	\item \textbf{Binary:} sign (1 bit)
	\[
	Q_B(x)= 
	\begin{cases}
	+\alpha,	& \text{if } x\geq 0\\
	-\alpha,	& \text{if } x < 0
	\end{cases}
	\]
	\item \textbf{Integer:} rounding (bit-width $k \ge 2$)
	\[
	Q_{I,k}(x) = \textrm{clip}\left(\textrm{round}\left(\frac{x}{\alpha}\right), -2^{k-1}+1, 2^{k-1}-1\right) \cdot \alpha
	\]
	where $\textrm{round}(x / \alpha)$ maps $x / \alpha$ to the closest integer value and $\textrm{clip}(x, l, u) = \max(\min(x, u), l)$, ensuring that $Q_{I,k}(x) \in \{-w_{\floor*{D/2}},\cdots, +w_{\floor*{D/2}}\}$. This quantization scheme is termed \emph{integer} quantization, because integer rounding is an essential component.
\end{enumerate}

\subsubsection{Dynamic range}
\label{sec:dynamic_range}

The dynamic range $\alpha$ \cite{uhlich2019differentiable,jain2019trained,esser2019learned} is used to map integer weights $\textbf{W}_q \in \mathbb{D}^D$, encoded with $k$ bits, to real-valued numbers. The integer weights $\textbf{W}_q$ are stored in addition to one real-valued number per layer, i.e. the dynamic range, and scaled according to that dynamic-range $\alpha$. This scaling operation typically increases the model performance considerably and is thus often used in practice. Note that the memory requirements for the dynamic range can be neglected, because only one 32 bit value is stored per layer. In this paper, we consider two methods for determining the dynamic range, i.e.

\begin{enumerate}[label=(\roman*)]
	\item \textbf{Statistics approach:} Maximum absolute weight value
	\[
	\alpha = \max(\abs{\textbf{W}})),
	\]
	where $\textbf{W}$ are the auxiliary weights when using quantization aware training with the STE.
	\item \textbf{Learned approach:} Learns $\alpha$ as additional model parameter using a gradient based optimization algorithm.
\end{enumerate}

\subsubsection{Fixed and learned bit-width}
\label{sec:fixed_learned_bitwidth}

Quantization to discrete values with multiple bits (e.g. rounding) can be performed using either a fixed or a learned bit-width. Fixed bit-width quantization typically uses one homogeneous bit-width $k$ for all layers, which is defined prior to training.

Recent works \cite{uhlich2019differentiable} support heterogeneous bit-widths\footnote[2]{Also known as mixed precision quantization in the literature.}, which can be learned alongside the model weights using the STE.
This approach enables the use of different bit-widths for each layer without introducing additional hyperparameters. Note that manual hyperparameter search of layer-wise bit-widths would span over a space that is exponential in the number of layers which is generally intractable. Essentially, the bit-width $k$ becomes a trainable parameter and is optimized either directly, or implicitly through learning the step size $\delta_w$ and the dynamic range $\alpha$. For training these additional model parameters, they are incorporated in the computation graph and trained via backpropagation using the STE.

When the bit-widths are trainable parameters, we use an additional loss term favoring fewer bits, i.e. the average bit-width of weights, activations, or both. By weighting this average bit-width loss by the corresponding frequencies, i.e. the number of weights or activations per layer, this average weighted bit-width loss becomes directly proportional to the resulting overall memory requirements. In order to properly scale this bit-width loss w.r.t. the performance loss (e.g. the mean squared error), an adaptive scaling factor can be used, e.g. dependent on the validation F1-Score.

\subsection{Training distributions over discrete weights}
\label{subsec:distr}

\begin{figure}
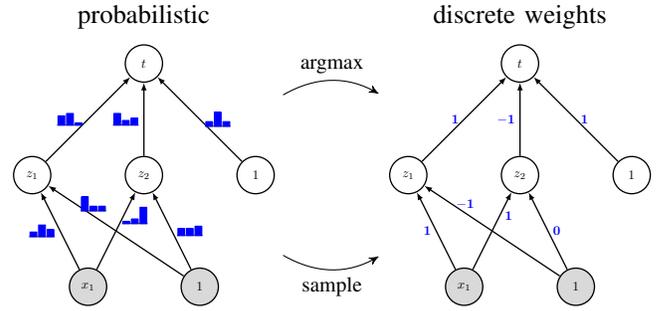

	\centering
	\footnotesize
		\include{big_picture_tikz}
	\caption{Overview of \emph{training distributions over discrete weights} for trained quantization. Model weights are replaced by distributions over discrete weights (left) and optimized using a gradient-based optimization algorithm. The discrete-valued model is then obtained by sampling or by selecting the most probable weights (right).}
	\label{fig:distr_training}
\end{figure}

An alternative approach to quantization aware training is obtained by training a distribution over discrete weights \cite{shayer2018learning,kingma2015reparameterization,roth2020discrete}.
Let $q_{\boldsymbol{\nu}}(\mathbf{W}_q)$ be a discrete distribution over the weights $\mathbf{W}_q$ governed by \emph{continuous} parameters $\boldsymbol{\nu}$.
Moreover, assume that the individual weights $w_q \in \mathbf{W}_q$ are independent such that $q_{\boldsymbol{\nu}}(\mathbf{W}_q)$ factorizes into a product of factors $q_{\nu_w}(w_q)$ for the individual weights $w_q$, each governed by its own parameters $\nu_w$.
Instead of learning the discrete weights directly, the idea is to first train the distribution $q_{\boldsymbol{\nu}}(\mathbf{W}_q)$ by optimizing a loss $\mathcal{L}(\boldsymbol{\nu})$ that is differentiable with respect to the distribution parameters $\boldsymbol{\nu}$.
After training, a DNN with discrete weights $\mathbf{W}_q$ is obtained by either selecting the most probable weights $\arg \max_{\mathbf{W}_q} q_{\boldsymbol{\nu}}(\mathbf{W}_q)$ or by sampling from $q_{\boldsymbol{\nu}}(\mathbf{W}_q)$.
This process is illustrated for ternary weights in Figure~\ref{fig:distr_training} where, intuitively, the probability bars at each connection correspond to the trainable parameters $\nu_w$.
An important property of weight distributions is that they allow us to obtain \emph{prediction uncertainties} by averaging the outputs of several DNNs whose weights are sampled from $q_{\boldsymbol{\nu}}(\mathbf{W}_q)$.

For the definition of the loss $\mathcal{L}(\boldsymbol{\nu})$, assume that we are given a loss over the weights $\mathcal{L}(\mathbf{W}_q)$.
We can then define a loss as expectation over $\mathcal{L}(\mathbf{W}_q)$ with respect to the distribution $q_{\boldsymbol{\nu}}(\mathbf{W}_q)$, i.e.,
\begin{equation}
  \mathcal{L}(\boldsymbol{\nu}) = \mathbb{E}_{q(\mathbf{W}_q|q_{\boldsymbol{\nu}})}[\mathcal{L}(\mathbf{W}_q)]  + \lambda r(\boldsymbol{\nu}),
\label{eq:l_prob}
\end{equation}
where $r(\boldsymbol{\nu})$ is a regularizer for the distribution parameters $\boldsymbol{\nu}$ and $\lambda \ge 0$ is a tunable hyperparameter.

However, the expectation in \eqref{eq:l_prob} is a sum over exponentially many terms, which is generally intractable.
We use a practical approximation for the expected loss based on the central limit theorem that has been widely used in the literature \cite{NIPS2014_5269,pmlr-v37-hernandez-lobatoc15,roth2020discrete,shayer2018learning,peters2018probabilistic}.
The central limit theorem states that the average over many independent random variables tends towards a Gaussian distribution.
This is particularly convenient for DNNs with weight distributions as each neuron computes a sum over many random variables.
This allows us to approximate the distribution of the $i$\textsuperscript{th} activation $a_i^{l+1}$ in layer $l+1$ by a Gaussian $\mathcal{N}(\mu_{a_i^{l+1}}, \sigma_{a_i^{l+1}}^2)$ with
\begin{equation}
  \mu_{a_i^{l+1}} = \sum_{j} \mathbb{E}[ w_{i,j}^{l+1} ] x_j^{l} \quad \text{and} \quad \sigma_{a_i^{l+1}}^2 = \sum_j \mathbb{V}[w_{i,j}^{l+1}](x_j^{l})^2,
\end{equation}
where $x_j^{l}$ is the $j$\textsuperscript{th} input from the previous layer.
In the next step, the Gaussian activation distributions are converted into deterministic values by a backpropagation-compatible sampling procedure known as the local reparameterization trick \cite{NIPS2015_5666,shayer2018learning,peters2018probabilistic}.

A repeated application of these steps, i.e., Gaussian approximation followed by the local reparameterization trick, leads to a tractable approximation of $\mathcal{L}(\boldsymbol{\nu})$.
Note that this approach can also be used for piecewise constant activation functions such as the sign function \cite{peters2018probabilistic,roth2020discrete}.
In this case, the local reparameterization trick must be applied \emph{after} the activation function to maintain a differentiable loss $\mathcal{L}(\boldsymbol{\nu})$.
This introduces some subtleties in the training process; most notably, the local reparameterization trick is applied at the \emph{discrete} sign activation distribution requiring methods such as the Gumbel softmax approximation \cite{jang2016categorical,maddison2016concrete}.

In practice, it is common to store the parameters $\boldsymbol{\nu}$ as unnormalized log-probabilities.
Shayer et al.\ \cite{shayer2018learning} proposed to use the squared $\ell^2$ norm as the regularizer $r(\boldsymbol{\nu})$.
This enforces the distribution $q_{\nu_w}(w_q)$ to become more uniform and, therefore, to exhibit increased variance and entropy.
Hence, the hyperparameter $\lambda$ controls the level of variability of $q_{\nu_w}(w_q)$ and, consequently, also the amount of prediction uncertainty obtained by sampling from $q_{\boldsymbol{\nu}}(\mathbf{W}_q)$.

For training, the distribution parameters $\boldsymbol{\nu}$ are initialized using real-valued weights $\mathbf{W}$ from a pre-trained DNN with the same architecture according to the method presented in \cite{roth2020discrete}.
In this paper, we consider training distributions over ternary weights with $\mathbb{D}^{3} = \{-\alpha,0,+\alpha\}$, requiring $k = 2$ bits per weight. In this context, the dynamic range $\alpha$ can be seen as a simple scaling factor for ternary integer weights $w \in \{-1,0,+1\}$. It is computed based on the real-valued weights $\mathbf{W}$ from a pre-trained DNN, or trained as a learnable parameter. Note that most Bayesian inspired approaches for quantization \cite{roth2020discrete,shayer2018learning, kingma2015reparameterization} use a dynamic range of $\alpha = 1$, hence no scaling is performed on the integer weights. 

\section{Experimental setup}
We use real-world FMCW/CS radar measurements combined with simulated interference to obtain input-output pairs for training CNN models in order to perform the denoising and interference mitigation tasks. The model is applied to the processed radar signal after the second DFT, i.e. the RD map. The overall goal is the correct detection of peaks in the RD map that correspond to real objects rather than clutter or noise.

\subsection{Data set} \label{sec:dataset}
The measurements were recorded in typical inner-city scenarios, where each measurement consists of 32 consecutive radar snapshots (RD maps). We used only the first receive antenna for constructing the data set. The radar signal contains reflections from static and moving objects as well as receiver noise. The simulated interference, that is added to the time domain measurement signal according to \eqref{eq:signal-model}, is generated by sampling uniformly from the ego radar and interferer radar transmit parameters. See \cite{Rock1907:Complex} for a detailed description of the simulation parameters and~\cite{9114627, toth2020analysis} for an extensive analysis of the used measurement signals. The data set splits for training, validation and testing contain 2500, 250 and 250 RD maps, respectively. Data set splits are strictly non-overlapping, hence, multiple RD maps belonging to snapshots within the same measurement cycle are never contained in multiple data set splits.

\emph{Validation of simulated interference:} For model training and evaluation we only use simulated interference signals because of the immense effort that is required for recording large amounts of CNN input-output pairs of synchronized real-world measurements with and without interference. In the remainder of this section, we qualitatively analyze the validity of our simulated interference signals in order to show, that they indeed have the same characteristics as interference measurements and thus qualify for the proxy task. Therefore, we recorded a real sensor interference measurement for comparison. It was conducted in a static environment, such that most of the object signal can be extracted in a post-processing step and only the inference signal plus noise remain. We used the same sensor with identical configuration parameters for the interference measurement and the inner-city measurement campaign.

The simulated signal is generated according to~\eqref{eq:fast_time_model}. While the RF transmit signal parameters of ego radar and interferer are known, the relative timing as well as the effects of the filters $h[n]$ are unknown. For an adequate reconstruction, the unknown parameters in ~\eqref{eq:fast_time_model} are estimated from the measured signal.

\begin{figure}[t]
	\tiny
	\centering
	\resizebox{\columnwidth}{!} {
		\centering
		\raisebox{6.5mm}{\rotatebox{90}{\textbf{Measured interference}}}
		\subfigure[]{
			\input{td_meas.tex}
			\label{fig:interf_td_meas}
		}
		\subfigure[]{
\begin{tikzpicture}

\begin{axis}[
width=2cm,
height=2cm,
scale only axis,
axis background/.style={fill=white!89.80392156862746!black},
axis line style={white},
point meta max=0,
point meta min=-30,
tick align=inside,
tick pos=left,
x grid style={white},
xlabel={Velocity [m/s]},
xmajorgrids,
xmin=-16, xmax=15.6666666666667,
xtick style={color=white!33.33333333333333!black},
y grid style={white},
ylabel={Distance [m]},
ymajorgrids,
ymin=6, ymax=60,
ytick style={color=white!33.33333333333333!black}
]
\addplot graphics [includegraphics cmd=\pgfimage,xmin=-16, xmax=15.6666666666667, ymin=6, ymax=60] {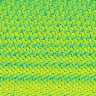};
\end{axis}

\end{tikzpicture}
			\label{fig:interf_rd_meas}
		}
		\subfigure[]{
\begin{tikzpicture}

\begin{axis}[
width=2cm,
height=2cm,
scale only axis,
axis background/.style={fill=white!89.80392156862746!black},
axis line style={white},
point meta max=0,
point meta min=-30,
tick align=inside,
tick pos=left,
x grid style={white},
xlabel={Velocity [m/s]},
xmajorgrids,
xmin=-16, xmax=15.6666666666667,
xtick style={color=white!33.33333333333333!black},
y grid style={white},
ylabel={Distance [m]},
ymajorgrids,
ymin=6, ymax=60,
ytick style={color=white!33.33333333333333!black}
]
\addplot graphics [includegraphics cmd=\pgfimage,xmin=-16, xmax=15.6666666666667, ymin=6, ymax=60] {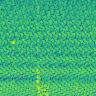};
\end{axis}

\end{tikzpicture}
			\label{fig:interf_rd_mix_meas}
		}
	}
	\resizebox{\columnwidth}{!} {
		\centering
		\raisebox{6.5mm}{\rotatebox{90}{\textbf{Simulated interference}}}
		\subfigure[]{
%
%
\definecolor{mycolor1}{rgb}{1.00000,0.00000,1.00000}%
\begin{tikzpicture}

\begin{axis}[%
width=2cm,,
height=2cm,
scale only axis,
xmin=0.600000000000023,
xmax=12.775,
xlabel style={font=\color{white!15!black}},
xlabel={Time [$\mu$s]},
ymin=-1200,
ymax=1200,
ylabel style={font=\color{white!15!black}, at={(-0.2, 0.5)}},
ylabel={Amplitude},
axis background/.style={fill=white},
axis x line*=bottom,
axis y line*=left,
xmajorgrids,
ymajorgrids
]
\addplot [color=red, semithick, forget plot]
  table[row sep=crcr]{%
0.600000000000023	0\\
3.54999999999995	0\\
3.57500000000005	0.131646233972106\\
3.625	0.0631336955920005\\
3.64999999999998	0.209963977771167\\
3.67499999999995	0.299808493386308\\
3.70000000000005	0.171067624280568\\
3.72500000000002	0.165607446882973\\
3.75	0.429311940755838\\
3.77499999999998	0.494771351375675\\
3.79999999999995	0.221262844249054\\
3.82500000000005	0.233661343237372\\
3.85000000000002	0.676752170894815\\
3.875	0.72428616430409\\
3.89999999999998	0.216731230566893\\
3.92499999999995	0.238282860777076\\
3.95000000000005	0.973498818016651\\
3.97500000000002	1.00421434590442\\
4	0.10711458743458\\
4.02499999999998	0.131632935661855\\
4.04999999999995	1.35421334983073\\
4.07500000000005	1.36473138943313\\
4.10000000000002	-0.186851052782345\\
4.125	-0.164028494281752\\
4.14999999999998	1.87959336252277\\
4.17499999999995	1.86206966530006\\
4.20000000000005	-0.79201236473898\\
4.22500000000002	-0.774108431371815\\
4.25	2.65815601043403\\
4.27499999999998	2.59957570771826\\
4.29999999999995	-1.91279546809778\\
4.32500000000005	-1.90068171817461\\
4.35000000000002	3.88378553850134\\
4.375	3.76442540019593\\
4.39999999999998	-3.88053015522519\\
4.42499999999995	-3.87038779306488\\
4.45000000000005	5.89986708010656\\
4.47500000000002	5.69041183514355\\
4.5	-7.23517677956102\\
4.52499999999998	-7.21414969363229\\
4.54999999999995	9.30725423235515\\
4.57500000000005	8.96471883748097\\
4.60000000000002	-12.8614596721008\\
4.625	-12.8004599763145\\
4.64999999999998	15.1470414735443\\
4.67499999999995	14.6013871267288\\
4.70000000000005	-22.2166458522806\\
4.72500000000002	-22.0333766034577\\
4.75	25.18636083094\\
4.77499999999998	24.2728015960141\\
4.79999999999995	-37.6219550311973\\
4.82500000000005	-37.0986329105783\\
4.85000000000002	42.1518874941569\\
4.875	40.641316356203\\
4.89999999999998	-62.259082698649\\
4.92499999999995	-61.4778187718291\\
4.95000000000005	69.3134962706945\\
4.97500000000002	68.3226513061945\\
5	-99.0407406861112\\
5.02499999999998	-101.621420379249\\
5.04999999999995	108.344851587449\\
5.07500000000005	116.586477102351\\
5.10000000000002	-146.882254164527\\
5.125	-170.654475081698\\
5.14999999999998	153.166150065368\\
5.17499999999995	203.566319047226\\
5.20000000000005	-190.600245198924\\
5.22500000000002	-291.250021353335\\
5.25	174.026450173466\\
5.27499999999998	353.877167939932\\
5.29999999999995	-177.493810305277\\
5.32500000000005	-480.354704211432\\
5.35000000000002	87.8869068161041\\
5.375	558.50127410024\\
5.39999999999998	9.21405959539914\\
5.42499999999995	-669.598018173723\\
5.45000000000005	-239.22862000795\\
5.47500000000002	648.552051839184\\
5.5	469.539065012968\\
5.52499999999998	-556.3880506172\\
5.54999999999995	-765.767686183856\\
5.57500000000005	217.474497340194\\
5.60000000000002	876.322011889491\\
5.625	212.148577778279\\
5.64999999999998	-779.542682023707\\
5.67499999999995	-701.553727575321\\
5.70000000000005	272.231625252584\\
5.72500000000002	837.76358517081\\
5.75	364.278833491479\\
5.77499999999998	-500.331879308815\\
5.79999999999995	-797.535452712444\\
5.82500000000005	-282.283905232715\\
5.85000000000002	493.335200015208\\
5.875	791.039163201635\\
5.89999999999998	343.162792556398\\
5.92499999999995	-432.801503143575\\
5.95000000000005	-830.170713543724\\
5.97500000000002	-510.405561699913\\
6	223.616300558814\\
6.02499999999998	769.25847046692\\
6.04999999999995	752.787545761508\\
6.07500000000005	244.099803145286\\
6.10000000000002	-391.557155944826\\
6.125	-761.833671865414\\
6.14999999999998	-661.572075151034\\
6.17499999999995	-167.229780927281\\
6.20000000000005	431.445254119365\\
6.22500000000002	824.36935772727\\
6.25	852.443430155623\\
6.27499999999998	546.761603362358\\
6.32500000000005	-397.189074836804\\
6.35000000000002	-671.78370680272\\
6.375	-680.578771046893\\
6.39999999999998	-447.368337067698\\
6.47500000000002	709.112887140168\\
6.5	937.213711061611\\
6.52499999999998	1009.6301683677\\
6.54999999999995	938.468254668296\\
6.57500000000005	760.788139610056\\
6.67499999999995	-187.584435251235\\
6.70000000000005	-340.075263912159\\
6.72500000000002	-439.505936168617\\
6.75	-490.337323613336\\
6.77499999999998	-500.644196053388\\
6.79999999999995	-481.058801418024\\
6.82500000000005	-442.552997863661\\
6.89999999999998	-298.105128503864\\
6.92499999999995	-260.800937383148\\
6.95000000000005	-235.333613101285\\
6.97500000000002	-223.456147073902\\
7	-225.98111184836\\
7.02499999999998	-242.732434254097\\
7.04999999999995	-272.575730990019\\
7.07500000000005	-313.375212338444\\
7.14999999999998	-457.489626087257\\
7.17499999999995	-490.566740058245\\
7.20000000000005	-501.022441834702\\
7.22500000000002	-477.706924480631\\
7.25	-410.692800295109\\
7.27499999999998	-293.063192814823\\
7.29999999999995	-121.970546901545\\
7.32500000000005	98.2600695878614\\
7.39999999999998	832.013471740996\\
7.42499999999995	977.549234735518\\
7.45000000000005	1002.03233242019\\
7.47500000000002	874.926566999989\\
7.5	596.811868302342\\
7.57500000000005	-551.124209940863\\
7.60000000000002	-708.164752254904\\
7.625	-604.501916310909\\
7.64999999999998	-250.554743693824\\
7.70000000000005	682.991121298471\\
7.72500000000002	885.660204110437\\
7.75	725.026033622364\\
7.77499999999998	232.361827013282\\
7.79999999999995	-368.690299349067\\
7.82500000000005	-751.062840211831\\
7.85000000000002	-681.415493135705\\
7.875	-182.445495958331\\
7.89999999999998	457.562612274608\\
7.92499999999995	826.732331569347\\
7.95000000000005	633.085014241139\\
8	-696.151647739183\\
8.02499999999998	-769.609324791801\\
8.04999999999995	-171.005000533679\\
8.07500000000005	570.175975324564\\
8.10000000000002	773.329472947614\\
8.125	244.238813757202\\
8.14999999999998	-530.980150824723\\
8.17499999999995	-794.078549963213\\
8.20000000000005	-222.953784001908\\
8.22500000000002	626.727530287908\\
8.25	756.446972232871\\
8.29999999999995	-879.313629795829\\
8.32500000000005	-505.686407228577\\
8.35000000000002	566.315552598073\\
8.375	787.659918438125\\
8.39999999999998	-184.679656745916\\
8.42499999999995	-860.901054957913\\
8.45000000000005	-206.689442364487\\
8.47500000000002	691.392020823359\\
8.5	398.672048507962\\
8.52499999999998	-520.043283919965\\
8.54999999999995	-522.746175918566\\
8.57500000000005	296.99368317997\\
8.60000000000002	495.48118925164\\
8.625	-175.249819431622\\
8.64999999999998	-474.200350983746\\
8.67499999999995	48.3749712714578\\
8.70000000000005	381.676539504891\\
8.72500000000002	-11.7658335699102\\
8.75	-334.732632431954\\
8.79999999999995	252.044868356889\\
8.82500000000005	34.1511844587296\\
8.85000000000002	-216.771589688037\\
8.875	-52.472893980698\\
8.89999999999998	157.777392374875\\
8.92499999999995	34.2813535076049\\
8.95000000000005	-136.027214222259\\
8.97500000000002	-40.5799139332596\\
9	96.6231146942096\\
9.02499999999998	23.803899814921\\
9.04999999999995	-83.7294849298036\\
9.07500000000005	-26.5100263037519\\
9.10000000000002	58.2971855364514\\
9.125	14.5978876328143\\
9.14999999999998	-50.5680845852825\\
9.17499999999995	-16.0780646961093\\
9.20000000000005	34.8614899187984\\
9.22500000000002	8.74864459841115\\
9.25	-29.9916728338795\\
9.27499999999998	-9.36514287010039\\
9.29999999999995	20.9037374607053\\
9.32500000000005	5.43399122594428\\
9.35000000000002	-17.4700041316887\\
9.375	-5.21012783817207\\
9.39999999999998	12.7366296919179\\
9.42499999999995	3.60393955705354\\
9.45000000000005	-9.92932510865228\\
9.47500000000002	-2.65637497666148\\
9.5	7.97794250480945\\
9.52499999999998	2.58548886035578\\
9.54999999999995	-5.41321158340872\\
9.57500000000005	-1.10384893346509\\
9.60000000000002	5.19181736088831\\
9.625	2.00315771020416\\
9.64999999999998	-2.72941326628313\\
9.67499999999995	-0.181184816329733\\
9.70000000000005	3.54133954832218\\
9.72500000000002	1.65018462055025\\
9.75	-1.15640598027437\\
9.77499999999998	0.345670971446225\\
9.79999999999995	2.54294480861404\\
9.82500000000005	1.41507401214926\\
9.85000000000002	-0.256854347158765\\
9.875	0.624213255066934\\
9.89999999999998	1.91838969584501\\
9.92499999999995	1.23866044521128\\
9.95000000000005	0.233377807555939\\
9.97500000000002	0.746169807153592\\
10	1.50695597604658\\
10.025	1.08854212503024\\
10.05	0.471504944751928\\
10.075	0.766118872136076\\
10.1	1.2137509715144\\
10.125	0.944350898788457\\
10.15	0.547341427281481\\
10.175	0.712869301652177\\
10.2	0.979723604549918\\
10.225	0.789548954032739\\
10.25	0.506928525934427\\
10.275	0.596623306832498\\
10.3	0.76477069177065\\
10.325	0.607239619061261\\
10.35	0.366696365513121\\
10.375	0.413711094206974\\
10.4	0\\
12.775	-0\\
};
\end{axis}
\end{tikzpicture}%
			\label{fig:interf_td_sim}
		}
		\subfigure[]{
\begin{tikzpicture}

\begin{axis}[
width=2cm,
height=2cm,
scale only axis,
axis background/.style={fill=white!89.80392156862746!black},
axis line style={white},
point meta max=0,
point meta min=-30,
tick align=inside,
tick pos=left,
x grid style={white},
xlabel={Velocity [m/s]},
xmajorgrids,
xmin=-16, xmax=15.6666666666667,
xtick style={color=white!33.33333333333333!black},
y grid style={white},
ylabel={Distance [m]},
ymajorgrids,
ymin=6, ymax=60,
ytick style={color=white!33.33333333333333!black}
]
\addplot graphics [includegraphics cmd=\pgfimage,xmin=-16, xmax=15.6666666666667, ymin=6, ymax=60] {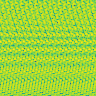};
\end{axis}

\end{tikzpicture}
			\label{fig:interf_rd_sim}
		}
		\subfigure[]{
\begin{tikzpicture}

\begin{axis}[
width=2cm,
height=2cm,
scale only axis,
axis background/.style={fill=white!89.80392156862746!black},
axis line style={white},
point meta max=0,
point meta min=-30,
tick align=inside,
tick pos=left,
x grid style={white},
xlabel={Velocity [m/s]},
xmajorgrids,
xmin=-16, xmax=15.6666666666667,
xtick style={color=white!33.33333333333333!black},
y grid style={white},
ylabel={Distance [m]},
ymajorgrids,
ymin=6, ymax=60,
ytick style={color=white!33.33333333333333!black}
]
\addplot graphics [includegraphics cmd=\pgfimage,xmin=-16, xmax=15.6666666666667, ymin=6, ymax=60] {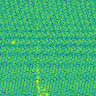};
\end{axis}

\end{tikzpicture}
			\label{fig:interf_rd_mix_sim}
		}
	}
	\resizebox{\columnwidth}{!} {
		\centering
		\resizebox{0.42\columnwidth}{!} {
			\subfigure[]{
\begin{tikzpicture}

\begin{axis}[
width=2cm,
height=2cm,
scale only axis,
axis background/.style={fill=white!89.80392156862746!black},
axis line style={white},
colorbar,
colorbar style={
	yticklabel style={
		text width=width("$-30$"),
		align=right
	},
	ylabel={dB}
},
colorbar/width=1.5mm,
colormap/viridis,
point meta max=0,
point meta min=-30,
tick align=inside,
tick pos=left,
x grid style={white},
xlabel={Velocity [m/s]},
xmajorgrids,
xmin=-16, xmax=15.6666666666667,
xtick style={color=white!33.33333333333333!black},
y grid style={white},
ylabel={Distance [m]},
ymajorgrids,
ymin=6, ymax=60,
ytick style={color=white!33.33333333333333!black}
]
\addplot graphics [includegraphics cmd=\pgfimage,xmin=-16, xmax=15.6666666666667, ymin=6, ymax=60] {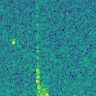};
\end{axis}

\end{tikzpicture}
				\label{fig:interf_rd_clean}
			}
		}
		\subfigure[]{
			\raisebox{4.7mm}{
				\includegraphics[width=4.2cm]{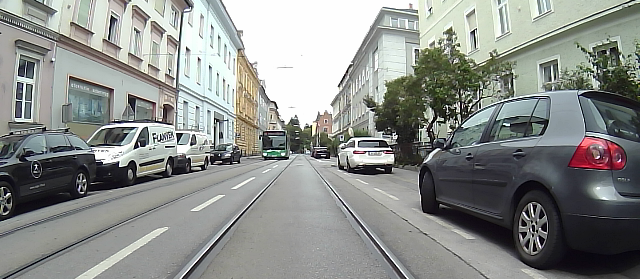}	
			}
			\label{fig:interf_cam}
		}
	}
	
	\caption{Measured (top) and simulated (middle) IF interference signal. From left to right, the top rows show the time domain signal of a ramp with one single interference burst (a,d), the RD map with interference only (b,e), and the RD map when mixed with an object signal (c,f). The object signal without interference (g) was recorded in the traffic measurement campaign and the corresponding camera snapshot is depicted in (h); it shows a bus in around 35 m distance that is approaching the measurement vehicle with around 12 m/s relative velocity, and some parking cars. The SNR in (g) is around 22 dB; it decreases to about 10 dB in (c) and (f).}
	\label{fig:int_meas}
\end{figure}

Figure~\ref{fig:int_meas} illustrates a recorded interference in the top row and its simulated reconstruction in the middle row. In Figures~\ref{fig:interf_td_meas} and \ref{fig:interf_td_sim}, we can see the IF signal of a single interfered ramp in time domain. The interference burst is visible from roughly 5-9{\textmu}s and incorporates a chirp-like form, as mentioned in Section~\ref{sec:signal_model}. The simulated signal matches the measurement relatively well in terms of its location, duration and general chirp-like form. However, differences can be found in the form of the envelope and in the signal's symmetry properties. This is a direct consequence of missing information about the analog receive filters in the hardware sensor. In particular, a model of their phase responses would be required, as a non-constant group delay causes such distortions of the envelope. Furthermore, also multipath reflections may contribute to such effects.

Besides these analog filters, the measurement is also pre-processed digitally by zero-phase high- and low-pass filtering. These digital filters are known and simulated, which is why Figure~\ref{fig:interf_td_sim} also differs from the ideal interference as illustrated in Figure~\ref{fig:ideal_burst_ex}.

One measurement of 128 ramps contains several such chirp bursts at different locations, this results in structured noise patterns on the RD map, as shown in Figures~\ref{fig:interf_rd_meas} and \ref{fig:interf_rd_sim} without adding object signals and in Figures~\ref{fig:interf_rd_mix_meas} and \ref{fig:interf_rd_mix_sim} when combined with the object signal from Figure~\ref{fig:interf_rd_clean}. Even though simulated and measured interference are not exactly identical, they both incorporate similar characteristics and patterns, and can hardly be distinguished in RD domain. This qualitative investigation shows, that simulated interference may well approximate real-world signals.

\subsection{Evaluation}
The \emph{F1-Score} is used as evaluation metric, which is defined as the harmonic mean of precision $p$ and recall $r$, i.e.:

\begin{equation}
	\mathrm{F_1} = 2 \frac{
		p \cdot r
	} {
		p + r
	}.
\end{equation}
The precision $p$ measures the ratio of correct object detections to the total number of object detections according to the \emph{interference mitigated RD map}, thus it considers the number of false-alarms. The recall $r$ defines the ratio of correct object detections to the total number of object detections in the \emph{ground truth data}, thus it considers the number of correctly identified object peaks.
The ground truth target detections were obtained by manually labeling the clean measurement RD maps without interference. A \emph{Cell Averaging Constant False Alarm Rate (CA-CFAR)} target detection algorithm \cite{scharf1991statistical} is used to automatically extract detections, hence peak locations, from the interference mitigated model outputs. Both the ground truth target detections and the CA-CFAR generated detections from interference mitigated RD maps are the basis for the F1-Score. We calculate the mean sample-wise F1-Score per model and report the mean and the standard deviation over three individually trained models with independent initialization if not stated otherwise.

\footnotetext[3]{All model names indicate their architecture using the scheme L\textless LAYERS\textgreater -C\textless CHANNELS\textgreater -\{A/B\}, where A or B indicates architecture A or B.}

\section{Experimental results}
First, we analyze performance effects due to binarization of CNN-based models from \cite{9114627} for interference mitigation. We consider binary weights and activations, and we illustrate the impact of the CNN architecture on the performance degradation due to quantization.

Based on high performing, structurally small and real-valued models we present quantization results with different fixed as well as learned bit-widths, which are determined layer-wise and learned in addition to the model weights. We compare a real-valued model, a quantized model and a selection of 'classical' interference mitigation methods in terms of F1-Scores. Also, we show a qualitative example of real-sensor interference mitigation.

Next, we consider ternary weight models and compare the results obtained by quantization aware training with models that are trained using discrete weight distributions. Then, we show that the training of discrete weight distributions can be used to produce uncertainty estimates of the predicted RD maps. Finally, we visualize the CNN model's learned filter kernels and activations.

\subsection{Binarization effects of CNN-based models for interference mitigation}

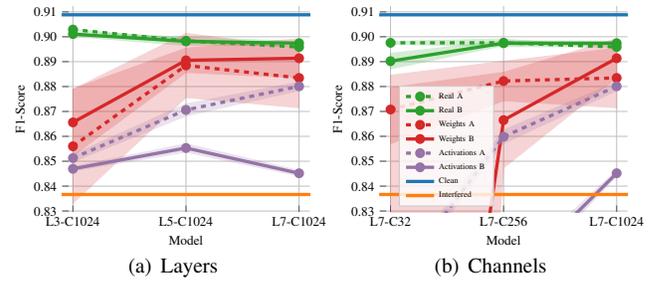
\begin{figure}[t]
	\centering
	\footnotesize
	\subfigure[Layers]{
		\resizebox {0.5\columnwidth} {!} {
\begin{tikzpicture}

\definecolor{color0}{rgb}{0.172549019607843,0.627450980392157,0.172549019607843}
\definecolor{color1}{rgb}{0.83921568627451,0.152941176470588,0.156862745098039}
\definecolor{color2}{HTML}{9673A6} 
\definecolor{color3}{rgb}{0.12156862745098,0.466666666666667,0.705882352941177}
\definecolor{color4}{rgb}{1,0.498039215686275,0.0549019607843137}

\begin{axis}[
scale only axis=true,
width=0.6\columnwidth,
height=0.5\columnwidth,
axis line style={white},
tick align=inside,
tick pos=left,
xlabel={Model},
xmajorgrids,
xmin=0.9, xmax=3.1,
xtick style={color=white!33.3333333333333!black},
xtick={1,2,3},
xticklabels={L3-C1024,L5-C1024,L7-C1024},
ylabel={F1-Score},
ymajorgrids,
ymin=0.829394605429512, ymax=0.912609964357914,
ytick style={color=white!33.3333333333333!black},
ytick={0.82,0.83,0.84,0.85,0.86,0.87,0.88,0.89,0.9,0.91,0.92},
yticklabels={0.82,0.83,0.84,0.85,0.86,0.87,0.88,0.89,0.90,0.91,0.92}
]
\path [draw=color0, fill=color0, opacity=0.2, very thin]
(axis cs:1,0.902897931604162)
--(axis cs:1,0.902897931604162)
--(axis cs:2,0.897604626763606)
--(axis cs:3,0.895901738534557)
--(axis cs:3,0.895901738534557)
--(axis cs:3,0.895901738534557)
--(axis cs:2,0.899118919805831)
--(axis cs:1,0.902897931604162)
--cycle;

\path [draw=color0, fill=color0, opacity=0.2, very thin]
(axis cs:1,0.902012058200272)
--(axis cs:1,0.900161424338658)
--(axis cs:2,0.896518099271671)
--(axis cs:3,0.896609350358013)
--(axis cs:3,0.898222277521526)
--(axis cs:3,0.898222277521526)
--(axis cs:2,0.899814448360316)
--(axis cs:1,0.902012058200272)
--cycle;

\path [draw=color1, fill=color1, opacity=0.2, very thin]
(axis cs:1,0.878822097602364)
--(axis cs:1,0.83317712174444)
--(axis cs:2,0.875582841001602)
--(axis cs:3,0.871494150197411)
--(axis cs:3,0.895419171185435)
--(axis cs:3,0.895419171185435)
--(axis cs:2,0.901396377151305)
--(axis cs:1,0.878822097602364)
--cycle;

\path [draw=color1, fill=color1, opacity=0.2, very thin]
(axis cs:1,0.879031305846938)
--(axis cs:1,0.852180134208572)
--(axis cs:2,0.885677187310018)
--(axis cs:3,0.883533569189472)
--(axis cs:3,0.899205849234899)
--(axis cs:3,0.899205849234899)
--(axis cs:2,0.89541217428892)
--(axis cs:1,0.879031305846938)
--cycle;

\path [draw=color2, fill=color2, opacity=0.2, very thin]
(axis cs:1,0.852695209947125)
--(axis cs:1,0.849829478455929)
--(axis cs:2,0.868192703504482)
--(axis cs:3,0.878408221125369)
--(axis cs:3,0.881791624885775)
--(axis cs:3,0.881791624885775)
--(axis cs:2,0.87301867410171)
--(axis cs:1,0.852695209947125)
--cycle;

\path [draw=color2, fill=color2, opacity=0.2, very thin]
(axis cs:1,0.84775907187048)
--(axis cs:1,0.846150111420446)
--(axis cs:2,0.853640943481085)
--(axis cs:3,0.8440879047742)
--(axis cs:3,0.846243028984415)
--(axis cs:3,0.846243028984415)
--(axis cs:2,0.856995651994883)
--(axis cs:1,0.84775907187048)
--cycle;

\addplot [line width=2pt, color3]
table {%
0.9 0.908827448042986
3.1 0.908827448042986
};
\addplot [line width=2pt, color4]
table {%
0.9 0.836633657524875
3.1 0.836633657524875
};
\addplot [line width=2pt, color0, dashed, mark=*, mark options={solid}]
table {%
1 0.902897931604162
2 0.898361773284719
3 0.895901738534557
};
\addplot [line width=2pt, color0, mark=*]
table {%
1 0.901086741269465
2 0.898166273815993
3 0.89741581393977
};
\addplot [line width=2pt, color1, dashed, mark=*, mark options={solid}]
table {%
1 0.855999609673402
2 0.888489609076454
3 0.883456660691423
};
\addplot [line width=2pt, color1, mark=*]
table {%
1 0.865605720027755
2 0.890544680799469
3 0.891369709212185
};
\addplot [line width=2pt, color2, dashed, mark=*, mark options={solid}]
table {%
1 0.851262344201527
2 0.870605688803096
3 0.880099923005572
};
\addplot [line width=2pt, color2, mark=*]
table {%
1 0.846954591645463
2 0.855318297737984
3 0.845165466879307
};
\end{axis}

\end{tikzpicture}
		}
		\label{fig:result1a}
	}
	\hspace{-0.6cm}
	\subfigure[Channels]{
		\resizebox {0.5\columnwidth} {!} {
\begin{tikzpicture}

\definecolor{color0}{rgb}{0.172549019607843,0.627450980392157,0.172549019607843}
\definecolor{color1}{rgb}{0.83921568627451,0.152941176470588,0.156862745098039}
\definecolor{color2}{HTML}{9673A6} 
\definecolor{color3}{rgb}{0.12156862745098,0.466666666666667,0.705882352941177}
\definecolor{color4}{rgb}{1,0.498039215686275,0.0549019607843137}

\begin{axis}[
scale only axis=true,
width=0.6\columnwidth,
height=0.5\columnwidth,
axis line style={white},
legend cell align={left},
legend style={fill opacity=0.8, draw opacity=1, text opacity=1, at={(0.08,0.03)}, anchor=south west, draw=white!80!black, font=\tiny},
tick align=inside,
tick pos=left,
xlabel={Model},
xmajorgrids,
xmin=0.9, xmax=3.1,
xtick style={color=white!33.3333333333333!black},
xtick={1,2,3},
xticklabels={L7-C32,L7-C256,L7-C1024},
ylabel={F1-Score},
ymajorgrids,
ymin=0.829394605429512, ymax=0.912609964357914,
ytick style={color=white!33.3333333333333!black},
ytick={0.82,0.83,0.84,0.85,0.86,0.87,0.88,0.89,0.9,0.91,0.92},
yticklabels={0.82,0.83,0.84,0.85,0.86,0.87,0.88,0.89,0.90,0.91,0.92},
]
\path [draw=color0, fill=color0, opacity=0.2, very thin]
(axis cs:1,0.898125065444409)
--(axis cs:1,0.897022575320038)
--(axis cs:2,0.896074539518329)
--(axis cs:3,0.895901738534557)
--(axis cs:3,0.895901738534557)
--(axis cs:3,0.895901738534557)
--(axis cs:2,0.899093199399536)
--(axis cs:1,0.898125065444409)
--cycle;

\path [draw=color0, fill=color0, opacity=0.2, very thin]
(axis cs:1,0.893321047732215)
--(axis cs:1,0.887081011773133)
--(axis cs:2,0.896502161533129)
--(axis cs:3,0.896609350358013)
--(axis cs:3,0.898222277521526)
--(axis cs:3,0.898222277521526)
--(axis cs:2,0.898363015516298)
--(axis cs:1,0.893321047732215)
--cycle;

\path [draw=color1, fill=color1, opacity=0.2, very thin]
(axis cs:1,0.884531466968341)
--(axis cs:1,0.857011232870117)
--(axis cs:2,0.874306752140633)
--(axis cs:3,0.871494150197411)
--(axis cs:3,0.895419171185435)
--(axis cs:3,0.895419171185435)
--(axis cs:2,0.890142676510463)
--(axis cs:1,0.884531466968341)
--cycle;

\path [draw=color1, fill=color1, opacity=0.2, very thin]
(axis cs:1,0.879528282110347)
--(axis cs:1,0.233961731863795)
--(axis cs:2,0.847323625628789)
--(axis cs:3,0.883533569189472)
--(axis cs:3,0.899205849234899)
--(axis cs:3,0.899205849234899)
--(axis cs:2,0.885687626967746)
--(axis cs:1,0.879528282110347)
--cycle;

\path [draw=color2, fill=color2, opacity=0.2, very thin]
(axis cs:1,0.802513271207556)
--(axis cs:1,0.797376403467372)
--(axis cs:2,0.857168406827616)
--(axis cs:3,0.878408221125369)
--(axis cs:3,0.881791624885775)
--(axis cs:3,0.881791624885775)
--(axis cs:2,0.862502987873036)
--(axis cs:1,0.802513271207556)
--cycle;

\path [draw=color2, fill=color2, opacity=0.2, very thin]
(axis cs:1,0.436100883897742)
--(axis cs:1,0.209943092338428)
--(axis cs:2,0.79558615278744)
--(axis cs:3,0.8440879047742)
--(axis cs:3,0.846243028984415)
--(axis cs:3,0.846243028984415)
--(axis cs:2,0.80447677362606)
--(axis cs:1,0.436100883897742)
--cycle;

\addplot [line width=2pt, color0, dashed, mark=*, mark options={solid}]
table {%
1 0.897573820382224
2 0.897583869458933
3 0.895901738534557
};
\addlegendentry{Real A}
\addplot [line width=2pt, color0, mark=*]
table {%
1 0.890201029752674
2 0.897432588524714
3 0.89741581393977
};
\addlegendentry{Real B}
\addplot [line width=2pt, color1, dashed, mark=*, mark options={solid}]
table {%
1 0.870771349919229
2 0.882224714325548
3 0.883456660691423
};
\addlegendentry{Weights A}
\addplot [line width=2pt, color1, mark=*]
table {%
1 0.556745006987071
2 0.866505626298268
3 0.891369709212185
};
\addlegendentry{Weights B}
\addplot [line width=2pt, color2, dashed, mark=*, mark options={solid}]
table {%
1 0.799944837337464
2 0.859835697350326
3 0.880099923005572
};
\addlegendentry{Activations A}
\addplot [line width=2pt, color2, mark=*]
table {%
1 0.323021988118085
2 0.80003146320675
3 0.845165466879307
};
\addlegendentry{Activations B}
\addplot [line width=2pt, color3]
table {%
	0.9 0.908827448042987
	3.1 0.908827448042987
};
\addlegendentry{Clean}
\addplot [line width=2pt, color4]
table {%
	0.9 0.836633657524875
	3.1 0.836633657524875
};
\addlegendentry{Interfered}
\end{axis}

\end{tikzpicture}
		}
		\label{fig:result1b}
	}
	\caption{Performance comparison of binarized models using different numbers of layers and channels. The reference F1-Score of the clean measurement data {\color{jstspblue}(Clean)} and the interfered data {\color{jstsporange}(Interfered)} are indicated as horizontal lines. For each model configuration on the x-axis, the solid line indicates the performance of a bottleneck-based architecture of channels (Architecture B, see Section~\ref{sec:modelarchB}). The dashed line uses the same number of channels $C = 1024$ in every layer except the last one, which consists of two channels (Architecture A, see Section~\ref{sec:modelarchA}).\protect\footnotemark[3]}
	\label{fig:result1ab}
\end{figure}

Figure \ref{fig:result1ab} shows a performance comparison of different model architectures and quantization strategies. Figure~\ref{fig:result1a} shows architectures with $C = 1024$ maximal channels and $L \in \{3, 5, 7\}$ layers, while Figure~\ref{fig:result1b} shows architectures with $L = 7$ layers and $C \in \{32, 256, 1024\}$ maximal channels. Both figures contain real-valued models (Real), models with binary weights (Weights) and models with binary activations (Activations). Note that only a model surpassing the score for interfered data (Interfered) yields an improvement. All quantized models were trained using quantization aware training with the STE and binary quantization (sign).

The real-valued baseline (Real) in Figure \ref{fig:result1a} performs well for both model architectures (A or B, see Section \ref{sec:modelarchA}) as well as for all considered numbers of layers $L$ and channels $C$. Models with binary weights (Weights) or activations (Activations) tend towards a better performance with a higher number of layers and model parameters. An exception is the model with binary activations and architecture B, where the minimal number of channels, i.e. the number of channels in the $L-1^{th}$ layer for a model with $L$ layers, seems to be the limiting factor. Generally, architecture B yields better results for binarized weights whereas architecture A is better suited for binarized activations.

Figure~\ref{fig:result1b} shows that the real-valued model (Real) is also robust with regard to different numbers of maximal channels $C \in \{32, 256, 1024\}$ for $L = 7$ layers. Binary weight (Weights) and binary activation (Activations) models on the other hand depend highly on the number of channels. The limiting factor is not only the minimal number of channels, but also the total number of channels. This can be observed by comparing architecture B with $C=32$ and $C=256$ maximal channels, where they both have 8 channels in layer $L-1$ but the model with an overall higher number of channels performs better. Models with binary activations require a very large number of channels and thus parameters in order to reach a high F1-Score.

In summary, we have shown that binary weight models can almost reach the performance of their real-valued equivalent given a high number of model parameters and especially channels. In the binary weight case, architecture B is preferable. For binary activations however, architecture A performs better. In any case, a large number of parameters is required in order to reach a high F1-Score.

\subsection{Performance-memory relation of different architectures and quantization strategies}

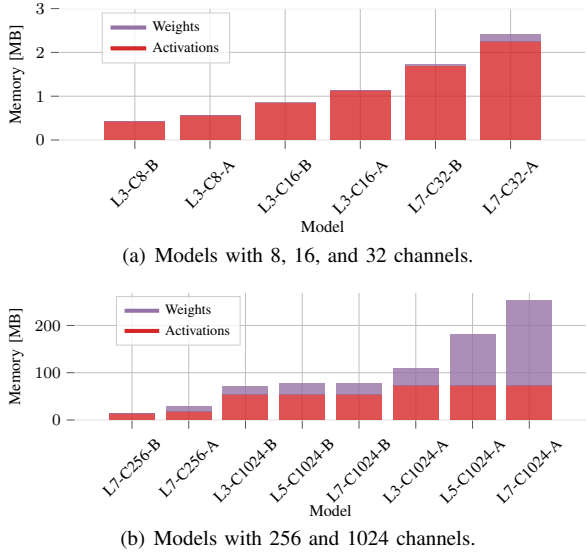
\begin{figure}[t]
	\centering
	\footnotesize
	\subfigure[Models with 8, 16, and 32 channels.]{
		\resizebox {0.9\columnwidth} {!} {
\begin{tikzpicture}

\definecolor{color0}{rgb}{0.83921568627451,0.152941176470588,0.156862745098039}  
\definecolor{color1}{HTML}{9673A6} 

\begin{axis}[
scale only axis=true,
width=\columnwidth,
height=0.25\columnwidth,
axis line style={white},
legend cell align={left},
legend style={fill opacity=0.8, draw opacity=1, text opacity=1, draw=white!80!black, at={(0.08,0.97)}, anchor=north west, font=\scriptsize},
legend entries={{Weights},{Activations}},
tick align=outside,
tick pos=left,
xlabel={Model},
x label style={at={(axis description cs:.5,-0.57)}},
xmajorgrids,
xmin=-1, xmax=6,
xtick style={color=white!33.3333333333333!black},
xtick={0,1,2,3,4,5},
xticklabel style = {rotate=45.0},
xticklabels={L3-C8-B,L3-C8-A,L3-C16-B,L3-C16-A,L7-C32-B,L7-C32-A},
ylabel={Memory [MB]},
ymajorgrids,
ymin=0, ymax=3,
ytick style={color=white!33.3333333333333!black}
]

\addlegendimage{solid, no markers, color1, line width=2pt}
\addlegendimage{solid, no markers, color0, line width=2pt}

\draw[draw=none,fill=color0,fill opacity=0.8,very thin] (axis cs:-0.4,0) rectangle (axis cs:0.4,0.421875);
\draw[draw=none,fill=color0,fill opacity=0.8,very thin] (axis cs:0.6,0) rectangle (axis cs:1.4,0.5625);
\draw[draw=none,fill=color0,fill opacity=0.8,very thin] (axis cs:1.6,0) rectangle (axis cs:2.4,0.84375);
\draw[draw=none,fill=color0,fill opacity=0.8,very thin] (axis cs:2.6,0) rectangle (axis cs:3.4,1.125);
\draw[draw=none,fill=color0,fill opacity=0.8,very thin] (axis cs:3.6,0) rectangle (axis cs:4.4,1.6875);
\draw[draw=none,fill=color0,fill opacity=0.8,very thin] (axis cs:4.6,0) rectangle (axis cs:5.4,2.25);
\draw[draw=none,fill=color1,fill opacity=0.8,very thin] (axis cs:-0.4,0.421875) rectangle (axis cs:0.4,0.423797607421875);
\draw[draw=none,fill=color1,fill opacity=0.8,very thin] (axis cs:0.6,0.5625) rectangle (axis cs:1.4,0.5657958984375);
\draw[draw=none,fill=color1,fill opacity=0.8,very thin] (axis cs:1.6,0.84375) rectangle (axis cs:2.4,0.84979248046875);
\draw[draw=none,fill=color1,fill opacity=0.8,very thin] (axis cs:2.6,1.125) rectangle (axis cs:3.4,1.135986328125);
\draw[draw=none,fill=color1,fill opacity=0.8,very thin] (axis cs:3.6,1.6875) rectangle (axis cs:4.4,1.71881103515625);
\draw[draw=none,fill=color1,fill opacity=0.8,very thin] (axis cs:4.6,2.25) rectangle (axis cs:5.4,2.43017578125);
\end{axis}
\end{tikzpicture}
		}
		\label{fig:result_mem1}
	}
	\subfigure[Models with 256 and 1024 channels.]{
		\resizebox {0.9\columnwidth} {!} {
\begin{tikzpicture}

\definecolor{color0}{rgb}{0.83921568627451,0.152941176470588,0.156862745098039}  
\definecolor{color1}{HTML}{9673A6} 

\begin{axis}[
scale only axis=true,
width=\columnwidth,
height=0.25\columnwidth,
axis line style={white},
legend cell align={left},
legend style={fill opacity=0.8, draw opacity=1, text opacity=1, draw=white!80!black, at={(0.08,0.97)}, anchor=north west, font=\scriptsize},
legend entries={{Weights},{Activations}},
tick align=outside,
tick pos=left,
xlabel={Model},
x label style={at={(axis description cs:.5,-0.62)}},
xmajorgrids,
xmin=5, xmax=14,
xtick style={color=white!33.3333333333333!black},
xtick={6,7,8,9,10,11,12,13},
xticklabel style = {rotate=45.0},
xticklabels={L7-C256-B,L7-C256-A,L3-C1024-B,L5-C1024-B,L7-C1024-B,L3-C1024-A,L5-C1024-A,L7-C1024-A},
ylabel={Memory [MB]},
ymajorgrids,
ymin=0, ymax=270,
ytick style={color=white!33.3333333333333!black}
]

\addlegendimage{solid, no markers, color1, line width=2pt}
\addlegendimage{solid, no markers, color0, line width=2pt}

\draw[draw=none,fill=color0,fill opacity=0.8,very thin] (axis cs:5.6,0) rectangle (axis cs:6.4,13.5);
\draw[draw=none,fill=color0,fill opacity=0.8,very thin] (axis cs:6.6,0) rectangle (axis cs:7.4,18);
\draw[draw=none,fill=color0,fill opacity=0.8,very thin] (axis cs:7.6,0) rectangle (axis cs:8.4,54);
\draw[draw=none,fill=color0,fill opacity=0.8,very thin] (axis cs:8.6,0) rectangle (axis cs:9.4,54);
\draw[draw=none,fill=color0,fill opacity=0.8,very thin] (axis cs:9.6,0) rectangle (axis cs:10.4,54);
\draw[draw=none,fill=color0,fill opacity=0.8,very thin] (axis cs:10.6,0) rectangle (axis cs:11.4,72);
\draw[draw=none,fill=color0,fill opacity=0.8,very thin] (axis cs:11.6,0) rectangle (axis cs:12.4,72);
\draw[draw=none,fill=color0,fill opacity=0.8,very thin] (axis cs:12.6,0) rectangle (axis cs:13.4,72);
\draw[draw=none,fill=color1,fill opacity=0.8,very thin] (axis cs:5.6,13.5) rectangle (axis cs:6.4,15.0166625976562);
\draw[draw=none,fill=color1,fill opacity=0.8,very thin] (axis cs:6.6,18) rectangle (axis cs:7.4,29.28515625);
\draw[draw=none,fill=color1,fill opacity=0.8,very thin] (axis cs:7.6,54) rectangle (axis cs:8.4,72.10546875);
\draw[draw=none,fill=color1,fill opacity=0.8,very thin] (axis cs:8.6,54) rectangle (axis cs:9.4,77.7041015625);
\draw[draw=none,fill=color1,fill opacity=0.8,very thin] (axis cs:9.6,54) rectangle (axis cs:10.4,78.049072265625);
\draw[draw=none,fill=color1,fill opacity=0.8,very thin] (axis cs:10.6,72) rectangle (axis cs:11.4,108.140625);
\draw[draw=none,fill=color1,fill opacity=0.8,very thin] (axis cs:11.6,72) rectangle (axis cs:12.4,180.140625);
\draw[draw=none,fill=color1,fill opacity=0.8,very thin] (axis cs:12.6,72) rectangle (axis cs:13.4,252.140625);
\end{axis}
\end{tikzpicture}
		}
		\label{fig:result_mem2}
	}
	\caption{Total memory requirements [MB] for real-valued models during the inference step. The total requirements consist of the memory for {\color{jstsppurple}weights} plus the memory for the two largest consecutive  layers of {\color{jstspred}activations}.\protect\footnotemark[3]}
	\label{fig:result_mem}
\end{figure}

The total memory requirements during the inference step stem from storing (i) model parameters and (ii) activations of two consecutive layers that need to be stored during the computation. For the sake of run time and energy efficient computations, these parameters and variables have to be stored in fast accessible on-chip memory simultaneously.

Figure~\ref{fig:result_mem} shows the total memory requirements per model architecture. All depicted models are real-valued and reach a similar F1-Score of $F1 \ge 0.89$. Models with few channels (e.g. 8, 16 or 32) have \emph{much smaller} memory requirements than models with many channels (e.g. 256 or 1024); note the different y-axis scales in Figures~\ref{fig:result_mem1} and \ref{fig:result_mem2}. Quantization reduces the memory footprint by a factor of up to 32, namely in the binary case. Since even the smallest models from Figure~\ref{fig:result_mem} reach a high F1-Score and bigger models could not surpass their memory efficiency even with binary quantization, only these small models can be used as base models for quantization.

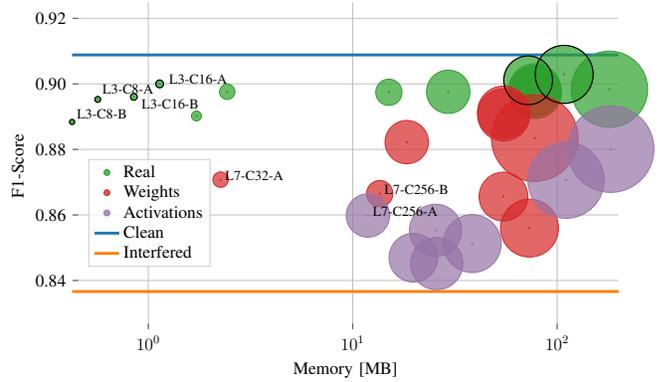
\begin{figure}[t]
	{
	\centering
	\large
	\resizebox{\columnwidth}{!}{
\begin{tikzpicture}

\definecolor{color0}{rgb}{0.172549019607843,0.627450980392157,0.172549019607843}  
\definecolor{color1}{rgb}{0.83921568627451,0.152941176470588,0.156862745098039}  
\definecolor{color2}{HTML}{9673A6} 
\definecolor{color3}{rgb}{0.12156862745098,0.466666666666667,0.705882352941177}
\definecolor{color4}{rgb}{1,0.498039215686275,0.0549019607843137}

\begin{axis}[
scale only axis=true,
width=1.5\columnwidth,
height=0.9\columnwidth,
legend cell align={left},
legend style={fill opacity=0.8, draw opacity=1, text opacity=1, draw=white!80!black, 
	at={(0.03,0.52)}, anchor=north west, font=\large},
legend entries={{Real},{Weights},{Activations},{Clean},{Interfered}},
axis line style={white},
log basis x={10},
tick align=inside,
tick pos=left,
xlabel={Memory [MB]},
xmajorgrids,
xmin=0.423881530761719, xmax=200,
xmode=log,
xtick style={color=white!33.3333333333333!black},
xtick={0.01,0.1,1,10,100,1000,10000},
xticklabels={\(\displaystyle {10^{-2}}\),\(\displaystyle {10^{-1}}\),\(\displaystyle {10^{0}}\),\(\displaystyle {10^{1}}\),\(\displaystyle {10^{2}}\),\(\displaystyle {10^{3}}\),\(\displaystyle {10^{4}}\)},
ylabel={F1-Score},
ymajorgrids,
ymin=0.825, ymax=0.925,
ytick style={color=white!33.3333333333333!black},
ytick={0.82,0.84,0.86,0.88,0.9,0.92,0.94},
yticklabels={0.82,0.84,0.86,0.88,0.90,0.92,0.94}
]

\addlegendimage{only marks, mark=*, draw=color0,fill=color0}
\addlegendimage{only marks, mark=*, draw=color1,fill=color1}
\addlegendimage{only marks, mark=*, draw=color2,fill=color2}
\addlegendimage{draw=color3,line width=2pt}
\addlegendimage{draw=color4,line width=2pt}

\addplot [scatter, only marks, mark=*, scatter/use mapped color={
	draw=color0,
	fill=color0,
}, fill=color0, fill opacity=0.7, colormap/viridis, visualization depends on={\thisrow{sizedata} \as\perpointmarksize}, scatter/@pre marker code/.append style={/tikz/mark size=\perpointmarksize}]
table{%
x                      y                      sizedata
1.71949005126953 0.890201029752674 3.46593742488784
2.43213653564453 0.897573820382224 5.34482333212864
15.0204849243164 0.897432588524714 9.08476217547867
29.3007888793945 0.897583869458933 14.9959309468778
78.064338684082 0.89741581393977 18.1157411545607
252.203132629395 0.895901738534557 29.9655737576612
72.1152420043945 0.901086741269465 16.8746688314898
108.156257629395 0.902897931604162 20.0561088512119
77.7182693481445 0.898166273815993 18.050277724276
180.179695129395 0.898361773284719 26.3767688320872
0.849952697753906 0.89602647761059 2.30815716391312
1.13623809814453 0.899994953380168 2.67119521399816
0.423881530761719 0.888376159413645 1.74799081460642
0.565925598144531 0.895253786576051 1.9900830119242
};
\addplot [scatter, only marks, mark=*, scatter/use mapped color={
	draw=color0,
	fill=color0,
},  fill=color0, fill opacity=1, visualization depends on={\thisrow{sizedata} \as\perpointmarksize}, scatter/@pre marker code/.append style={/tikz/mark size=\perpointmarksize}]
table{%
	x                      y                      sizedata
	1.71949005126953 0.890201029752674 0.4
	2.43213653564453 0.897573820382224 0.4
	15.0204849243164 0.897432588524714 0.4
	29.3007888793945 0.897583869458933 0.4
	78.064338684082 0.89741581393977 0.4
	252.203132629395 0.895901738534557 0.4
	72.1152420043945 0.901086741269465 0.4
	108.156257629395 0.902897931604162 0.4
	77.7182693481445 0.898166273815993 0.4
	180.179695129395 0.898361773284719 0.4
	0.849952697753906 0.89602647761059 0.4
	1.13623809814453 0.899994953380168 0.4
	0.423881530761719 0.888376159413645 0.4
	0.565925598144531 0.895253786576051 0.4
};
\addplot [scatter, only marks, mark=*, scatter/use mapped color={
	draw=color1,
	fill=color1,
}, fill=color1, fill opacity=0.7, colormap/viridis, visualization depends on={\thisrow{sizedata} \as\perpointmarksize}, scatter/@pre marker code/.append style={/tikz/mark size=\perpointmarksize}]
table{%
x                      y                      sizedata
1.68915748596191 0.556745006987071 3.46593742488784
2.25759124755859 0.870771349919229 5.34482333212864
13.5512180328369 0.866505626298268 9.08476217547867
18.368293762207 0.882224714325548 14.9959309468778
54.7667999267578 0.891369709212185 18.1157411545607
77.6919021606445 0.883456660691423 29.9655737576612
54.6096267700195 0.865605720027755 16.8746688314898
73.2131423950195 0.855999609673402 20.0561088512119
54.7549209594727 0.890544680799469 18.050277724276
0.844631195068359 0.772597789906985 2.30815716391312
};
\addplot [scatter, only marks, mark=*, scatter/use mapped color={
	draw=color1,
	fill=color1,
}, fill=color1, fill opacity=1, visualization depends on={\thisrow{sizedata} \as\perpointmarksize}, scatter/@pre marker code/.append style={/tikz/mark size=\perpointmarksize}]
table{%
	x y sizedata
	1.68915748596191 0.556745006987071 0.4
	2.25759124755859 0.870771349919229 0.4
	13.5512180328369 0.866505626298268 0.4
	18.368293762207 0.882224714325548 0.4
	54.7667999267578 0.891369709212185 0.4
	77.6919021606445 0.883456660691423 0.4
	54.6096267700195 0.865605720027755 0.4
	73.2131423950195 0.855999609673402 0.4
	54.7549209594727 0.890544680799469 0.4
	0.844631195068359 0.772597789906985 0.4
};
\addplot [scatter, only marks, mark=*, scatter/use mapped color={
	draw=color2,
	fill=color2,
}, fill=color2, fill opacity=0.7, colormap/viridis, visualization depends on={\thisrow{sizedata} \as\perpointmarksize}, scatter/@pre marker code/.append style={/tikz/mark size=\perpointmarksize}]
table{%
x                      y                      sizedata
0.137458801269531 0.323021988118085 3.46593742488784
0.287605285644531 0.799944837337464 5.34482333212864
1.94235992431641 0.80003146320675 9.08476217547867
11.8632888793945 0.859835697350326 14.9959309468778
25.751838684082 0.845165466879307 18.1157411545607
182.453132629395 0.880099923005572 29.9655737576612
19.8027420043945 0.846954591645463 16.8746688314898
38.4062576293945 0.851262344201527 20.0561088512119
25.4057693481445 0.855318297737984 18.050277724276
110.429695129395 0.870605688803096 26.3767688320872
};
\addplot [scatter, only marks, mark=*, scatter/use mapped color={
	draw=color2,
	fill=color2,
}, fill=color2, fill opacity=1, colormap/viridis,visualization depends on={\thisrow{sizedata} \as \perpointmarksize}, scatter/@pre marker code/.append style={/tikz/mark size=\perpointmarksize}]
table {
	x                      y                      sizedata
	0.137458801269531 0.323021988118085 0.4
	0.287605285644531 0.799944837337464 0.4
	1.94235992431641 0.80003146320675 0.4
	11.8632888793945 0.859835697350326 0.4
	25.751838684082 0.845165466879307 0.4
	182.453132629395 0.880099923005572 0.4
	19.8027420043945 0.846954591645463 0.4
	38.4062576293945 0.851262344201527 0.4
	25.4057693481445 0.855318297737984 0.4
	110.429695129395 0.870605688803096 0.4
};
\addplot [line width=2pt, color3]
table {%
0.423881530761719 0.908827448042987
200 0.908827448042987
};
\addplot [line width=2pt, color4]
table {%
0.423881530761719 0.836633657524875
200 0.836633657524875
};
\addplot [scatter, thick, only marks, mark=o, scatter/use mapped color={
	draw=black,
},  color=black, colormap/viridis, visualization depends on={\thisrow{sizedata} \as \perpointmarksize}, scatter/@pre marker code/.append style={/tikz/mark size=\perpointmarksize}]
table{%
	x                      y                      sizedata
	0.423881530761719 0.888376159413645 1.74799081460642
	0.565925598144531 0.895253786576051 1.9900830119242
	0.849952697753906 0.89602647761059 2.30815716391312
	1.13623809814453 0.899994953380168 2.67119521399816
	72.1152420043945 0.901086741269465 16.8746688314898
	108.156257629395 0.902897931604162 20.0561088512119
};
\draw (axis cs:0.423881530761719,0.8895) node[
  scale=0.8,
  anchor=base west,
  text=black,
  rotate=0.0,
  node on layer=front,
]{L3-C8-B};
\draw (axis cs:0.565925598144531,0.897) node[
  scale=0.8,
  anchor=base west,
  text=black,
  rotate=0.0,
  node on layer=front,
]{L3-C8-A};
\draw (axis cs:0.88,0.8925) node[
  scale=0.8,
  anchor=base west,
  text=black,
  rotate=0.0,
  node on layer=front,
]{L3-C16-B};
\draw (axis cs:1.2,0.899994953380168) node[
  scale=0.8,
  anchor=base west,
  text=black,
  rotate=0.0,
  node on layer=front,
]{L3-C16-A};
\draw (axis cs:2.25759124755859,0.870771349919229) node[
  scale=0.8,
  anchor=base west,
  text=black,
  rotate=0.0,
  node on layer=front,
]{L7-C32-A};
\draw (axis cs:13.5512180328369,0.866505626298268) node[
  scale=0.8,
  anchor=base west,
  text=black,
  rotate=0.0,
  node on layer=front,
]{L7-C256-B};
\draw (axis cs:11.8632888793945,0.859835697350326) node[
  scale=0.8,
  anchor=base west,
  text=black,
  rotate=0.0,
  node on layer=front,
]{L7-C256-A};
\end{axis}

\end{tikzpicture}
	}
	\caption{Average F1-Score vs. memory requirements in megabytes [MB] during the inference step of real-valued {\color{jstspgreen}(Real)}, binary weight {\color{jstspred}(Weights)} and binary activation {\color{jstsppurple}(Activations)} models. The circle size corresponds to the number of million operations (MOPS) required during the inference step. The Pareto optimal points are marked using black borderlines; they all belong to real-valued models. The smallest models for each category are annotated\protect\footnotemark[3]; see TABLE \ref{tab:1c} for details of annotated models.}
	\label{fig:result1c}
	}
\end{figure}

\begin{table}[t]
	\begin{center}
		\scriptsize
		\setlength\tabcolsep{5pt}
		\begin{tabular}{lrrrrrrr}\toprule
			\multicolumn{1}{c}{Model} & \multicolumn{1}{c}{Param.} & \multicolumn{1}{c}{QT} & \multicolumn{1}{c}{Weights} & \multicolumn{1}{c}{Act.} & \multicolumn{1}{c}{Total} & \multicolumn{1}{c}{MOPS} & \multicolumn{1}{c}{F1-Score}\\
			&&& \multicolumn{1}{c}{[MB]} & \multicolumn{1}{c}{[MB]} & \multicolumn{1}{c}{[MB]} &&\\
			\midrule
			L3-C8-B & 504 & {\color{jstspgreen}R} & 0.002 & 0.42 & 0.42 & 5 & 0.888 \\
			L3-C8-A & 864 & {\color{jstspgreen}R} & 0.003 & 0.56 & 0.57 & 8 & 0.895 \\
			L3-C16-B & 1584 & {\color{jstspgreen}R} & 0.006 & 0.84 & 0.85 & 15 & 0.896 \\
			L3-C16-A & 2880 & {\color{jstspgreen}R} & 0.011 & 1.12 & 1.14 & 27 & 0.900 \\
			L7-C32-A & 47k & {\color{jstspred}W} & 0.006 & 2.25 & 2.26 & 441 & 0.871 \\
			L7-C256-B & 398k & {\color{jstspred}W} & 0.047 & 13.50 & 13.55 & 3.7k & 0.867 \\
			L7-C256-A & 2958k & {\color{jstsppurple}A} & 11.285 & 0.56 & 11.86 & 27k & 0.860 \\
			\bottomrule
		\end{tabular}
		\caption{Memory usage and performance details for models annotated\protect\footnotemark[3] in Figure~\ref{fig:result1c}. The quantization type QT can be {\color{jstspgreen}R} for real-valued models, {\color{jstspred}W} for binary weights, or {\color{jstsppurple}A} for binary activations (sign). }
		\label{tab:1c}
	\end{center}
\end{table}

Figure~\ref{fig:result1c} illustrates the performance to memory relation for real-valued models, models with binary weights and models with binary activations. TABLE~\ref{tab:1c} lists details of the smallest models per quantization type. The results clearly show that models with binarized weights or activations reaching an acceptable F1-Score require more memory than a real-valued alternative with fewer parameters. All Pareto optimal points correspond to real-valued models. Already particularly small real-valued models reach a high F1-Score of $F1 > 0.89$, such as the model denoted as 'L3-C16-B' which has $L = 3$ layers and $C_l=[16,8,2]$ channels in these layers. We thus conclude, that considered binarized CNNs are \emph{not} suited for radar denoising and interference mitigation. Instead, we investigate quantization with multiple bits for weights and activations which could be used to further reduce the memory requirements of small real-valued models. We choose the model with $L = 3$ layers and $C_l=[16, 8, 2]$ channels as a base model for all further experiments, it yields an F1-Score of $F1 = 0.8960$.

\subsection{Quantization with multiple bits using fixed and learned bit-widths}

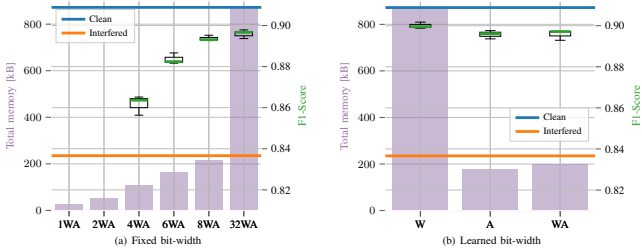
\begin{figure}[t]
	{
	\centering
	\footnotesize
	\resizebox{0.49\columnwidth}{!}{
		\subfigure[Fixed bit-width]{
\begin{tikzpicture}

\definecolor{color0}{HTML}{9673A6} 
\definecolor{color1}{rgb}{0.172549019607843,0.627450980392157,0.172549019607843}
\definecolor{color2}{rgb}{0.12156862745098,0.466666666666667,0.705882352941177}
\definecolor{color3}{rgb}{1,0.498039215686275,0.0549019607843137}

\begin{axis}[
scale only axis=true,
width=0.6\columnwidth,
height=0.6\columnwidth,
axis line style={white},
tick align=outside,
tick pos=left,
xmajorgrids,
xmin=0.5, xmax=6.5,
xtick style={color=white!33.3333333333333!black,font=\tiny},
xtick={1,2,3,4,5,6},
xticklabels={1WA,2WA,4WA,6WA,8WA,32WA},
ylabel={Total memory [kB]},
ylabel style={color=color0},
ymajorgrids,
ymin=0, ymax=900,
ytick style={color=white!33.3333333333333!black}
]
\draw[draw=none,fill=color0,fill opacity=0.5,very thin] (axis cs:0.6,0) rectangle (axis cs:1.4,27.357421875);
\draw[draw=none,fill=color0,fill opacity=0.5,very thin] (axis cs:1.6,0) rectangle (axis cs:2.4,54.55078125);
\draw[draw=none,fill=color0,fill opacity=0.5,very thin] (axis cs:2.6,0) rectangle (axis cs:3.4,108.9375);
\draw[draw=none,fill=color0,fill opacity=0.5,very thin] (axis cs:3.6,0) rectangle (axis cs:4.4,163.32421875);
\draw[draw=none,fill=color0,fill opacity=0.5,very thin] (axis cs:4.6,0) rectangle (axis cs:5.4,217.7109375);
\draw[draw=none,fill=color0,fill opacity=0.5,very thin] (axis cs:5.6,0) rectangle (axis cs:6.4,870.3515625);
\end{axis}

\begin{axis}[
scale only axis=true,
width=0.6\columnwidth,
height=0.6\columnwidth,
axis line style={white},
axis y line=right,
legend cell align={left},
legend style={fill opacity=0.8, draw opacity=1, text opacity=1, at={(0.02,0.96)}, anchor=north west, draw=white!80!black, font=\scriptsize},
tick align=outside,
xmajorgrids,
xmin=0.5, xmax=6.5,
xtick pos=left,
xtick style={color=white!33.3333333333333!black},
xtick={1,2,3,4,5,6},
xticklabels={1WA,2WA,4WA,6WA,8WA,32WA},
ylabel={F1-Score},
ylabel style={color=color1},
ymajorgrids,
ymin=0.81, ymax=0.912,
ytick pos=right,
ytick style={color=white!33.3333333333333!black},
ytick={0.8,0.82,0.84,0.86,0.88,0.9,0.92},
yticklabels={0.80,0.82,0.84,0.86,0.88,0.90,0.92}
]
\addplot [black, forget plot]
table {%
1 0
1 0
};
\addplot [black, forget plot]
table {%
1 0
1 0
};
\addplot [black, forget plot]
table {%
0.875 0
1.125 0
};
\addplot [black, forget plot]
table {%
0.875 0
1.125 0
};
\addplot [black, forget plot]
table {%
2 0.724102604931597
2 0.724067482158539
};
\addplot [black, forget plot]
table {%
2 0.727548151004027
2 0.7309585743034
};
\addplot [black, forget plot]
table {%
1.875 0.724067482158539
2.125 0.724067482158539
};
\addplot [black, forget plot]
table {%
1.875 0.7309585743034
2.125 0.7309585743034
};
\addplot [black, forget plot]
table {%
3 0.860000051320365
3 0.856341720640426
};
\addplot [black, forget plot]
table {%
3 0.864417985222416
3 0.865177588444528
};
\addplot [black, forget plot]
table {%
2.875 0.856341720640426
3.125 0.856341720640426
};
\addplot [black, forget plot]
table {%
2.875 0.865177588444528
3.125 0.865177588444528
};
\addplot [black, forget plot]
table {%
4 0.882029383878705
4 0.881664347471184
};
\addplot [black, forget plot]
table {%
4 0.88452925871811
4 0.886664097149992
};
\addplot [black, forget plot]
table {%
3.875 0.881664347471184
4.125 0.881664347471184
};
\addplot [black, forget plot]
table {%
3.875 0.886664097149992
4.125 0.886664097149992
};
\addplot [black, forget plot]
table {%
5 0.892895973099222
5 0.89266161782335
};
\addplot [black, forget plot]
table {%
5 0.894190336083145
5 0.895250343791197
};
\addplot [black, forget plot]
table {%
4.875 0.89266161782335
5.125 0.89266161782335
};
\addplot [black, forget plot]
table {%
4.875 0.895250343791197
5.125 0.895250343791197
};
\addplot [black, forget plot]
table {%
6 0.894994219860701
6 0.893634678040802
};
\addplot [black, forget plot]
table {%
6 0.896953341114642
6 0.897859861721293
};
\addplot [black, forget plot]
table {%
5.875 0.893634678040802
6.125 0.893634678040802
};
\addplot [black, forget plot]
table {%
5.875 0.897859861721293
6.125 0.897859861721293
};
\addplot [line width=2pt, color2]
table {%
0.5 0.908827448042987
6.5 0.908827448042987
};
\addlegendentry{Clean}
\addplot [line width=2pt, color3]
table {%
0.5 0.836633657524875
6.5 0.836633657524875
};
\addlegendentry{Interfered}
\path [draw=black]
(axis cs:0.75,0)
--(axis cs:1.25,0)
--(axis cs:1.25,0)
--(axis cs:0.75,0)
--(axis cs:0.75,0)
--cycle;
\path [draw=black]
(axis cs:1.75,0.724102604931597)
--(axis cs:2.25,0.724102604931597)
--(axis cs:2.25,0.727548151004027)
--(axis cs:1.75,0.727548151004027)
--(axis cs:1.75,0.724102604931597)
--cycle;
\path [draw=black]
(axis cs:2.75,0.860000051320365)
--(axis cs:3.25,0.860000051320365)
--(axis cs:3.25,0.864417985222416)
--(axis cs:2.75,0.864417985222416)
--(axis cs:2.75,0.860000051320365)
--cycle;
\path [draw=black]
(axis cs:3.75,0.882029383878705)
--(axis cs:4.25,0.882029383878705)
--(axis cs:4.25,0.88452925871811)
--(axis cs:3.75,0.88452925871811)
--(axis cs:3.75,0.882029383878705)
--cycle;
\path [draw=black]
(axis cs:4.75,0.892895973099222)
--(axis cs:5.25,0.892895973099222)
--(axis cs:5.25,0.894190336083145)
--(axis cs:4.75,0.894190336083145)
--(axis cs:4.75,0.892895973099222)
--cycle;
\path [draw=black]
(axis cs:5.75,0.894994219860701)
--(axis cs:6.25,0.894994219860701)
--(axis cs:6.25,0.896953341114642)
--(axis cs:5.75,0.896953341114642)
--(axis cs:5.75,0.894994219860701)
--cycle;
\addplot [color1, forget plot, line width=2pt]
table {%
0.75 0
1.25 0
};
\addplot [semithick, color1, forget plot, line width=2pt]
table {%
1.75 0.724137727704655
2.25 0.724137727704655
};
\addplot [semithick, color1, forget plot, line width=2pt]
table {%
2.75 0.863658382000304
3.25 0.863658382000304
};
\addplot [semithick, color1, forget plot, line width=2pt]
table {%
3.75 0.882394420286227
4.25 0.882394420286227
};
\addplot [semithick, color1, forget plot, line width=2pt]
table {%
4.75 0.893130328375094
5.25 0.893130328375094
};
\addplot [semithick, color1, forget plot, line width=2pt]
table {%
5.75 0.89669028731551
6.25 0.89669028731551
};
\end{axis}

\end{tikzpicture}
			\label{fig:result1e1}
		}
	}
	\resizebox{0.49\columnwidth}{!}{
		\subfigure[Learned bit-width]{
\begin{tikzpicture}

\definecolor{color0}{HTML}{9673A6} 
\definecolor{color1}{rgb}{0.172549019607843,0.627450980392157,0.172549019607843}
\definecolor{color2}{rgb}{0.12156862745098,0.466666666666667,0.705882352941177}
\definecolor{color3}{rgb}{1,0.498039215686275,0.0549019607843137}

\begin{axis}[
scale only axis=true,
width=0.6\columnwidth,
height=0.6\columnwidth,
axis line style={white},
tick align=outside,
tick pos=left,
xmajorgrids,
xmin=0.5, xmax=3.5,
xtick style={color=white!33.3333333333333!black},
xtick={1,2,3},
xticklabels={W,A,WA},
ylabel={Total memory [kB]},
ylabel style={color=color0},
ymajorgrids,
ymin=0, ymax=900,
ytick style={color=white!33.3333333333333!black}
]
\draw[draw=none,fill=color0,fill opacity=0.5,very thin] (axis cs:0.6,0) rectangle (axis cs:1.4,864.83203125);
\draw[draw=none,fill=color0,fill opacity=0.5,very thin] (axis cs:1.6,0) rectangle (axis cs:2.4,177.3515625);
\draw[draw=none,fill=color0,fill opacity=0.5,very thin] (axis cs:2.6,0) rectangle (axis cs:3.4,198.990234375);
\end{axis}

\begin{axis}[
scale only axis=true,
width=0.6\columnwidth,
height=0.6\columnwidth,
axis line style={white},
axis y line=right,
legend cell align={left},
legend style={fill opacity=0.8, draw opacity=1, text opacity=1, at={(0.97,0.33)}, anchor=south east, draw=white!80!black, font=\scriptsize},
tick align=outside,
xmajorgrids,
xmin=0.5, xmax=3.5,
xtick pos=left,
xtick style={color=white!33.3333333333333!black},
xtick={1,2,3},
xticklabels={W,A,WA},
ylabel={F1-Score},
ylabel style={color=color1},
ymajorgrids,
ymin=0.81, ymax=0.912,
ytick pos=right,
ytick style={color=white!33.3333333333333!black},
ytick={0.8,0.82,0.84,0.86,0.88,0.9,0.92},
yticklabels={0.80,0.82,0.84,0.86,0.88,0.90,0.92}
]
\addplot [black, forget plot]
table {%
1 0.899165984317292
1 0.8988046039711
};
\addplot [black, forget plot]
table {%
1 0.900630230582204
1 0.901733096500924
};
\addplot [black, forget plot]
table {%
0.925 0.8988046039711
1.075 0.8988046039711
};
\addplot [black, forget plot]
table {%
0.925 0.901733096500924
1.075 0.901733096500924
};
\addplot [black, forget plot]
table {%
2 0.894828887175329
2 0.893594054918086
};
\addplot [black, forget plot]
table {%
2 0.896836916089156
2 0.897610112745739
};
\addplot [black, forget plot]
table {%
1.925 0.893594054918086
2.075 0.893594054918086
};
\addplot [black, forget plot]
table {%
1.925 0.897610112745739
2.075 0.897610112745739
};
\addplot [black, forget plot]
table {%
3 0.89502608283126
3 0.892905827371154
};
\addplot [black, forget plot]
table {%
3 0.897178603586299
3 0.89721086888123
};
\addplot [black, forget plot]
table {%
2.925 0.892905827371154
3.075 0.892905827371154
};
\addplot [black, forget plot]
table {%
2.925 0.89721086888123
3.075 0.89721086888123
};
\addplot [line width=2pt, color2]
table {%
0.5 0.908827448042987
3.5 0.908827448042987
};
\addlegendentry{Clean}
\addplot [line width=2pt, color3]
table {%
0.5 0.836633657524875
3.5 0.836633657524875
};
\addlegendentry{Interfered}
\path [draw=black]
(axis cs:0.85,0.899165984317292)
--(axis cs:1.15,0.899165984317292)
--(axis cs:1.15,0.900630230582204)
--(axis cs:0.85,0.900630230582204)
--(axis cs:0.85,0.899165984317292)
--cycle;
\path [draw=black]
(axis cs:1.85,0.894828887175329)
--(axis cs:2.15,0.894828887175329)
--(axis cs:2.15,0.896836916089156)
--(axis cs:1.85,0.896836916089156)
--(axis cs:1.85,0.894828887175329)
--cycle;
\path [draw=black]
(axis cs:2.85,0.89502608283126)
--(axis cs:3.15,0.89502608283126)
--(axis cs:3.15,0.897178603586299)
--(axis cs:2.85,0.897178603586299)
--(axis cs:2.85,0.89502608283126)
--cycle;
\addplot [semithick, color1, forget plot, line width=2pt]
table {%
0.85 0.899527364663483
1.15 0.899527364663483
};
\addplot [semithick, color1, forget plot, line width=2pt]
table {%
1.85 0.896063719432573
2.15 0.896063719432573
};
\addplot [semithick, color1, forget plot, line width=2pt]
table {%
2.85 0.897146338291367
3.15 0.897146338291367
};
\end{axis}

\end{tikzpicture}
			\label{fig:result1e2}
		}
	}	
	\caption{Average F1-Score and memory requirements [kB] for models with fixed and learned bit-widths. Fixed bit-widths (left) are used for weights as well as activations (WA). Learned bit-widths (right) are optimized for either weights (W), activations (A) or both (WA). The model has $L=3$ layers and $C=[16,8,2]$ channels. Details about memory requirements, performance and average bit-widths are listed in TABLE~\ref{tab:1d}.}
	\label{fig:result1e}
}
\end{figure}

\begin{table}[t]
	\centering
	\scriptsize
	\setlength\tabcolsep{5pt}
		\begin{tabular}{c|rr|rr|r|r}\toprule
			\multicolumn{1}{c}{QT} & \multicolumn{2}{c}{Weights} & \multicolumn{2}{c}{Activations} & \multicolumn{1}{c}{Total} & \multicolumn{1}{c}{F1-Score}\\
			\cmidrule{2-3} \cmidrule{4-5}
			\multicolumn{1}{c}{} & \multicolumn{1}{c}{[kB]} & \multicolumn{1}{c}{[Bits]} & \multicolumn{1}{c}{[kB]} & \multicolumn{1}{c}{[Bits]} & \multicolumn{1}{c}{[kB]} & \\ \midrule
			\multicolumn{7}{l}{\textbf{Fixed bit-width}} \\ \midrule
			 4WA & $0.77$ & 4 & $108$ & 4 & $109$ & $0.8617 \pm 0.004$ \\
			 6WA & $1.16$ & 6 & $162$ & 6 & $163$ & $0.8836 \pm 0.002$ \\
			 8WA  & $1.55$ & 8 & $216$ & 8 & $218$ & $0.8937 \pm 0.001$ \\
			 32WA & $6.19$ & 32 & $864$ & 32 & $870$ & $0.8960 \pm 0.002$ \\  \midrule
			\multicolumn{7}{l}{\textbf{Learned bit-width}} \\ \midrule
			 W & $0.67$ & 4.3 & $864$ & 32.0 & $865$ & $0.9000 \pm 0.001$ \\
			 A & $6.19$ & 32.0 & $171$ & 6.5 & $177$ & $0.8958 \pm 0.002$ \\
			 WA & $0.83$ & 5.0 & $198$ & 7.5 & $198.99$ & $0.8958 \pm 0.002$ \\
			\bottomrule
		\end{tabular}
	\caption{Memory, performance and bit-width details for results shown in Figure~\ref{fig:result1e}. The quantization type QT can be W for weights, A for activations, or WA for weights and activations. The number of bits for fixed bit-widths is indicated in front of the quantization type. The learned bit-width is stated in average bits per value over all layers. The memory usage is stated in kilobytes [kB] and the F1-Score is listed as 'mean $\pm$ standard deviation' over three independently trained models. The model has $L = 3$ layers and $C_l=[16,8,2]$ channels.}
	\label{tab:1d}
\end{table}

We aim to further reduce the memory size of the real-valued model with $L = 3$ layers and $C=[16, 8, 2]$ channels without substantial performance degradation. We use the \emph{Brevitas} framework \cite{brevitas} to (i) analyze quantization performance with different fixed bit-widths $k \in \{1, 2, 4, 6, 8, 32\}$ for weights as well as activations and (ii) to learn the bit-width of model weights and activations per layer. In any case, we use integer quantization and determine the dynamic range as the maximum absolute value over the real-valued auxiliary weights (see Section~\ref{sec:dynamic_range}). Figure~\ref{fig:result1e} shows the performance of these different quantization strategies, where Figure~\ref{fig:result1e1} contains models with fixed bit-widths including the real-valued base model, and Figure~\ref{fig:result1e2} contains models with learned bit-widths, using the following quantization options:
\begin{enumerate}[label=(\roman*)]
	\item \textbf{Weights (W):} Quantized weights with learned bit-widths and real-valued activations.
	\item \textbf{Activations (A):} Real-valued weights and quantized activations with learned bit-widths.
	\item \textbf{Weights+Activations (WA):} Quantized weights and activations with learned bit-widths.
\end{enumerate}
TABLE~\ref{tab:1d} lists corresponding details regarding the performance and memory requirements for weights and activations, including average bit-widths.

Models with a fixed number of $k=1$ or $k=2$ bits are not suited for the task and do not even reach the F1-Score of data without mitigation. With $k=4$, $k=6$, and $k=8$ bits the performance increases steadily and almost reaches the real-valued score with only 8 bits. The resulting memory saving with 8-bit weights and activations is approximately 75\% compared to the real-valued baseline.

For quantization strategies using learned bit-widths, we consider the smallest bit-widths reached without significant performance degradation per category (Weights, Activations, Weights+Activations). The results shown in Figure~\ref{fig:result1e2} and their corresponding details stated in TABLE~\ref{tab:1d} show the effectiveness of quantized activations over quantized weights for the considered task of RD interference mitigation. When optimizing only the weight bit-widths and keeping full precision activations, we reach an average learned bit-width of 4.3 bits over all layers. Nonetheless, the overall memory reduction is minimal. When quantizing only activations on the other hand, an average bit-width of 6.5 is reached which reduces the overall memory requirements during the inference step considerably. With 177.35 kB, we achieve a memory reduction of approximately 80\% for the model with learned bit-width activations compared to the real-valued baseline. This behavior stems from the large number of activations in two consecutive layers compared to the total number of model weights. We have this inequality in number of values per activation layer versus model weights due to our task and the associated model.

\begin{figure*}[t]
	\hfill
	\begin{minipage}[t]{.31\linewidth}
		\centering
		\footnotesize
		\resizebox {\textwidth} {!} {
			\input{cdf_classical_f1.tex}
		}
		\caption{CDF comparison of the sample-wise F1-Score between the real-valued CNN model (CNN-Real), the 8 bit quantized CNN model (CNN-8WA), and the three classical methods zeroing, IMAT and Ramp filtering.}
		\label{fig:cdf-classical}
	\end{minipage}%
	\hfill
	\begin{minipage}[t]{.31\linewidth}
		\centering
		\footnotesize
		\resizebox{\textwidth}{!}{
\begin{tikzpicture}

\definecolor{color0}{rgb}{0.172549019607843,0.627450980392157,0.172549019607843}
\definecolor{color1}{rgb}{0.83921568627451,0.152941176470588,0.156862745098039}
\definecolor{color2}{rgb}{0.12156862745098,0.466666666666667,0.705882352941177}
\definecolor{color3}{rgb}{1,0.498039215686275,0.0549019607843137}

\pgfplotsset{
	compat=1.11,
	legend image code/.code={
		\draw[mark repeat=2,mark phase=2]
		plot coordinates {
			(0cm,0cm)
			(0.15cm,0cm)        
			(0.3cm,0cm)         
		};%
	}
}

\begin{axis}[
scale only axis=true,
width=\textwidth,
height=0.75\textwidth,
axis line style={white},
legend cell align={left},
legend style={fill opacity=0.8, draw opacity=1, text opacity=1, at={(0.01,0.08)},
	 anchor=south west, draw=white!80!black, font=\scriptsize},
legend entries={{Quantization aware training (STE)},{Discrete weight distributions},{Clean},{Interfered}},
tick align=outside,
tick pos=left,
xlabel={Factor for scaling the dynamic range},
xmajorgrids,
xmin=0.5, xmax=3.5,
xtick style={color=white!33.3333333333333!black},
xtick={1,2,3},
xticklabels={None,Statistics,Learned},
ylabel={F1-Score},
ymajorgrids,
ymin=0.81, ymax=0.912,
ytick style={color=white!33.3333333333333!black},
ytick={0.8,0.82,0.84,0.86,0.88,0.9,0.92},
yticklabels={0.80,0.82,0.84,0.86,0.88,0.90,0.92}
]

\addlegendimage{draw=color1, line width=2pt}
\addlegendimage{draw=color0, line width=2pt}
\addlegendimage{draw=color2, line width=2pt}
\addlegendimage{draw=color3, line width=2pt}

\addplot [black, forget plot]
table {%
1 0.86658129817688
1 0.863275351898895
};
\addplot [black, forget plot]
table {%
1 0.875968658313156
1 0.882050072171447
};
\addplot [black, forget plot]
table {%
0.925 0.863275351898895
1.075 0.863275351898895
};
\addplot [black, forget plot]
table {%
0.925 0.882050072171447
1.075 0.882050072171447
};
\addplot [black, forget plot]
table {%
2 0.88671198478356
2 0.88358250226221
};
\addplot [black, forget plot]
table {%
2 0.891123707923546
2 0.892405948542181
};
\addplot [black, forget plot]
table {%
1.925 0.88358250226221
2.075 0.88358250226221
};
\addplot [black, forget plot]
table {%
1.925 0.892405948542181
2.075 0.892405948542181
};
\addplot [black, forget plot]
table {%
3 0.887902372446183
3 0.886433651763762
};
\addplot [black, forget plot]
table {%
3 0.891258955169717
3 0.89314681721083
};
\addplot [black, forget plot]
table {%
2.925 0.886433651763762
3.075 0.886433651763762
};
\addplot [black, forget plot]
table {%
2.925 0.89314681721083
3.075 0.89314681721083
};
\addplot [black, forget plot]
table {%
1 0
1 0
};
\addplot [black, forget plot]
table {%
1 0
1 0
};
\addplot [black, forget plot]
table {%
0.925 0
1.075 0
};
\addplot [black, forget plot]
table {%
0.925 0
1.075 0
};
\addplot [black, forget plot]
table {%
2 0.893926669162087
2 0.893762441276736
};
\addplot [black, forget plot]
table {%
2 0.895198182584275
2 0.896305468121113
};
\addplot [black, forget plot]
table {%
1.925 0.893762441276736
2.075 0.893762441276736
};
\addplot [black, forget plot]
table {%
1.925 0.896305468121113
2.075 0.896305468121113
};
\addplot [black, forget plot]
table {%
3 0.814074165487459
3 0.813857112992121
};
\addplot [black, forget plot]
table {%
3 0.835264071889457
3 0.856236925796116
};
\addplot [black, forget plot]
table {%
2.925 0.813857112992121
3.075 0.813857112992121
};
\addplot [black, forget plot]
table {%
2.925 0.856236925796116
3.075 0.856236925796116
};
\addplot [line width=2pt, color2]
table {%
0.5 0.908827448042987
3.5 0.908827448042987
};
\addplot [line width=2pt, color3]
table {%
0.5 0.836633657524875
3.5 0.836633657524875
};
\path [draw=black]
(axis cs:0.85,0.86658129817688)
--(axis cs:1.15,0.86658129817688)
--(axis cs:1.15,0.875968658313156)
--(axis cs:0.85,0.875968658313156)
--(axis cs:0.85,0.86658129817688)
--cycle;
\path [draw=black]
(axis cs:1.85,0.88671198478356)
--(axis cs:2.15,0.88671198478356)
--(axis cs:2.15,0.891123707923546)
--(axis cs:1.85,0.891123707923546)
--(axis cs:1.85,0.88671198478356)
--cycle;
\path [draw=black]
(axis cs:2.85,0.887902372446183)
--(axis cs:3.15,0.887902372446183)
--(axis cs:3.15,0.891258955169717)
--(axis cs:2.85,0.891258955169717)
--(axis cs:2.85,0.887902372446183)
--cycle;
\path [draw=black]
(axis cs:0.85,0)
--(axis cs:1.15,0)
--(axis cs:1.15,0)
--(axis cs:0.85,0)
--(axis cs:0.85,0)
--cycle;
\path [draw=black]
(axis cs:1.85,0.893926669162087)
--(axis cs:2.15,0.893926669162087)
--(axis cs:2.15,0.895198182584275)
--(axis cs:1.85,0.895198182584275)
--(axis cs:1.85,0.893926669162087)
--cycle;
\path [draw=black]
(axis cs:2.85,0.814074165487459)
--(axis cs:3.15,0.814074165487459)
--(axis cs:3.15,0.835264071889457)
--(axis cs:2.85,0.835264071889457)
--(axis cs:2.85,0.814074165487459)
--cycle;
\addplot [color0, forget plot, line width=2pt]
table {%
0.85 0.869887244454865
1.15 0.869887244454865
};
\addplot [color0, forget plot, line width=2pt]
table {%
1.85 0.88984146730491
2.15 0.88984146730491
};
\addplot [color0, forget plot, line width=2pt]
table {%
2.85 0.889371093128605
3.15 0.889371093128605
};
\addplot [color1, forget plot, line width=2pt]
table {%
0.85 0
1.15 0
};
\addplot [color1, forget plot, line width=2pt]
table {%
1.85 0.894090897047438
2.15 0.894090897047438
};
\addplot [color1, forget plot, line width=2pt]
table {%
2.85 0.814291217982798
3.15 0.814291217982798
};
\end{axis}

\end{tikzpicture}
		}
		\caption{Performance comparison between quantization aware training with the STE and discrete weight distributions. Different factors for scaling the dynamic range are depicted on the x-axis, i.e. \emph{None}, computed from \emph{statistics} and \emph{learned} as parameter.}
		\label{fig:result1g}
	\end{minipage}%
	\hfill
	\begin{minipage}[t]{.31\linewidth}
		\centering
		\footnotesize
		\resizebox{\textwidth}{!}{
			\input{performance_paper1j.tex}
		}
		\caption{F1-Score with different regularization weight $\lambda$. The discrete model weights are obtained by the most probable weights {\color{jstspgreen}(MP)}, sampling once {\color{jstspred}(S1)} or sampling 100 times {\color{jstsppurple}(S100)}.}
		\label{fig:result1j}
	\end{minipage}%
	\hfill
\end{figure*}

We observe that quantizing only activations using a learned bit-width results in the highest memory reduction without substantial performance degradation. The learned activation bit-width using quantized activations is one bit lower than the learned activation bit-width when using quantized weights and activations. In the latter case, local minima in the combined loss function are reached, that hinder further optimization. The small difference in learned activation bit-widths has a stronger effect on the overall memory requirements than weight quantization.

In comparison to the fixed 8 bit model for weights and activations (8WA), we have a memory reduction for the model with learned bit-width activations (A) of around 40 kB with a slightly better F1-Score (+ 0.0021). However, this relatively small improvement might be negligible in practice, because of the implementation overhead for heterogeneous bit-widths. To fully exploit a heterogeneous bit-width, a specialized hardware with custom arithmetic units would be required. Therefore, we recommend to use 8 bits for all weights and activations, which allows the use of standard integer processing units.

\subsection{Performance comparison with classical interference mitigation methods}

\begin{table}[t]
	\centering
	\scriptsize
	\setlength\tabcolsep{8pt}
	\begin{tabular}{@{}crrrrr@{}}\toprule
		Method & F1-Score & \multicolumn{3}{c}{Parameters} & MOPS\\
		\cmidrule{3-5}
		&& Count & Bits & [kB] & \\ \midrule
		CNN-Real & 0.8960 & 1584 & 32 & 6.19 & 15.33 \\
		CNN-8WA & 0.8937 & 1584 & 8 & 1.55 & 15.33 \\
		Zeroing & 0.8515  & - & - & - & 2.83 \\
		RFmin & 0.8754  & - & - & - & 3.13 \\
		IMAT & 0.8604  & - & - & - & 3.88 \\
	\end{tabular}
	\caption{Comparison of the average F1-Score and resource requirements between the real-valued CNN model (CNN-Real), the 8 bit quantized CNN model (CNN-8WA), and the three classical methods zeroing, IMAT and Ramp filtering.}
	\label{tab:classical}
\end{table}

We compare our CNN-based models in the real-valued setting (Real) as well as in the 8 bit quantized setting (8WA) with the classical and state-of-the-art interference mitigation methods zeroing \cite{Fischer}, \emph{Iterative method with adaptive thresholding (IMAT)} \cite{Bechter2017a} and \emph{Ramp filtering} \cite{WAG18}; see \cite{fu2017complex,toth2020analysis} for an overview of these methods. Zeroing and IMAT highly depend on an interference detection step, which influences their performance considerably. In our experiments we identified time-domain samples incorporating interference with approximately 90\% accuracy; this interference detection rate seems feasible in practice. Note, that IMAT is even more susceptible to false detections than zeroing, and that Ramp filtering as well as the CNN-based models do not depend on such an explicit interference detection step at all.

Figure~\ref{fig:cdf-classical} shows the empirical \emph{cumulative density function (CDF)} of their evaluated per-sample F1-Score values. The 'clean' measurement and interfered signals are included as reference.
All three classical methods, namely zeroing, IMAT and Ramp filtering, improve the F1-Score when applied to the measurement signal with interference. Zeroing and IMAT yield very similar F1-Scores. Ramp filtering outperforms both other classical methods, particularly for samples with strong interference (see black magnification). See \cite{toth2020analysis} for a detailed performance analysis of classical interference mitigation methods.

The CNN-based models, both with real-valued as well as 8 bit quantized weights and activations, are competitive with the classical methods for all considered interference levels and even outperform the best classical method, namely Ramp filtering. The CDF shape indicates that the CNN-based models are robust with respect to different interference patterns and levels. This is indicated by the CDF's narrow form and high values for the lowest F1-Scores per CNN model (see gray magnification). The lowest F1-Scores are approximately $F1_{\textrm{low,8WA}} = 0.57$ and $F1_{\textrm{low,Real}} = 0.60$ for the 8 bit quantized (CNN-8WA) and real-valued (CNN-Real) CNNs respectively. For comparison, the measurements with and without interference yield a lowest F1-Score of $F1_{\textrm{low,Interfered}} = 0$ and $F1_{\textrm{low,Clean}} = 0.63$.
The CDF of the 8 bit quantized CNN is very similar to the real-valued model’s CDF, showing that we can reduce the memory footprint by a factor of four without impairing performance.

TABLE~\ref{tab:classical} shows the average sample-wise F1-Score along with required parameters and million operations (MOPS) for the real-valued CNN (CNN-Real), the 8 bit quantized CNN (CNN-8WA), and the three classical methods zeroing, IMAT and Ramp filtering. The superior F1-Score of the CNN-based methods is at the expense of hardware resources. The CNN model parameters have to be stored and the models require around five times more MOPS than the classical methods. However, the quantized model improves these expenses considerably. Even though the CNN models have the same number of parameters (1584) and MOPS (15.33), only one fourth of memory is required to store the 8 bit quantized model in comparison to the real-valued CNN model. Also the energy and time consumption can be reduced in the quantized case, because pure integer arithmetic can be used for the convolution operations which might lead to a faster computation depending on the hardware.

\subsection{CNN interference mitigation on a real-sensor interference}
\begin{figure}[t]
	\centering
	\footnotesize
	\resizebox{\columnwidth}{!}{
		\subfigure[Object signal]{
\begin{tikzpicture}

\begin{axis}[
scale only axis=true,
width=2.9cm,
height=2.9cm,
axis background/.style={fill=white!89.80392156862746!black},
axis line style={white},
colormap/viridis,
point meta max=0,
point meta min=-30,
tick align=inside,
tick pos=left,
x grid style={white},
xlabel={Velocity [m/s]},
xmajorgrids,
xmin=-16, xmax=15.6666666666667,
xtick style={color=white!33.33333333333333!black},
y grid style={white},
ylabel={Distance [m]},
ymajorgrids,
ymin=6, ymax=60,
ytick style={color=white!33.33333333333333!black}
]
\addplot graphics [includegraphics cmd=\pgfimage,xmin=-16, xmax=15.6666666666667, ymin=6, ymax=60] {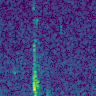};
\end{axis}

\end{tikzpicture}
			\label{fig:example_cnn_out_clean}
		}
		\hspace{-2mm}
		\subfigure[With interference]{
\begin{tikzpicture}

\begin{axis}[
scale only axis=true,
width=2.9cm,
height=2.9cm,
axis background/.style={fill=white!89.80392156862746!black},
axis line style={white},
colormap/viridis,
point meta max=0,
point meta min=-30,
tick align=inside,
tick pos=left,
x grid style={white},
xlabel={Velocity [m/s]},
xmajorgrids,
xmin=-16, xmax=15.6666666666667,
xtick style={color=white!33.33333333333333!black},
y grid style={white},
ylabel={Distance [m]},
ymajorgrids,
ymin=6, ymax=60,
ytick style={color=white!33.33333333333333!black}
]
\addplot graphics [includegraphics cmd=\pgfimage,xmin=-16, xmax=15.6666666666667, ymin=6, ymax=60] {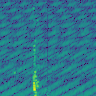};
\end{axis}

\end{tikzpicture}
			\label{fig:example_cnn_out_interf}
		}
		\hspace{-2mm}
		\subfigure[After CNN mitigation]{
\begin{tikzpicture}

\begin{axis}[
scale only axis=true,
width=2.9cm,
height=2.9cm,
axis background/.style={fill=white!89.80392156862746!black},
axis line style={white},
colorbar,
colorbar/width=1.5mm,
colorbar style={
	yticklabel style={
		text width=width("$-30$"),
		align=right
	},
	ylabel={dB}
},
colormap/viridis,
point meta max=0,
point meta min=-30,
tick align=inside,
tick pos=left,
x grid style={white},
xlabel={Velocity [m/s]},
xmajorgrids,
xmin=-16, xmax=15.6666666666667,
xtick style={color=white!33.33333333333333!black},
y grid style={white},
ylabel={Distance [m]},
ymajorgrids,
ymin=6, ymax=60,
ytick style={color=white!33.33333333333333!black}
]
\addplot graphics [includegraphics cmd=\pgfimage,xmin=-16, xmax=15.6666666666667, ymin=6, ymax=60] {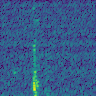};
\end{axis}

\end{tikzpicture}
			\label{fig:example_cnn_out_prediction}
		}
	}
	\caption{RD magnitude spectra in decibel [dB] without interference (SNR = 32~dB) combined with real-sensor interference (SNR = 19~dB) and after CNN interference mitigation (SNR = 26~dB).}
	\label{fig:example_cnn_out}
\end{figure}

In order to illustrate successful CNN interference mitigation performance on real-sensor interference signals, despite the lack of a large set of such measurements, we show quantitative results of an exemplary RD map in Figure~\ref{fig:example_cnn_out}. The interference signal is obtained using real-sensor measurements and combined with an object signal from the inner-city measurement campaign, as described in Section~\ref{sec:dataset}. Several object peaks are visible in the 'clean' measurement RD map (Figure~\ref{fig:example_cnn_out_clean}, SNR = 32~dB). The interfered RD map (Figure~\ref{fig:example_cnn_out_interf}, SNR = 19~dB) contains distinctive interference patterns and complicates the identification of object peaks. After interference mitigation using the real-valued CNN model, which was solely trained on simulated interference signals, most of these patterns are removed and the noise floor is damped, while the object peaks are retained (Figure~\ref{fig:example_cnn_out_prediction}, SNR = 26~dB). This example illustrates the potential of our approach, while more quantitative results and in-the-wild-interference signals are to be evaluated. The generalization from simulated to real-sensor interference signals is promising, but using real-sensor interference signals during training or for transfer learning has even more potential in terms of real-world interference mitigation.

\subsection{Quantization aware training vs. training distributions over discrete weights}

Here, we consider training discrete distributions over ternary weights as described in Section~\ref{subsec:distr}. In contrast to quantized DNN models that only produce point estimates, weight distributions additionally allow us to obtain uncertainty estimates over model predictions, i.e. RD uncertainty maps. We use the CNN model with $L=3$ layers, $C_l=[16,8,2]$ channels, real-valued ReLU activations and ternary weights $W \in \{-\alpha,0,+\alpha\}$ requiring 2 bits each.\footnote[4]{Quantizing activations with multiple bits using this technique goes beyond the scope of this work and is not considered at this point.} For comparison we use quantization aware training with the STE and integer quantization with a bit-width of $k = 2$; this also results in ternary weights $W \in \{-\alpha,0,+\alpha\}$.
Figure~\ref{fig:result1g} shows the performance comparison between quantization aware training with the STE and the training of discrete weight distributions with three different methods of determining the dynamic range $\alpha$. For ternary weights that are symmetric around zero, the dynamic range $\alpha$ is equal to the step size $\delta_w$ and can be seen as a simple scaling factor that is multiplied to integer ternary weights $W \in \{-1, 0, 1\}$. For the calculation of the scaling factor $\alpha$ we consider the three methods as described in Section~\ref{sec:dynamic_range}:
\begin{description}
	\item \textbf{None:} no scaling factor, i.e. $\alpha = 1$
	\item \textbf{Statistics:} the maximal absolute value of real-valued auxiliary weights (quantization aware training) or pre-trained weights (discrete distributions)
	\item \textbf{Learned:} scaling factor learned as a model parameter
\end{description}
Quantization aware training does not learn the interference mitigation task at all when no scaling factor $\alpha$ is used; this results from a combination of the used weight initialization (small values close to zero) and quantization type (integer rounding). Training discrete distributions over weights $W \in \{-1, 0, 1\}$ on the other hand yields a high F1-Score of $F1 = 0.87$ without scaling. With a scaling factor calculated from weight statistics, both quantization aware training and training with discrete distributions increase performance substantially and reach an F1-Score $\ge 0.89$. For training discrete distributions a learned scaling factor performs similar to the one from statistics, but for quantization aware training the learned scaling factor worsens the results considerably.

The choice of the dynamic range $\alpha$ has a high impact on the model performance. We propose to use the maximal absolute weight per layer. According to our results this choice of dynamic range calculation is robust, has a small computation overhead and results in a high performance, independent of the training approach. Training distributions over discrete weights yields competitive results in comparison to quantization aware training with the STE for the considered weight quantization. With regard to the scaling factor, training distributions over discrete weights even seems to be more robust.

\subsection{Uncertainty maps and effects of the distribution regularization term}

Figure~\ref{fig:result1j} shows different values for $\lambda$, i.e. the contribution of the distribution regularization term to the network loss (see Section~\ref{subsec:distr}). We use the same model as before, i.e. a CNN with $L = 3$ layers and $C_l=[16,8,2]$ channels, and use the maximum absolute weight as scaling factor $\alpha$ per layer. The results illustrate the effect of $\lambda$ for three different variants of retrieving the weight values:
\begin{description}
	\item[MP:] most probable weights
	\item[S1:] weights are sampled from the weight distribution
	\item[S100:] weights are sampled 100 times from the weight distribution and their average is used
\end{description}
Each variant that involves sampling (S1 and S100) is computed and evaluated 100 times in order to capture the variance of model predictions based on these sampled weights.

Our results show that for large $\lambda$ the performance decreases whereas there is a significant drop with $\lambda \ge 5 \cdot 10^{-8}$. The MP weights always perform best, whereas the difference to S1 and S100 increases with a higher regularization term. Both S1 and S100 tend to have a higher variance with a higher regularization term. S100 is always better than S1 but it does not outperform the most probable weights MP.

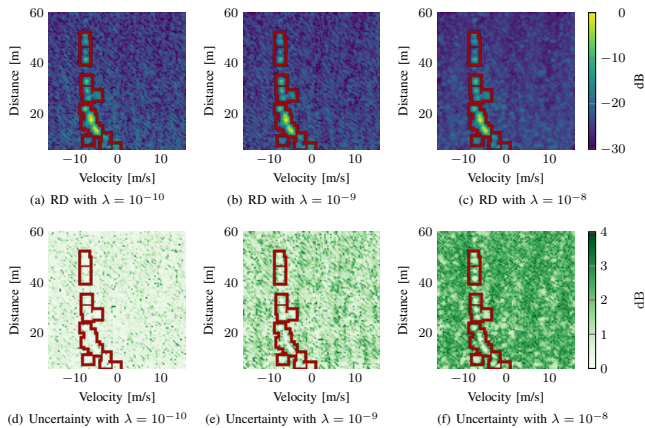
\begin{figure}[t]
	\centering
	\footnotesize
	\resizebox{\columnwidth}{!}{
		\subfigure[RD with $\lambda=10^{-10}$]{
\begin{tikzpicture}

\begin{axis}[
scale only axis=true,
width=2.9cm,
height=2.9cm,
axis background/.style={fill=white!89.8039215686275!black},
axis line style={white},
colormap/viridis,
point meta max=0,
point meta min=-30,
tick align=inside,
tick pos=left,
x grid style={white},
xlabel={Velocity [m/s]},
xmajorgrids,
xmin=-16, xmax=15.6666666666667,
xtick style={color=white!33.3333333333333!black},
y grid style={white},
ylabel={Distance [m]},
ymajorgrids,
ymin=6, ymax=60,
ytick style={color=white!33.3333333333333!black}
]
\addplot graphics [includegraphics cmd=\pgfimage,xmin=-16, xmax=15.6666666666667, ymin=6, ymax=60] {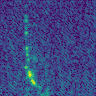};
\addplot graphics [includegraphics cmd=\pgfimage,xmin=-16, xmax=15.6666666666667, ymin=6, ymax=60] {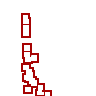};
\end{axis}

\end{tikzpicture}
			\label{fig:result1k1_mean}
		}
		\hspace{-2mm}
		\subfigure[RD with $\lambda=10^{-9}$]{
\begin{tikzpicture}

\begin{axis}[
scale only axis=true,
width=2.9cm,
height=2.9cm,
axis background/.style={fill=white!89.8039215686275!black},
axis line style={white},
colormap/viridis,
point meta max=0,
point meta min=-30,
tick align=inside,
tick pos=left,
x grid style={white},
xlabel={Velocity [m/s]},
xmajorgrids,
xmin=-16, xmax=15.6666666666667,
xtick style={color=white!33.3333333333333!black},
y grid style={white},
ylabel={Distance [m]},
ymajorgrids,
ymin=6, ymax=60,
ytick style={color=white!33.3333333333333!black}
]
\addplot graphics [includegraphics cmd=\pgfimage,xmin=-16, xmax=15.6666666666667, ymin=6, ymax=60] {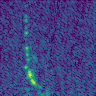};
\addplot graphics [includegraphics cmd=\pgfimage,xmin=-16, xmax=15.6666666666667, ymin=6, ymax=60] {ref_object_overlay_sharp.png};
\end{axis}

\end{tikzpicture}
			\label{fig:result1k2_mean}
		}
		\hspace{-2mm}
		\subfigure[RD with $\lambda=10^{-8}$]{
\begin{tikzpicture}

\begin{axis}[
scale only axis=true,
width=2.9cm,
height=2.9cm,
axis background/.style={fill=white!89.8039215686275!black},
axis line style={white},
colorbar,
colorbar/width=1.5mm,
colorbar style={
	yticklabel style={
		text width=width("$-30$"),
		align=right
	},
	ylabel={dB}
},
colormap/viridis,
point meta max=0,
point meta min=-30,
tick align=inside,
tick pos=left,
x grid style={white},
xlabel={Velocity [m/s]},
xmajorgrids,
xmin=-16, xmax=15.6666666666667,
xtick style={color=white!33.3333333333333!black},
y grid style={white},
ylabel={Distance [m]},
ymajorgrids,
ymin=6, ymax=60,
ytick style={color=white!33.3333333333333!black}
]
\addplot graphics [includegraphics cmd=\pgfimage,xmin=-16, xmax=15.6666666666667, ymin=6, ymax=60] {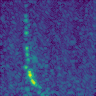};
\addplot graphics [includegraphics cmd=\pgfimage,xmin=-16, xmax=15.6666666666667, ymin=6, ymax=60] {ref_object_overlay_sharp.png};
\end{axis}

\end{tikzpicture}
			\label{fig:result1k3_mean}
		}
	}
	
	\resizebox{\columnwidth}{!}{
		\subfigure[Uncertainty with $\lambda=10^{-10}$]{
\begin{tikzpicture}

\begin{axis}[
scale only axis=true,
width=2.9cm,
height=2.9cm,
axis background/.style={fill=white!89.8039215686275!black},
axis line style={white},
colormap={mymap}{[1pt]
  rgb(0pt)=(0.968627450980392,0.988235294117647,0.96078431372549);
  rgb(1pt)=(0.898039215686275,0.96078431372549,0.87843137254902);
  rgb(2pt)=(0.780392156862745,0.913725490196078,0.752941176470588);
  rgb(3pt)=(0.631372549019608,0.850980392156863,0.607843137254902);
  rgb(4pt)=(0.454901960784314,0.768627450980392,0.462745098039216);
  rgb(5pt)=(0.254901960784314,0.670588235294118,0.364705882352941);
  rgb(6pt)=(0.137254901960784,0.545098039215686,0.270588235294118);
  rgb(7pt)=(0,0.427450980392157,0.172549019607843);
  rgb(8pt)=(0,0.266666666666667,0.105882352941176)
},
point meta max=4,
point meta min=0,
tick align=inside,
tick pos=left,
x grid style={white},
xlabel={Velocity [m/s]},
xmajorgrids,
xmin=-16, xmax=15.6666666666667,
xtick style={color=white!33.3333333333333!black},
y grid style={white},
ylabel={Distance [m]},
ymajorgrids,
ymin=6, ymax=60,
ytick style={color=white!33.3333333333333!black}
]
\addplot graphics [includegraphics cmd=\pgfimage,xmin=-16, xmax=15.6666666666667, ymin=6, ymax=60] {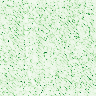};
\addplot graphics [includegraphics cmd=\pgfimage,xmin=-16, xmax=15.6666666666667, ymin=6, ymax=60] {ref_object_overlay_sharp.png};
\end{axis}

\end{tikzpicture}
			\label{fig:result1k1_std}
		}
		\hspace{-2mm}
		\subfigure[Uncertainty with $\lambda=10^{-9}$]{
\begin{tikzpicture}

\begin{axis}[
scale only axis=true,
width=2.9cm,
height=2.9cm,
axis background/.style={fill=white!89.8039215686275!black},
axis line style={white},
colormap={mymap}{[1pt]
  rgb(0pt)=(0.968627450980392,0.988235294117647,0.96078431372549);
  rgb(1pt)=(0.898039215686275,0.96078431372549,0.87843137254902);
  rgb(2pt)=(0.780392156862745,0.913725490196078,0.752941176470588);
  rgb(3pt)=(0.631372549019608,0.850980392156863,0.607843137254902);
  rgb(4pt)=(0.454901960784314,0.768627450980392,0.462745098039216);
  rgb(5pt)=(0.254901960784314,0.670588235294118,0.364705882352941);
  rgb(6pt)=(0.137254901960784,0.545098039215686,0.270588235294118);
  rgb(7pt)=(0,0.427450980392157,0.172549019607843);
  rgb(8pt)=(0,0.266666666666667,0.105882352941176)
},
point meta max=4,
point meta min=0,
tick align=inside,
tick pos=left,
x grid style={white},
xlabel={Velocity [m/s]},
xmajorgrids,
xmin=-16, xmax=15.6666666666667,
xtick style={color=white!33.3333333333333!black},
y grid style={white},
ylabel={Distance [m]},
ymajorgrids,
ymin=6, ymax=60,
ytick style={color=white!33.3333333333333!black}
]
\addplot graphics [includegraphics cmd=\pgfimage,xmin=-16, xmax=15.6666666666667, ymin=6, ymax=60] {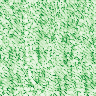};
\addplot graphics [includegraphics cmd=\pgfimage,xmin=-16, xmax=15.6666666666667, ymin=6, ymax=60] {ref_object_overlay_sharp.png};
\end{axis}

\end{tikzpicture}
			\label{fig:result1k2_std}
		}
		\hspace{-2mm}
		\subfigure[Uncertainty with $\lambda=10^{-8}$]{
\begin{tikzpicture}

\begin{axis}[
scale only axis=true,
width=2.9cm,
height=2.9cm,
axis background/.style={fill=white!89.8039215686275!black},
axis line style={white},
colorbar,
colorbar/width=1.5mm,
colorbar style={
	yticklabel style={
		text width=width("$-30$"),
		align=left
	},
	ylabel={$\mathrm{dB}$}
},
colormap={mymap}{[1pt]
  rgb(0pt)=(0.968627450980392,0.988235294117647,0.96078431372549);
  rgb(1pt)=(0.898039215686275,0.96078431372549,0.87843137254902);
  rgb(2pt)=(0.780392156862745,0.913725490196078,0.752941176470588);
  rgb(3pt)=(0.631372549019608,0.850980392156863,0.607843137254902);
  rgb(4pt)=(0.454901960784314,0.768627450980392,0.462745098039216);
  rgb(5pt)=(0.254901960784314,0.670588235294118,0.364705882352941);
  rgb(6pt)=(0.137254901960784,0.545098039215686,0.270588235294118);
  rgb(7pt)=(0,0.427450980392157,0.172549019607843);
  rgb(8pt)=(0,0.266666666666667,0.105882352941176)
},
point meta max=4,
point meta min=0,
tick align=inside,
tick pos=left,
x grid style={white},
xlabel={Velocity [m/s]},
xmajorgrids,
xmin=-16, xmax=15.6666666666667,
xtick style={color=white!33.3333333333333!black},
y grid style={white},
ylabel={Distance [m]},
ymajorgrids,
ymin=6, ymax=60,
ytick style={color=white!33.3333333333333!black}
]
\addplot graphics [includegraphics cmd=\pgfimage,xmin=-16, xmax=15.6666666666667, ymin=6, ymax=60] {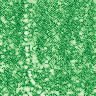};
\addplot graphics [includegraphics cmd=\pgfimage,xmin=-16, xmax=15.6666666666667, ymin=6, ymax=60] {ref_object_overlay_sharp.png};
\end{axis}

\end{tikzpicture}
			\label{fig:result1k3_std}
		}
	}
	\caption{Interference mitigated RD predictions (top) and corresponding uncertainty estimates (bottom) with increasing influence (left to right) of the distribution regularization term with $\lambda=\{10^{-10},10^{-9},10^{-8}\}$. Manually labeled object peaks are indicated in red.}
	\label{fig:result1k_uncertainty}
\end{figure}

\begin{figure}[t]
	\centering
	\footnotesize
	\resizebox{0.8\columnwidth}{!}{
		\subfigure[Object signal]{
\begin{tikzpicture}

\begin{axis}[
scale only axis=true,
width=2.9cm,
height=2.9cm,
axis background/.style={fill=white!89.8039215686275!black},
axis line style={white},
colormap/viridis,
point meta max=0,
point meta min=-30,
tick align=inside,
tick pos=left,
x grid style={white},
xlabel={Velocity [m/s]},
xmajorgrids,
xmin=-16, xmax=15.6666666666667,
xtick style={color=white!33.3333333333333!black},
y grid style={white},
ylabel={Distance [m]},
ymajorgrids,
ymin=6, ymax=60,
ytick style={color=white!33.3333333333333!black}
]
\addplot graphics [includegraphics cmd=\pgfimage,xmin=-16, xmax=15.6666666666667, ymin=6, ymax=60] {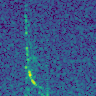};
\addplot graphics [includegraphics cmd=\pgfimage,xmin=-16, xmax=15.6666666666667, ymin=6, ymax=60] {ref_object_overlay_sharp.png};
\end{axis}

\end{tikzpicture}
			\label{fig:result1kc}
		}
		\subfigure[With interference]{
\begin{tikzpicture}

\begin{axis}[
scale only axis=true,
width=2.9cm,
height=2.9cm,
axis background/.style={fill=white!89.8039215686275!black},
axis line style={white},
colorbar,
colorbar style={
	yticklabel style={
		text width=width("$-30$"),
		align=right
	},
	ylabel={dB}
},
colorbar/width=1.5mm,
colormap/viridis,
point meta max=0,
point meta min=-30,
tick align=inside,
tick pos=left,
x grid style={white},
xlabel={Velocity [m/s]},
xmajorgrids,
xmin=-16, xmax=15.6666666666667,
xtick style={color=white!33.3333333333333!black},
y grid style={white},
ylabel={Distance [m]},
ymajorgrids,
ymin=6, ymax=60,
ytick style={color=white!33.3333333333333!black}
]
\addplot graphics [includegraphics cmd=\pgfimage,xmin=-16, xmax=15.6666666666667, ymin=6, ymax=60] {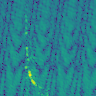};
\addplot graphics [includegraphics cmd=\pgfimage,xmin=-16, xmax=15.6666666666667, ymin=6, ymax=60] {ref_object_overlay_sharp.png};
\end{axis}

\end{tikzpicture}
			\label{fig:result1ki}
		}
	}
	\caption{RD magnitude spectra with and without interference. Manually labeled object peaks are indicated in red.}
	\label{fig:result1k_ref}
\end{figure}

Figure~\ref{fig:result1k_uncertainty} shows the RD predictions and their corresponding uncertainty estimates, i.e. the standard deviation of RD log magnitudes, of S100 over 100 evaluations for different $\lambda \in \{10^{-10},10^{-9},10^{-8}\}$. Figure~\ref{fig:result1k_ref} shows the corresponding 'clean' measurement and the measurement with interference as reference. All RD predictions and uncertainty maps contain an overlay with the 'ground truth' object detections, i.e. the manual labels as considered for evaluation.

As already shown in the last experiment, we achieve the highest interference mitigation performance with a low $\lambda$. Accordingly, the average RD predictions of S100 over 100 evaluations contain well-suppressed interference patterns with $\lambda = 10^{-10}$. For larger $\lambda$ values they contain slightly more noise, interference and smoothed object peaks. Also the standard deviation over the 100 RD log magnitude predictions increases with a higher regularization term, while the object peaks have an extremely low uncertainty independent of $\lambda$. The interference patterns, as contained in the measurement with interference in Figure~\ref{fig:result1k_ref}, become more apparent with higher output uncertainty as shown in Figure~\ref{fig:result1k_uncertainty} in the right column.

\subsection{Visualization and interpretation of learned filter kernels}

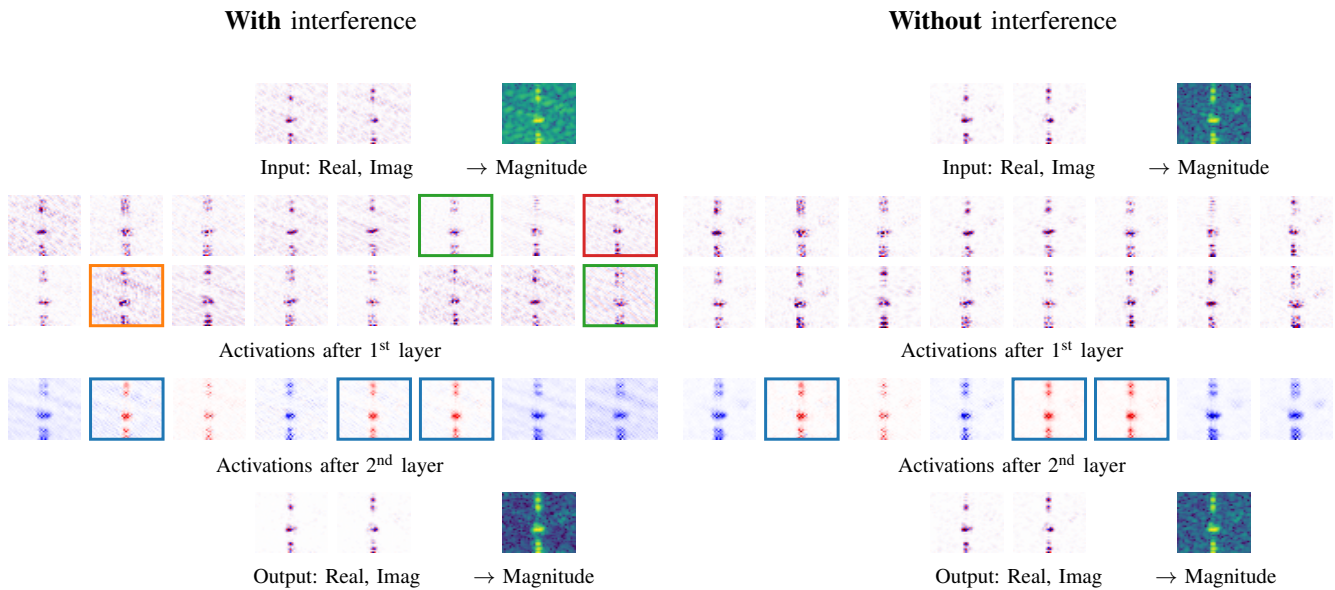
\begin{figure*}
	\centering
	\footnotesize
	{\normalsize
		\begin{tabularx}{\textwidth}{XcXXcX}
			& \textbf{With} interference &&& \textbf{Without} interference & \\
			&&&&&
		\end{tabularx}
	}
\subfigure{
		\resizebox{0.98\columnwidth}{!}{
			\begin{tikzpicture}
\begin{groupplot}[
	group style={group size=8 by 1},
	width=\columnwidth
]\nextgroupplot[
axis background/.style={fill=white!89.80392156862746!black},
axis line style={white},
tick align=outside,
x grid style={white},
xmajorticks=false,
xmin=-6.66666666666667, xmax=6.33333333333333,
xtick style={color=white!33.33333333333333!black},
y grid style={white},
ymajorticks=false,
ymin=6, ymax=28.1684210526316,
ytick style={color=white!33.33333333333333!black},
opacity=0.0,
]
\addplot graphics [includegraphics cmd=\pgfimage,xmin=-6.66666666666667, xmax=6.33333333333333, ymin=6, ymax=28.1684210526316] {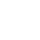};

\nextgroupplot[
axis background/.style={fill=white!89.80392156862746!black},
axis line style={white},
tick align=outside,
x grid style={white},
xmajorticks=false,
xmin=-6.66666666666667, xmax=6.33333333333333,
xtick style={color=white!33.33333333333333!black},
y grid style={white},
ymajorticks=false,
ymin=6, ymax=28.1684210526316,
ytick style={color=white!33.33333333333333!black},
opacity=0.0,
]
\addplot graphics [includegraphics cmd=\pgfimage,xmin=-6.66666666666667, xmax=6.33333333333333, ymin=6, ymax=28.1684210526316] {visualization_activations_placeholder.png};

\nextgroupplot[
axis background/.style={fill=white!89.80392156862746!black},
axis line style={white},
tick align=outside,
x grid style={white},
xmajorticks=false,
xmin=-6.66666666666667, xmax=6.33333333333333,
xtick style={color=white!33.33333333333333!black},
y grid style={white},
ymajorticks=false,
ymin=6, ymax=28.1684210526316,
ytick style={color=white!33.33333333333333!black},
opacity=0.0,
]
\addplot graphics [includegraphics cmd=\pgfimage,xmin=-6.66666666666667, xmax=6.33333333333333, ymin=6, ymax=28.1684210526316] {visualization_activations_placeholder.png};

\nextgroupplot[
axis background/.style={fill=white!89.80392156862746!black},
axis line style={white},
tick align=outside,
x grid style={white},
xmajorticks=false,
xmin=-6.66666666666667, xmax=6.33333333333333,
xtick style={color=white!33.33333333333333!black},
y grid style={white},
ymajorticks=false,
ymin=6, ymax=28.1684210526316,
ytick style={color=white!33.33333333333333!black},
]
\addplot graphics [includegraphics cmd=\pgfimage,xmin=-6.66666666666667, xmax=6.33333333333333, ymin=6, ymax=28.1684210526316] {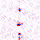};

\nextgroupplot[
axis background/.style={fill=white!89.80392156862746!black},
axis line style={white},
tick align=outside,
x grid style={white},
xmajorticks=false,
xmin=-6.66666666666667, xmax=6.33333333333333,
xtick style={color=white!33.33333333333333!black},
y grid style={white},
ymajorticks=false,
ymin=6, ymax=28.1684210526316,
ytick style={color=white!33.33333333333333!black},
]
\addplot graphics [includegraphics cmd=\pgfimage,xmin=-6.66666666666667, xmax=6.33333333333333, ymin=6, ymax=28.1684210526316] {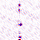};

\nextgroupplot[
axis background/.style={fill=white!89.80392156862746!black},
axis line style={white},
tick align=outside,
x grid style={white},
xmajorticks=false,
xmin=-6.66666666666667, xmax=6.33333333333333,
xtick style={color=white!33.33333333333333!black},
y grid style={white},
ymajorticks=false,
ymin=6, ymax=28.1684210526316,
ytick style={color=white!33.33333333333333!black},
opacity=0.0,
]
\addplot graphics [includegraphics cmd=\pgfimage,xmin=-6.66666666666667, xmax=6.33333333333333, ymin=6, ymax=28.1684210526316] {visualization_activations_placeholder.png};

\nextgroupplot[
axis background/.style={fill=white!89.80392156862746!black},
axis line style={white},
tick align=outside,
x grid style={white},
xmajorticks=false,
xmin=-6.66666666666667, xmax=6.33333333333333,
xtick style={color=white!33.33333333333333!black},
y grid style={white},
ymajorticks=false,
ymin=6, ymax=28.1684210526316,
ytick style={color=white!33.33333333333333!black},
]
\addplot graphics [includegraphics cmd=\pgfimage,xmin=-6.66666666666667, xmax=6.33333333333333, ymin=6, ymax=28.1684210526316] {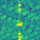};

\nextgroupplot[
axis background/.style={fill=white!89.80392156862746!black},
axis line style={white},
tick align=outside,
x grid style={white},
xmajorticks=false,
xmin=-6.66666666666667, xmax=6.33333333333333,
xtick style={color=white!33.33333333333333!black},
y grid style={white},
ymajorticks=false,
ymin=6, ymax=28.1684210526316,
ytick style={color=white!33.33333333333333!black},
opacity=0.0,
]
\addplot graphics [includegraphics cmd=\pgfimage,xmin=-6.66666666666667, xmax=6.33333333333333, ymin=6, ymax=28.1684210526316] {visualization_activations_placeholder.png};

\end{groupplot}

\end{tikzpicture}
		}
}
\subfigure{
		\resizebox{0.98\columnwidth}{!}{
			\begin{tikzpicture}
\begin{groupplot}[
	group style={group size=8 by 1},
	width=\columnwidth
]\nextgroupplot[
axis background/.style={fill=white!89.80392156862746!black},
axis line style={white},
tick align=outside,
x grid style={white},
xmajorticks=false,
xmin=-6.66666666666667, xmax=6.33333333333333,
xtick style={color=white!33.33333333333333!black},
y grid style={white},
ymajorticks=false,
ymin=6, ymax=28.1684210526316,
ytick style={color=white!33.33333333333333!black},
opacity=0.0,
]
\addplot graphics [includegraphics cmd=\pgfimage,xmin=-6.66666666666667, xmax=6.33333333333333, ymin=6, ymax=28.1684210526316] {visualization_activations_placeholder.png};

\nextgroupplot[
axis background/.style={fill=white!89.80392156862746!black},
axis line style={white},
tick align=outside,
x grid style={white},
xmajorticks=false,
xmin=-6.66666666666667, xmax=6.33333333333333,
xtick style={color=white!33.33333333333333!black},
y grid style={white},
ymajorticks=false,
ymin=6, ymax=28.1684210526316,
ytick style={color=white!33.33333333333333!black},
opacity=0.0,
]
\addplot graphics [includegraphics cmd=\pgfimage,xmin=-6.66666666666667, xmax=6.33333333333333, ymin=6, ymax=28.1684210526316] {visualization_activations_placeholder.png};

\nextgroupplot[
axis background/.style={fill=white!89.80392156862746!black},
axis line style={white},
tick align=outside,
x grid style={white},
xmajorticks=false,
xmin=-6.66666666666667, xmax=6.33333333333333,
xtick style={color=white!33.33333333333333!black},
y grid style={white},
ymajorticks=false,
ymin=6, ymax=28.1684210526316,
ytick style={color=white!33.33333333333333!black},
opacity=0.0,
]
\addplot graphics [includegraphics cmd=\pgfimage,xmin=-6.66666666666667, xmax=6.33333333333333, ymin=6, ymax=28.1684210526316] {visualization_activations_placeholder.png};

\nextgroupplot[
axis background/.style={fill=white!89.80392156862746!black},
axis line style={white},
tick align=outside,
x grid style={white},
xmajorticks=false,
xmin=-6.66666666666667, xmax=6.33333333333333,
xtick style={color=white!33.33333333333333!black},
y grid style={white},
ymajorticks=false,
ymin=6, ymax=28.1684210526316,
ytick style={color=white!33.33333333333333!black},
]
\addplot graphics [includegraphics cmd=\pgfimage,xmin=-6.66666666666667, xmax=6.33333333333333, ymin=6, ymax=28.1684210526316] {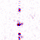};

\nextgroupplot[
axis background/.style={fill=white!89.80392156862746!black},
axis line style={white},
tick align=outside,
x grid style={white},
xmajorticks=false,
xmin=-6.66666666666667, xmax=6.33333333333333,
xtick style={color=white!33.33333333333333!black},
y grid style={white},
ymajorticks=false,
ymin=6, ymax=28.1684210526316,
ytick style={color=white!33.33333333333333!black},
]
\addplot graphics [includegraphics cmd=\pgfimage,xmin=-6.66666666666667, xmax=6.33333333333333, ymin=6, ymax=28.1684210526316] {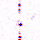};

\nextgroupplot[
axis background/.style={fill=white!89.80392156862746!black},
axis line style={white},
tick align=outside,
x grid style={white},
xmajorticks=false,
xmin=-6.66666666666667, xmax=6.33333333333333,
xtick style={color=white!33.33333333333333!black},
y grid style={white},
ymajorticks=false,
ymin=6, ymax=28.1684210526316,
ytick style={color=white!33.33333333333333!black},
opacity=0.0,
]
\addplot graphics [includegraphics cmd=\pgfimage,xmin=-6.66666666666667, xmax=6.33333333333333, ymin=6, ymax=28.1684210526316] {visualization_activations_placeholder.png};

\nextgroupplot[
axis background/.style={fill=white!89.80392156862746!black},
axis line style={white},
tick align=outside,
x grid style={white},
xmajorticks=false,
xmin=-6.66666666666667, xmax=6.33333333333333,
xtick style={color=white!33.33333333333333!black},
y grid style={white},
ymajorticks=false,
ymin=6, ymax=28.1684210526316,
ytick style={color=white!33.33333333333333!black},
]
\addplot graphics [includegraphics cmd=\pgfimage,xmin=-6.66666666666667, xmax=6.33333333333333, ymin=6, ymax=28.1684210526316] {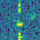};

\nextgroupplot[
axis background/.style={fill=white!89.80392156862746!black},
axis line style={white},
tick align=outside,
x grid style={white},
xmajorticks=false,
xmin=-6.66666666666667, xmax=6.33333333333333,
xtick style={color=white!33.33333333333333!black},
y grid style={white},
ymajorticks=false,
ymin=6, ymax=28.1684210526316,
ytick style={color=white!33.33333333333333!black},
opacity=0.0,
]
\addplot graphics [includegraphics cmd=\pgfimage,xmin=-6.66666666666667, xmax=6.33333333333333, ymin=6, ymax=28.1684210526316] {visualization_activations_placeholder.png};

\end{groupplot}
\end{tikzpicture}
		}
}
{
	\begin{tabularx}{\textwidth}{XcXXcX}
		& \hspace{25mm}Input: Real, Imag\hspace{7mm}$\rightarrow$ Magnitude &&& \hspace{25mm}Input: Real, Imag\hspace{7mm}$\rightarrow$ Magnitude &
	\end{tabularx}
}
\subfigure{
		\resizebox{0.98\columnwidth}{!}{
			\begin{tikzpicture}

\definecolor{color0}{rgb}{0.172549019607843,0.627450980392157,0.172549019607843}
\definecolor{color1}{rgb}{0.83921568627451,0.152941176470588,0.156862745098039}
\definecolor{color3}{rgb}{1,0.498039215686275,0.0549019607843137}

\begin{groupplot}[
	group style={group size=8 by 2},
	width=\columnwidth
]\nextgroupplot[
axis background/.style={fill=white!89.80392156862746!black},
axis line style={white},
tick align=outside,
x grid style={white},
xmajorticks=false,
xmin=-6.66666666666667, xmax=6.33333333333333,
xtick style={color=white!33.33333333333333!black},
y grid style={white},
ymajorticks=false,
ymin=6, ymax=28.1684210526316,
ytick style={color=white!33.33333333333333!black},
]
\addplot graphics [includegraphics cmd=\pgfimage,xmin=-6.66666666666667, xmax=6.33333333333333, ymin=6, ymax=28.1684210526316] {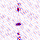};

\nextgroupplot[
axis background/.style={fill=white!89.80392156862746!black},
axis line style={white},
tick align=outside,
x grid style={white},
xmajorticks=false,
xmin=-6.66666666666667, xmax=6.33333333333333,
xtick style={color=white!33.33333333333333!black},
y grid style={white},
ymajorticks=false,
ymin=6, ymax=28.1684210526316,
ytick style={color=white!33.33333333333333!black},
]
\addplot graphics [includegraphics cmd=\pgfimage,xmin=-6.66666666666667, xmax=6.33333333333333, ymin=6, ymax=28.1684210526316] {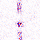};

\nextgroupplot[
axis background/.style={fill=white!89.80392156862746!black},
axis line style={white},
tick align=outside,
x grid style={white},
xmajorticks=false,
xmin=-6.66666666666667, xmax=6.33333333333333,
xtick style={color=white!33.33333333333333!black},
y grid style={white},
ymajorticks=false,
ymin=6, ymax=28.1684210526316,
ytick style={color=white!33.33333333333333!black},
]
\addplot graphics [includegraphics cmd=\pgfimage,xmin=-6.66666666666667, xmax=6.33333333333333, ymin=6, ymax=28.1684210526316] {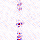};

\nextgroupplot[
axis background/.style={fill=white!89.80392156862746!black},
axis line style={white},
tick align=outside,
x grid style={white},
xmajorticks=false,
xmin=-6.66666666666667, xmax=6.33333333333333,
xtick style={color=white!33.33333333333333!black},
y grid style={white},
ymajorticks=false,
ymin=6, ymax=28.1684210526316,
ytick style={color=white!33.33333333333333!black},
]
\addplot graphics [includegraphics cmd=\pgfimage,xmin=-6.66666666666667, xmax=6.33333333333333, ymin=6, ymax=28.1684210526316] {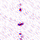};

\nextgroupplot[
axis background/.style={fill=white!89.80392156862746!black},
axis line style={white},
tick align=outside,
x grid style={white},
xmajorticks=false,
xmin=-6.66666666666667, xmax=6.33333333333333,
xtick style={color=white!33.33333333333333!black},
y grid style={white},
ymajorticks=false,
ymin=6, ymax=28.1684210526316,
ytick style={color=white!33.33333333333333!black},
]
\addplot graphics [includegraphics cmd=\pgfimage,xmin=-6.66666666666667, xmax=6.33333333333333, ymin=6, ymax=28.1684210526316] {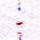};

\nextgroupplot[
axis background/.style={fill=white!89.80392156862746!black},
set layers=axis on top,
axis line style={color0, line width=3mm},
tick align=outside,
x grid style={white},
xmajorticks=false,
xmin=-6.66666666666667, xmax=6.33333333333333,
xtick style={color=white!33.33333333333333!black},
y grid style={white},
ymajorticks=false,
ymin=6, ymax=28.1684210526316,
ytick style={color=white!33.33333333333333!black},
]
\addplot graphics [includegraphics cmd=\pgfimage,xmin=-6.66666666666667, xmax=6.33333333333333, ymin=6, ymax=28.1684210526316] {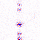};

\nextgroupplot[
axis background/.style={fill=white!89.80392156862746!black},
axis line style={white},
tick align=outside,
x grid style={white},
xmajorticks=false,
xmin=-6.66666666666667, xmax=6.33333333333333,
xtick style={color=white!33.33333333333333!black},
y grid style={white},
ymajorticks=false,
ymin=6, ymax=28.1684210526316,
ytick style={color=white!33.33333333333333!black},
]
\addplot graphics [includegraphics cmd=\pgfimage,xmin=-6.66666666666667, xmax=6.33333333333333, ymin=6, ymax=28.1684210526316] {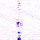};

\nextgroupplot[
axis background/.style={fill=white!89.80392156862746!black},
set layers=axis on top,
axis line style={color1, line width=3mm},
tick align=outside,
x grid style={white},
xmajorticks=false,
xmin=-6.66666666666667, xmax=6.33333333333333,
xtick style={color=white!33.33333333333333!black},
y grid style={white},
ymajorticks=false,
ymin=6, ymax=28.1684210526316,
ytick style={color=white!33.33333333333333!black},
]
\addplot graphics [includegraphics cmd=\pgfimage,xmin=-6.66666666666667, xmax=6.33333333333333, ymin=6, ymax=28.1684210526316] {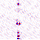};

\nextgroupplot[
axis background/.style={fill=white!89.80392156862746!black},
axis line style={white},
tick align=outside,
x grid style={white},
xmajorticks=false,
xmin=-6.66666666666667, xmax=6.33333333333333,
xtick style={color=white!33.33333333333333!black},
y grid style={white},
ymajorticks=false,
ymin=6, ymax=28.1684210526316,
ytick style={color=white!33.33333333333333!black},
]
\addplot graphics [includegraphics cmd=\pgfimage,xmin=-6.66666666666667, xmax=6.33333333333333, ymin=6, ymax=28.1684210526316] {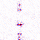};

\nextgroupplot[
axis background/.style={fill=white!89.80392156862746!black},
set layers=axis on top,
axis line style={color3, line width=3mm},
tick align=outside,
x grid style={white},
xmajorticks=false,
xmin=-6.66666666666667, xmax=6.33333333333333,
xtick style={color=white!33.33333333333333!black},
y grid style={white},
ymajorticks=false,
ymin=6, ymax=28.1684210526316,
ytick style={color=white!33.33333333333333!black},
]
\addplot graphics [includegraphics cmd=\pgfimage,xmin=-6.66666666666667, xmax=6.33333333333333, ymin=6, ymax=28.1684210526316] {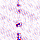};

\nextgroupplot[
axis background/.style={fill=white!89.80392156862746!black},
axis line style={white},
tick align=outside,
x grid style={white},
xmajorticks=false,
xmin=-6.66666666666667, xmax=6.33333333333333,
xtick style={color=white!33.33333333333333!black},
y grid style={white},
ymajorticks=false,
ymin=6, ymax=28.1684210526316,
ytick style={color=white!33.33333333333333!black},
]
\addplot graphics [includegraphics cmd=\pgfimage,xmin=-6.66666666666667, xmax=6.33333333333333, ymin=6, ymax=28.1684210526316] {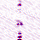};

\nextgroupplot[
axis background/.style={fill=white!89.80392156862746!black},
axis line style={white},
tick align=outside,
x grid style={white},
xmajorticks=false,
xmin=-6.66666666666667, xmax=6.33333333333333,
xtick style={color=white!33.33333333333333!black},
y grid style={white},
ymajorticks=false,
ymin=6, ymax=28.1684210526316,
ytick style={color=white!33.33333333333333!black},
]
\addplot graphics [includegraphics cmd=\pgfimage,xmin=-6.66666666666667, xmax=6.33333333333333, ymin=6, ymax=28.1684210526316] {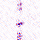};

\nextgroupplot[
axis background/.style={fill=white!89.80392156862746!black},
axis line style={white},
tick align=outside,
x grid style={white},
xmajorticks=false,
xmin=-6.66666666666667, xmax=6.33333333333333,
xtick style={color=white!33.33333333333333!black},
y grid style={white},
ymajorticks=false,
ymin=6, ymax=28.1684210526316,
ytick style={color=white!33.33333333333333!black},
]
\addplot graphics [includegraphics cmd=\pgfimage,xmin=-6.66666666666667, xmax=6.33333333333333, ymin=6, ymax=28.1684210526316] {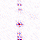};

\nextgroupplot[
axis background/.style={fill=white!89.80392156862746!black},
axis line style={white},
tick align=outside,
x grid style={white},
xmajorticks=false,
xmin=-6.66666666666667, xmax=6.33333333333333,
xtick style={color=white!33.33333333333333!black},
y grid style={white},
ymajorticks=false,
ymin=6, ymax=28.1684210526316,
ytick style={color=white!33.33333333333333!black},
]
\addplot graphics [includegraphics cmd=\pgfimage,xmin=-6.66666666666667, xmax=6.33333333333333, ymin=6, ymax=28.1684210526316] {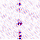};

\nextgroupplot[
axis background/.style={fill=white!89.80392156862746!black},
axis line style={white},
tick align=outside,
x grid style={white},
xmajorticks=false,
xmin=-6.66666666666667, xmax=6.33333333333333,
xtick style={color=white!33.33333333333333!black},
y grid style={white},
ymajorticks=false,
ymin=6, ymax=28.1684210526316,
ytick style={color=white!33.33333333333333!black},
]
\addplot graphics [includegraphics cmd=\pgfimage,xmin=-6.66666666666667, xmax=6.33333333333333, ymin=6, ymax=28.1684210526316] {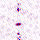};

\nextgroupplot[
axis background/.style={fill=white!89.80392156862746!black},
set layers=axis on top,
axis line style={color0, line width=3mm},
tick align=outside,
x grid style={white},
xmajorticks=false,
xmin=-6.66666666666667, xmax=6.33333333333333,
xtick style={color=white!33.33333333333333!black},
y grid style={white},
ymajorticks=false,
ymin=6, ymax=28.1684210526316,
ytick style={color=white!33.33333333333333!black},
]
\addplot graphics [includegraphics cmd=\pgfimage,xmin=-6.66666666666667, xmax=6.33333333333333, ymin=6, ymax=28.1684210526316] {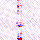};

\end{groupplot}
\end{tikzpicture}
		}
}
\subfigure{
		\resizebox{0.98\columnwidth}{!}{
			\begin{tikzpicture}
\begin{groupplot}[
	group style={group size=8 by 2},
	width=\columnwidth
]\nextgroupplot[
axis background/.style={fill=white!89.80392156862746!black},
axis line style={white},
tick align=outside,
x grid style={white},
xmajorticks=false,
xmin=-6.66666666666667, xmax=6.33333333333333,
xtick style={color=white!33.33333333333333!black},
y grid style={white},
ymajorticks=false,
ymin=6, ymax=28.1684210526316,
ytick style={color=white!33.33333333333333!black},
]
\addplot graphics [includegraphics cmd=\pgfimage,xmin=-6.66666666666667, xmax=6.33333333333333, ymin=6, ymax=28.1684210526316] {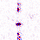};

\nextgroupplot[
axis background/.style={fill=white!89.80392156862746!black},
axis line style={white},
tick align=outside,
x grid style={white},
xmajorticks=false,
xmin=-6.66666666666667, xmax=6.33333333333333,
xtick style={color=white!33.33333333333333!black},
y grid style={white},
ymajorticks=false,
ymin=6, ymax=28.1684210526316,
ytick style={color=white!33.33333333333333!black},
]
\addplot graphics [includegraphics cmd=\pgfimage,xmin=-6.66666666666667, xmax=6.33333333333333, ymin=6, ymax=28.1684210526316] {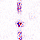};

\nextgroupplot[
axis background/.style={fill=white!89.80392156862746!black},
axis line style={white},
tick align=outside,
x grid style={white},
xmajorticks=false,
xmin=-6.66666666666667, xmax=6.33333333333333,
xtick style={color=white!33.33333333333333!black},
y grid style={white},
ymajorticks=false,
ymin=6, ymax=28.1684210526316,
ytick style={color=white!33.33333333333333!black},
]
\addplot graphics [includegraphics cmd=\pgfimage,xmin=-6.66666666666667, xmax=6.33333333333333, ymin=6, ymax=28.1684210526316] {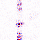};

\nextgroupplot[
axis background/.style={fill=white!89.80392156862746!black},
axis line style={white},
tick align=outside,
x grid style={white},
xmajorticks=false,
xmin=-6.66666666666667, xmax=6.33333333333333,
xtick style={color=white!33.33333333333333!black},
y grid style={white},
ymajorticks=false,
ymin=6, ymax=28.1684210526316,
ytick style={color=white!33.33333333333333!black},
]
\addplot graphics [includegraphics cmd=\pgfimage,xmin=-6.66666666666667, xmax=6.33333333333333, ymin=6, ymax=28.1684210526316] {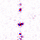};

\nextgroupplot[
axis background/.style={fill=white!89.80392156862746!black},
axis line style={white},
tick align=outside,
x grid style={white},
xmajorticks=false,
xmin=-6.66666666666667, xmax=6.33333333333333,
xtick style={color=white!33.33333333333333!black},
y grid style={white},
ymajorticks=false,
ymin=6, ymax=28.1684210526316,
ytick style={color=white!33.33333333333333!black},
]
\addplot graphics [includegraphics cmd=\pgfimage,xmin=-6.66666666666667, xmax=6.33333333333333, ymin=6, ymax=28.1684210526316] {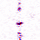};

\nextgroupplot[
axis background/.style={fill=white!89.80392156862746!black},
axis line style={white},
tick align=outside,
x grid style={white},
xmajorticks=false,
xmin=-6.66666666666667, xmax=6.33333333333333,
xtick style={color=white!33.33333333333333!black},
y grid style={white},
ymajorticks=false,
ymin=6, ymax=28.1684210526316,
ytick style={color=white!33.33333333333333!black},
]
\addplot graphics [includegraphics cmd=\pgfimage,xmin=-6.66666666666667, xmax=6.33333333333333, ymin=6, ymax=28.1684210526316] {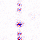};

\nextgroupplot[
axis background/.style={fill=white!89.80392156862746!black},
axis line style={white},
tick align=outside,
x grid style={white},
xmajorticks=false,
xmin=-6.66666666666667, xmax=6.33333333333333,
xtick style={color=white!33.33333333333333!black},
y grid style={white},
ymajorticks=false,
ymin=6, ymax=28.1684210526316,
ytick style={color=white!33.33333333333333!black},
]
\addplot graphics [includegraphics cmd=\pgfimage,xmin=-6.66666666666667, xmax=6.33333333333333, ymin=6, ymax=28.1684210526316] {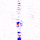};

\nextgroupplot[
axis background/.style={fill=white!89.80392156862746!black},
axis line style={white},
tick align=outside,
x grid style={white},
xmajorticks=false,
xmin=-6.66666666666667, xmax=6.33333333333333,
xtick style={color=white!33.33333333333333!black},
y grid style={white},
ymajorticks=false,
ymin=6, ymax=28.1684210526316,
ytick style={color=white!33.33333333333333!black},
]
\addplot graphics [includegraphics cmd=\pgfimage,xmin=-6.66666666666667, xmax=6.33333333333333, ymin=6, ymax=28.1684210526316] {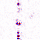};

\nextgroupplot[
axis background/.style={fill=white!89.80392156862746!black},
axis line style={white},
tick align=outside,
x grid style={white},
xmajorticks=false,
xmin=-6.66666666666667, xmax=6.33333333333333,
xtick style={color=white!33.33333333333333!black},
y grid style={white},
ymajorticks=false,
ymin=6, ymax=28.1684210526316,
ytick style={color=white!33.33333333333333!black},
]
\addplot graphics [includegraphics cmd=\pgfimage,xmin=-6.66666666666667, xmax=6.33333333333333, ymin=6, ymax=28.1684210526316] {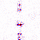};

\nextgroupplot[
axis background/.style={fill=white!89.80392156862746!black},
axis line style={white},
tick align=outside,
x grid style={white},
xmajorticks=false,
xmin=-6.66666666666667, xmax=6.33333333333333,
xtick style={color=white!33.33333333333333!black},
y grid style={white},
ymajorticks=false,
ymin=6, ymax=28.1684210526316,
ytick style={color=white!33.33333333333333!black},
]
\addplot graphics [includegraphics cmd=\pgfimage,xmin=-6.66666666666667, xmax=6.33333333333333, ymin=6, ymax=28.1684210526316] {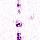};

\nextgroupplot[
axis background/.style={fill=white!89.80392156862746!black},
axis line style={white},
tick align=outside,
x grid style={white},
xmajorticks=false,
xmin=-6.66666666666667, xmax=6.33333333333333,
xtick style={color=white!33.33333333333333!black},
y grid style={white},
ymajorticks=false,
ymin=6, ymax=28.1684210526316,
ytick style={color=white!33.33333333333333!black},
]
\addplot graphics [includegraphics cmd=\pgfimage,xmin=-6.66666666666667, xmax=6.33333333333333, ymin=6, ymax=28.1684210526316] {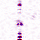};

\nextgroupplot[
axis background/.style={fill=white!89.80392156862746!black},
axis line style={white},
tick align=outside,
x grid style={white},
xmajorticks=false,
xmin=-6.66666666666667, xmax=6.33333333333333,
xtick style={color=white!33.33333333333333!black},
y grid style={white},
ymajorticks=false,
ymin=6, ymax=28.1684210526316,
ytick style={color=white!33.33333333333333!black},
]
\addplot graphics [includegraphics cmd=\pgfimage,xmin=-6.66666666666667, xmax=6.33333333333333, ymin=6, ymax=28.1684210526316] {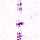};

\nextgroupplot[
axis background/.style={fill=white!89.80392156862746!black},
axis line style={white},
tick align=outside,
x grid style={white},
xmajorticks=false,
xmin=-6.66666666666667, xmax=6.33333333333333,
xtick style={color=white!33.33333333333333!black},
y grid style={white},
ymajorticks=false,
ymin=6, ymax=28.1684210526316,
ytick style={color=white!33.33333333333333!black},
]
\addplot graphics [includegraphics cmd=\pgfimage,xmin=-6.66666666666667, xmax=6.33333333333333, ymin=6, ymax=28.1684210526316] {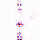};

\nextgroupplot[
axis background/.style={fill=white!89.80392156862746!black},
axis line style={white},
tick align=outside,
x grid style={white},
xmajorticks=false,
xmin=-6.66666666666667, xmax=6.33333333333333,
xtick style={color=white!33.33333333333333!black},
y grid style={white},
ymajorticks=false,
ymin=6, ymax=28.1684210526316,
ytick style={color=white!33.33333333333333!black},
]
\addplot graphics [includegraphics cmd=\pgfimage,xmin=-6.66666666666667, xmax=6.33333333333333, ymin=6, ymax=28.1684210526316] {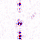};

\nextgroupplot[
axis background/.style={fill=white!89.80392156862746!black},
axis line style={white},
tick align=outside,
x grid style={white},
xmajorticks=false,
xmin=-6.66666666666667, xmax=6.33333333333333,
xtick style={color=white!33.33333333333333!black},
y grid style={white},
ymajorticks=false,
ymin=6, ymax=28.1684210526316,
ytick style={color=white!33.33333333333333!black},
]
\addplot graphics [includegraphics cmd=\pgfimage,xmin=-6.66666666666667, xmax=6.33333333333333, ymin=6, ymax=28.1684210526316] {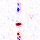};

\nextgroupplot[
axis background/.style={fill=white!89.80392156862746!black},
axis line style={white},
tick align=outside,
x grid style={white},
xmajorticks=false,
xmin=-6.66666666666667, xmax=6.33333333333333,
xtick style={color=white!33.33333333333333!black},
y grid style={white},
ymajorticks=false,
ymin=6, ymax=28.1684210526316,
ytick style={color=white!33.33333333333333!black},
]
\addplot graphics [includegraphics cmd=\pgfimage,xmin=-6.66666666666667, xmax=6.33333333333333, ymin=6, ymax=28.1684210526316] {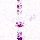};

\end{groupplot}
\end{tikzpicture}
		}
}
{
	\begin{tabularx}{\textwidth}{XcXXcX}
		& Activations after 1\textsuperscript{st} layer &&& Activations after 1\textsuperscript{st} layer &
	\end{tabularx}
}
\subfigure{
		\resizebox{0.98\columnwidth}{!}{
			\begin{tikzpicture}
	
\definecolor{blue0}{rgb}{0.12156862745098,0.466666666666667,0.705882352941177}

\begin{groupplot}[
	group style={group size=8 by 1},
	width=\columnwidth
]\nextgroupplot[
axis background/.style={fill=white!89.80392156862746!black},
axis line style={white},
tick align=outside,
x grid style={white},
xmajorticks=false,
xmin=-6.66666666666667, xmax=6.33333333333333,
xtick style={color=white!33.33333333333333!black},
y grid style={white},
ymajorticks=false,
ymin=6, ymax=28.1684210526316,
ytick style={color=white!33.33333333333333!black},
]
\addplot graphics [includegraphics cmd=\pgfimage,xmin=-6.66666666666667, xmax=6.33333333333333, ymin=6, ymax=28.1684210526316] {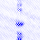};

\nextgroupplot[
axis background/.style={fill=white!89.80392156862746!black},
set layers=axis on top,
axis line style={blue0, line width=3mm},
tick align=outside,
x grid style={white},
xmajorticks=false,
xmin=-6.66666666666667, xmax=6.33333333333333,
xtick style={color=white!33.33333333333333!black},
y grid style={white},
ymajorticks=false,
ymin=6, ymax=28.1684210526316,
ytick style={color=white!33.33333333333333!black},
]
\addplot graphics [includegraphics cmd=\pgfimage,xmin=-6.66666666666667, xmax=6.33333333333333, ymin=6, ymax=28.1684210526316] {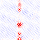};

\nextgroupplot[
axis background/.style={fill=white!89.80392156862746!black},
axis line style={white},
tick align=outside,
x grid style={white},
xmajorticks=false,
xmin=-6.66666666666667, xmax=6.33333333333333,
xtick style={color=white!33.33333333333333!black},
y grid style={white},
ymajorticks=false,
ymin=6, ymax=28.1684210526316,
ytick style={color=white!33.33333333333333!black},
]
\addplot graphics [includegraphics cmd=\pgfimage,xmin=-6.66666666666667, xmax=6.33333333333333, ymin=6, ymax=28.1684210526316] {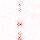};

\nextgroupplot[
axis background/.style={fill=white!89.80392156862746!black},
axis line style={white},
tick align=outside,
x grid style={white},
xmajorticks=false,
xmin=-6.66666666666667, xmax=6.33333333333333,
xtick style={color=white!33.33333333333333!black},
y grid style={white},
ymajorticks=false,
ymin=6, ymax=28.1684210526316,
ytick style={color=white!33.33333333333333!black},
]
\addplot graphics [includegraphics cmd=\pgfimage,xmin=-6.66666666666667, xmax=6.33333333333333, ymin=6, ymax=28.1684210526316] {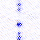};

\nextgroupplot[
axis background/.style={fill=white!89.80392156862746!black},
set layers=axis on top,
axis line style={blue0, line width=3mm},
tick align=outside,
x grid style={white},
xmajorticks=false,
xmin=-6.66666666666667, xmax=6.33333333333333,
xtick style={color=white!33.33333333333333!black},
y grid style={white},
ymajorticks=false,
ymin=6, ymax=28.1684210526316,
ytick style={color=white!33.33333333333333!black},
]
\addplot graphics [includegraphics cmd=\pgfimage,xmin=-6.66666666666667, xmax=6.33333333333333, ymin=6, ymax=28.1684210526316] {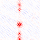};

\nextgroupplot[
axis background/.style={fill=white!89.80392156862746!black},
set layers=axis on top,
axis line style={blue0, line width=3mm},
tick align=outside,
x grid style={white},
xmajorticks=false,
xmin=-6.66666666666667, xmax=6.33333333333333,
xtick style={color=white!33.33333333333333!black},
y grid style={white},
ymajorticks=false,
ymin=6, ymax=28.1684210526316,
ytick style={color=white!33.33333333333333!black},
]
\addplot graphics [includegraphics cmd=\pgfimage,xmin=-6.66666666666667, xmax=6.33333333333333, ymin=6, ymax=28.1684210526316] {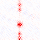};

\nextgroupplot[
axis background/.style={fill=white!89.80392156862746!black},
axis line style={white},
tick align=outside,
x grid style={white},
xmajorticks=false,
xmin=-6.66666666666667, xmax=6.33333333333333,
xtick style={color=white!33.33333333333333!black},
y grid style={white},
ymajorticks=false,
ymin=6, ymax=28.1684210526316,
ytick style={color=white!33.33333333333333!black},
]
\addplot graphics [includegraphics cmd=\pgfimage,xmin=-6.66666666666667, xmax=6.33333333333333, ymin=6, ymax=28.1684210526316] {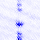};

\nextgroupplot[
axis background/.style={fill=white!89.80392156862746!black},
axis line style={white},
tick align=outside,
x grid style={white},
xmajorticks=false,
xmin=-6.66666666666667, xmax=6.33333333333333,
xtick style={color=white!33.33333333333333!black},
y grid style={white},
ymajorticks=false,
ymin=6, ymax=28.1684210526316,
ytick style={color=white!33.33333333333333!black},
]
\addplot graphics [includegraphics cmd=\pgfimage,xmin=-6.66666666666667, xmax=6.33333333333333, ymin=6, ymax=28.1684210526316] {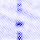};

\end{groupplot}
\end{tikzpicture}
		}
}
\subfigure{
		\resizebox{0.98\columnwidth}{!}{
			\begin{tikzpicture}
	
\definecolor{blue0}{rgb}{0.12156862745098,0.466666666666667,0.705882352941177}

\begin{groupplot}[
	group style={group size=8 by 1},
	width=\columnwidth
]\nextgroupplot[
axis background/.style={fill=white!89.80392156862746!black},
axis line style={white},
tick align=outside,
x grid style={white},
xmajorticks=false,
xmin=-6.66666666666667, xmax=6.33333333333333,
xtick style={color=white!33.33333333333333!black},
y grid style={white},
ymajorticks=false,
ymin=6, ymax=28.1684210526316,
ytick style={color=white!33.33333333333333!black},
]
\addplot graphics [includegraphics cmd=\pgfimage,xmin=-6.66666666666667, xmax=6.33333333333333, ymin=6, ymax=28.1684210526316] {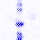};

\nextgroupplot[
axis background/.style={fill=white!89.80392156862746!black},
set layers=axis on top,
axis line style={blue0, line width=3mm},
tick align=outside,
x grid style={white},
xmajorticks=false,
xmin=-6.66666666666667, xmax=6.33333333333333,
xtick style={color=white!33.33333333333333!black},
y grid style={white},
ymajorticks=false,
ymin=6, ymax=28.1684210526316,
ytick style={color=white!33.33333333333333!black},
]
\addplot graphics [includegraphics cmd=\pgfimage,xmin=-6.66666666666667, xmax=6.33333333333333, ymin=6, ymax=28.1684210526316] {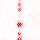};

\nextgroupplot[
axis background/.style={fill=white!89.80392156862746!black},
axis line style={white},
tick align=outside,
x grid style={white},
xmajorticks=false,
xmin=-6.66666666666667, xmax=6.33333333333333,
xtick style={color=white!33.33333333333333!black},
y grid style={white},
ymajorticks=false,
ymin=6, ymax=28.1684210526316,
ytick style={color=white!33.33333333333333!black},
]
\addplot graphics [includegraphics cmd=\pgfimage,xmin=-6.66666666666667, xmax=6.33333333333333, ymin=6, ymax=28.1684210526316] {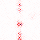};

\nextgroupplot[
axis background/.style={fill=white!89.80392156862746!black},
axis line style={white},
tick align=outside,
x grid style={white},
xmajorticks=false,
xmin=-6.66666666666667, xmax=6.33333333333333,
xtick style={color=white!33.33333333333333!black},
y grid style={white},
ymajorticks=false,
ymin=6, ymax=28.1684210526316,
ytick style={color=white!33.33333333333333!black},
]
\addplot graphics [includegraphics cmd=\pgfimage,xmin=-6.66666666666667, xmax=6.33333333333333, ymin=6, ymax=28.1684210526316] {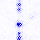};

\nextgroupplot[
axis background/.style={fill=white!89.80392156862746!black},
set layers=axis on top,
axis line style={blue0, line width=3mm},
tick align=outside,
x grid style={white},
xmajorticks=false,
xmin=-6.66666666666667, xmax=6.33333333333333,
xtick style={color=white!33.33333333333333!black},
y grid style={white},
ymajorticks=false,
ymin=6, ymax=28.1684210526316,
ytick style={color=white!33.33333333333333!black},
]
\addplot graphics [includegraphics cmd=\pgfimage,xmin=-6.66666666666667, xmax=6.33333333333333, ymin=6, ymax=28.1684210526316] {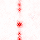};

\nextgroupplot[
axis background/.style={fill=white!89.80392156862746!black},
set layers=axis on top,
axis line style={blue0, line width=3mm},
tick align=outside,
x grid style={white},
xmajorticks=false,
xmin=-6.66666666666667, xmax=6.33333333333333,
xtick style={color=white!33.33333333333333!black},
y grid style={white},
ymajorticks=false,
ymin=6, ymax=28.1684210526316,
ytick style={color=white!33.33333333333333!black},
]
\addplot graphics [includegraphics cmd=\pgfimage,xmin=-6.66666666666667, xmax=6.33333333333333, ymin=6, ymax=28.1684210526316] {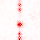};

\nextgroupplot[
axis background/.style={fill=white!89.80392156862746!black},
axis line style={white},
tick align=outside,
x grid style={white},
xmajorticks=false,
xmin=-6.66666666666667, xmax=6.33333333333333,
xtick style={color=white!33.33333333333333!black},
y grid style={white},
ymajorticks=false,
ymin=6, ymax=28.1684210526316,
ytick style={color=white!33.33333333333333!black},
]
\addplot graphics [includegraphics cmd=\pgfimage,xmin=-6.66666666666667, xmax=6.33333333333333, ymin=6, ymax=28.1684210526316] {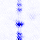};

\nextgroupplot[
axis background/.style={fill=white!89.80392156862746!black},
axis line style={white},
tick align=outside,
x grid style={white},
xmajorticks=false,
xmin=-6.66666666666667, xmax=6.33333333333333,
xtick style={color=white!33.33333333333333!black},
y grid style={white},
ymajorticks=false,
ymin=6, ymax=28.1684210526316,
ytick style={color=white!33.33333333333333!black},
]
\addplot graphics [includegraphics cmd=\pgfimage,xmin=-6.66666666666667, xmax=6.33333333333333, ymin=6, ymax=28.1684210526316] {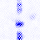};

\end{groupplot}
\end{tikzpicture}
		}
}
{
	\begin{tabularx}{\textwidth}{XcXXcX}
		& Activations after 2\textsuperscript{nd} layer &&& Activations after 2\textsuperscript{nd} layer &
	\end{tabularx}
}
\subfigure{
		\resizebox{0.98\columnwidth}{!}{
			\begin{tikzpicture}
\begin{groupplot}[
	group style={group size=8 by 1},
	width=\columnwidth
]\nextgroupplot[
axis background/.style={fill=white!89.80392156862746!black},
axis line style={white},
tick align=outside,
x grid style={white},
xmajorticks=false,
xmin=-6.66666666666667, xmax=6.33333333333333,
xtick style={color=white!33.33333333333333!black},
y grid style={white},
ymajorticks=false,
ymin=6, ymax=28.1684210526316,
ytick style={color=white!33.33333333333333!black},
opacity=0.0,
]
\addplot graphics [includegraphics cmd=\pgfimage,xmin=-6.66666666666667, xmax=6.33333333333333, ymin=6, ymax=28.1684210526316] {visualization_activations_placeholder.png};

\nextgroupplot[
axis background/.style={fill=white!89.80392156862746!black},
axis line style={white},
tick align=outside,
x grid style={white},
xmajorticks=false,
xmin=-6.66666666666667, xmax=6.33333333333333,
xtick style={color=white!33.33333333333333!black},
y grid style={white},
ymajorticks=false,
ymin=6, ymax=28.1684210526316,
ytick style={color=white!33.33333333333333!black},
opacity=0.0,
]
\addplot graphics [includegraphics cmd=\pgfimage,xmin=-6.66666666666667, xmax=6.33333333333333, ymin=6, ymax=28.1684210526316] {visualization_activations_placeholder.png};

\nextgroupplot[
axis background/.style={fill=white!89.80392156862746!black},
axis line style={white},
tick align=outside,
x grid style={white},
xmajorticks=false,
xmin=-6.66666666666667, xmax=6.33333333333333,
xtick style={color=white!33.33333333333333!black},
y grid style={white},
ymajorticks=false,
ymin=6, ymax=28.1684210526316,
ytick style={color=white!33.33333333333333!black},
opacity=0.0,
]
\addplot graphics [includegraphics cmd=\pgfimage,xmin=-6.66666666666667, xmax=6.33333333333333, ymin=6, ymax=28.1684210526316] {visualization_activations_placeholder.png};

\nextgroupplot[
axis background/.style={fill=white!89.80392156862746!black},
axis line style={white},
tick align=outside,
x grid style={white},
xmajorticks=false,
xmin=-6.66666666666667, xmax=6.33333333333333,
xtick style={color=white!33.33333333333333!black},
y grid style={white},
ymajorticks=false,
ymin=6, ymax=28.1684210526316,
ytick style={color=white!33.33333333333333!black},
]
\addplot graphics [includegraphics cmd=\pgfimage,xmin=-6.66666666666667, xmax=6.33333333333333, ymin=6, ymax=28.1684210526316] {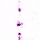};

\nextgroupplot[
axis background/.style={fill=white!89.80392156862746!black},
axis line style={white},
tick align=outside,
x grid style={white},
xmajorticks=false,
xmin=-6.66666666666667, xmax=6.33333333333333,
xtick style={color=white!33.33333333333333!black},
y grid style={white},
ymajorticks=false,
ymin=6, ymax=28.1684210526316,
ytick style={color=white!33.33333333333333!black},
]
\addplot graphics [includegraphics cmd=\pgfimage,xmin=-6.66666666666667, xmax=6.33333333333333, ymin=6, ymax=28.1684210526316] {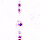};

\nextgroupplot[
axis background/.style={fill=white!89.80392156862746!black},
axis line style={white},
tick align=outside,
x grid style={white},
xmajorticks=false,
xmin=-6.66666666666667, xmax=6.33333333333333,
xtick style={color=white!33.33333333333333!black},
y grid style={white},
ymajorticks=false,
ymin=6, ymax=28.1684210526316,
ytick style={color=white!33.33333333333333!black},
opacity=0.0,
]
\addplot graphics [includegraphics cmd=\pgfimage,xmin=-6.66666666666667, xmax=6.33333333333333, ymin=6, ymax=28.1684210526316] {visualization_activations_placeholder.png};

\nextgroupplot[
axis background/.style={fill=white!89.80392156862746!black},
axis line style={white},
tick align=outside,
x grid style={white},
xmajorticks=false,
xmin=-6.66666666666667, xmax=6.33333333333333,
xtick style={color=white!33.33333333333333!black},
y grid style={white},
ymajorticks=false,
ymin=6, ymax=28.1684210526316,
ytick style={color=white!33.33333333333333!black},
]
\addplot graphics [includegraphics cmd=\pgfimage,xmin=-6.66666666666667, xmax=6.33333333333333, ymin=6, ymax=28.1684210526316] {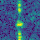};

\nextgroupplot[
axis background/.style={fill=white!89.80392156862746!black},
axis line style={white},
tick align=outside,
x grid style={white},
xmajorticks=false,
xmin=-6.66666666666667, xmax=6.33333333333333,
xtick style={color=white!33.33333333333333!black},
y grid style={white},
ymajorticks=false,
ymin=6, ymax=28.1684210526316,
ytick style={color=white!33.33333333333333!black},
opacity=0.0,
]
\addplot graphics [includegraphics cmd=\pgfimage,xmin=-6.66666666666667, xmax=6.33333333333333, ymin=6, ymax=28.1684210526316] {visualization_activations_placeholder.png};

\end{groupplot}
\end{tikzpicture}
		}
}
\subfigure{
		\resizebox{0.98\columnwidth}{!}{
			\begin{tikzpicture}
\begin{groupplot}[
	group style={group size=8 by 1},
	width=\columnwidth
]\nextgroupplot[
axis background/.style={fill=white!89.80392156862746!black},
axis line style={white},
tick align=outside,
x grid style={white},
xmajorticks=false,
xmin=-6.66666666666667, xmax=6.33333333333333,
xtick style={color=white!33.33333333333333!black},
y grid style={white},
ymajorticks=false,
ymin=6, ymax=28.1684210526316,
ytick style={color=white!33.33333333333333!black},
opacity=0.0,
]
\addplot graphics [includegraphics cmd=\pgfimage,xmin=-6.66666666666667, xmax=6.33333333333333, ymin=6, ymax=28.1684210526316] {visualization_activations_placeholder.png};

\nextgroupplot[
axis background/.style={fill=white!89.80392156862746!black},
axis line style={white},
tick align=outside,
x grid style={white},
xmajorticks=false,
xmin=-6.66666666666667, xmax=6.33333333333333,
xtick style={color=white!33.33333333333333!black},
y grid style={white},
ymajorticks=false,
ymin=6, ymax=28.1684210526316,
ytick style={color=white!33.33333333333333!black},
opacity=0.0,
]
\addplot graphics [includegraphics cmd=\pgfimage,xmin=-6.66666666666667, xmax=6.33333333333333, ymin=6, ymax=28.1684210526316] {visualization_activations_placeholder.png};

\nextgroupplot[
axis background/.style={fill=white!89.80392156862746!black},
axis line style={white},
tick align=outside,
x grid style={white},
xmajorticks=false,
xmin=-6.66666666666667, xmax=6.33333333333333,
xtick style={color=white!33.33333333333333!black},
y grid style={white},
ymajorticks=false,
ymin=6, ymax=28.1684210526316,
ytick style={color=white!33.33333333333333!black},
opacity=0.0,
]
\addplot graphics [includegraphics cmd=\pgfimage,xmin=-6.66666666666667, xmax=6.33333333333333, ymin=6, ymax=28.1684210526316] {visualization_activations_placeholder.png};

\nextgroupplot[
axis background/.style={fill=white!89.80392156862746!black},
axis line style={white},
tick align=outside,
x grid style={white},
xmajorticks=false,
xmin=-6.66666666666667, xmax=6.33333333333333,
xtick style={color=white!33.33333333333333!black},
y grid style={white},
ymajorticks=false,
ymin=6, ymax=28.1684210526316,
ytick style={color=white!33.33333333333333!black},
]
\addplot graphics [includegraphics cmd=\pgfimage,xmin=-6.66666666666667, xmax=6.33333333333333, ymin=6, ymax=28.1684210526316] {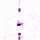};

\nextgroupplot[
axis background/.style={fill=white!89.80392156862746!black},
axis line style={white},
tick align=outside,
x grid style={white},
xmajorticks=false,
xmin=-6.66666666666667, xmax=6.33333333333333,
xtick style={color=white!33.33333333333333!black},
y grid style={white},
ymajorticks=false,
ymin=6, ymax=28.1684210526316,
ytick style={color=white!33.33333333333333!black},
]
\addplot graphics [includegraphics cmd=\pgfimage,xmin=-6.66666666666667, xmax=6.33333333333333, ymin=6, ymax=28.1684210526316] {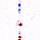};

\nextgroupplot[
axis background/.style={fill=white!89.80392156862746!black},
axis line style={white},
tick align=outside,
x grid style={white},
xmajorticks=false,
xmin=-6.66666666666667, xmax=6.33333333333333,
xtick style={color=white!33.33333333333333!black},
y grid style={white},
ymajorticks=false,
ymin=6, ymax=28.1684210526316,
ytick style={color=white!33.33333333333333!black},
opacity=0.0,
]
\addplot graphics [includegraphics cmd=\pgfimage,xmin=-6.66666666666667, xmax=6.33333333333333, ymin=6, ymax=28.1684210526316] {visualization_activations_placeholder.png};

\nextgroupplot[
axis background/.style={fill=white!89.80392156862746!black},
axis line style={white},
tick align=outside,
x grid style={white},
xmajorticks=false,
xmin=-6.66666666666667, xmax=6.33333333333333,
xtick style={color=white!33.33333333333333!black},
y grid style={white},
ymajorticks=false,
ymin=6, ymax=28.1684210526316,
ytick style={color=white!33.33333333333333!black},
]
\addplot graphics [includegraphics cmd=\pgfimage,xmin=-6.66666666666667, xmax=6.33333333333333, ymin=6, ymax=28.1684210526316] {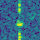};

\nextgroupplot[
axis background/.style={fill=white!89.80392156862746!black},
axis line style={white},
tick align=outside,
x grid style={white},
xmajorticks=false,
xmin=-6.66666666666667, xmax=6.33333333333333,
xtick style={color=white!33.33333333333333!black},
y grid style={white},
ymajorticks=false,
ymin=6, ymax=28.1684210526316,
ytick style={color=white!33.33333333333333!black},
opacity=0.0,
]
\addplot graphics [includegraphics cmd=\pgfimage,xmin=-6.66666666666667, xmax=6.33333333333333, ymin=6, ymax=28.1684210526316] {visualization_activations_placeholder.png};

\end{groupplot}
\end{tikzpicture}
		}
}
{
	\begin{tabularx}{\textwidth}{XcXXcX}
		& \hspace{25mm}Output: Real, Imag\hspace{7mm}$\rightarrow$ Magnitude &&& \hspace{25mm}Output: Real, Imag\hspace{7mm}$\rightarrow$ Magnitude &
	\end{tabularx}
}
	\caption{Visualization of RD map patches of inputs, outputs and activations after each layer for the real-valued CNN model. The left and right side show patches for an example with and without interference, respectively. The log magnitude spectra of inputs and outputs are shown only for illustration purpose and they are not directly used by the model. The color intensity of activations correlates with their magnitude and the color map scale is identical for corresponding activation patches on the left and right side. Red, white and blue indicate positive, zero and negative values, respectively.}
	\label{fig:visualization_activations}
\end{figure*}

\begin{figure}
	\centering
	\footnotesize
	\subfigure[1\textsuperscript{st} Layer]{
		\resizebox{\columnwidth}{!}{
			\input{weights_l1_id0.tex}
		}
	}
	\subfigure[3\textsuperscript{rd} Layer]{
		\resizebox{\columnwidth}{!}{
\begin{tikzpicture}

\begin{groupplot}[
	group style={group size=16 by 1},
	width=\columnwidth
	]
\nextgroupplot[
axis background/.style={fill=white!89.80392156862746!black},
axis line style={white},
tick align=outside,
x grid style={white},
xmajorticks=false,
xmin=-0.5, xmax=2.5,
xtick style={color=white!33.33333333333333!black},
y grid style={white},
ymajorticks=false,
ymin=-0.5, ymax=2.5,
ytick style={color=white!33.33333333333333!black}
]
\addplot graphics [includegraphics cmd=\pgfimage,xmin=-0.5, xmax=2.5, ymin=2.5, ymax=-0.5] {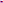};

\nextgroupplot[
axis background/.style={fill=white!89.80392156862746!black},
axis line style={white},
tick align=outside,
x grid style={white},
xmajorticks=false,
xmin=-0.5, xmax=2.5,
xtick style={color=white!33.33333333333333!black},
y grid style={white},
ymajorticks=false,
ymin=-0.5, ymax=2.5,
ytick style={color=white!33.33333333333333!black}
]
\addplot graphics [includegraphics cmd=\pgfimage,xmin=-0.5, xmax=2.5, ymin=2.5, ymax=-0.5] {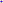};

\nextgroupplot[
axis background/.style={fill=white!89.80392156862746!black},
axis line style={white},
tick align=outside,
x grid style={white},
xmajorticks=false,
xmin=-0.5, xmax=2.5,
xtick style={color=white!33.33333333333333!black},
y grid style={white},
ymajorticks=false,
ymin=-0.5, ymax=2.5,
ytick style={color=white!33.33333333333333!black}
]
\addplot graphics [includegraphics cmd=\pgfimage,xmin=-0.5, xmax=2.5, ymin=2.5, ymax=-0.5] {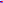};

\nextgroupplot[
axis background/.style={fill=white!89.80392156862746!black},
axis line style={white},
tick align=outside,
x grid style={white},
xmajorticks=false,
xmin=-0.5, xmax=2.5,
xtick style={color=white!33.33333333333333!black},
y grid style={white},
ymajorticks=false,
ymin=-0.5, ymax=2.5,
ytick style={color=white!33.33333333333333!black}
]
\addplot graphics [includegraphics cmd=\pgfimage,xmin=-0.5, xmax=2.5, ymin=2.5, ymax=-0.5] {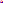};

\nextgroupplot[
axis background/.style={fill=white!89.80392156862746!black},
axis line style={white},
tick align=outside,
x grid style={white},
xmajorticks=false,
xmin=-0.5, xmax=2.5,
xtick style={color=white!33.33333333333333!black},
y grid style={white},
ymajorticks=false,
ymin=-0.5, ymax=2.5,
ytick style={color=white!33.33333333333333!black}
]
\addplot graphics [includegraphics cmd=\pgfimage,xmin=-0.5, xmax=2.5, ymin=2.5, ymax=-0.5] {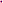};

\nextgroupplot[
axis background/.style={fill=white!89.80392156862746!black},
axis line style={white},
tick align=outside,
x grid style={white},
xmajorticks=false,
xmin=-0.5, xmax=2.5,
xtick style={color=white!33.33333333333333!black},
y grid style={white},
ymajorticks=false,
ymin=-0.5, ymax=2.5,
ytick style={color=white!33.33333333333333!black}
]
\addplot graphics [includegraphics cmd=\pgfimage,xmin=-0.5, xmax=2.5, ymin=2.5, ymax=-0.5] {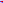};

\nextgroupplot[
axis background/.style={fill=white!89.80392156862746!black},
axis line style={white},
tick align=outside,
x grid style={white},
xmajorticks=false,
xmin=-0.5, xmax=2.5,
xtick style={color=white!33.33333333333333!black},
y grid style={white},
ymajorticks=false,
ymin=-0.5, ymax=2.5,
ytick style={color=white!33.33333333333333!black}
]
\addplot graphics [includegraphics cmd=\pgfimage,xmin=-0.5, xmax=2.5, ymin=2.5, ymax=-0.5] {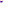};

\nextgroupplot[
axis background/.style={fill=white!89.80392156862746!black},
axis line style={white},
tick align=outside,
x grid style={white},
xmajorticks=false,
xmin=-0.5, xmax=2.5,
xtick style={color=white!33.33333333333333!black},
y grid style={white},
ymajorticks=false,
ymin=-0.5, ymax=2.5,
ytick style={color=white!33.33333333333333!black}
]
\addplot graphics [includegraphics cmd=\pgfimage,xmin=-0.5, xmax=2.5, ymin=2.5, ymax=-0.5] {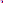};

\nextgroupplot[
axis background/.style={fill=white!89.80392156862746!black},
axis line style={white},
tick align=outside,
x grid style={white},
xmajorticks=false,
xmin=-0.5, xmax=2.5,
xtick style={color=white!33.33333333333333!black},
y grid style={white},
ymajorticks=false,
ymin=-0.5, ymax=2.5,
ytick style={color=white!33.33333333333333!black}
]
\addplot graphics [includegraphics cmd=\pgfimage,xmin=-0.5, xmax=2.5, ymin=2.5, ymax=-0.5] {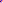};

\nextgroupplot[
axis background/.style={fill=white!89.80392156862746!black},
axis line style={white},
tick align=outside,
x grid style={white},
xmajorticks=false,
xmin=-0.5, xmax=2.5,
xtick style={color=white!33.33333333333333!black},
y grid style={white},
ymajorticks=false,
ymin=-0.5, ymax=2.5,
ytick style={color=white!33.33333333333333!black}
]
\addplot graphics [includegraphics cmd=\pgfimage,xmin=-0.5, xmax=2.5, ymin=2.5, ymax=-0.5] {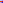};

\nextgroupplot[
axis background/.style={fill=white!89.80392156862746!black},
axis line style={white},
tick align=outside,
x grid style={white},
xmajorticks=false,
xmin=-0.5, xmax=2.5,
xtick style={color=white!33.33333333333333!black},
y grid style={white},
ymajorticks=false,
ymin=-0.5, ymax=2.5,
ytick style={color=white!33.33333333333333!black}
]
\addplot graphics [includegraphics cmd=\pgfimage,xmin=-0.5, xmax=2.5, ymin=2.5, ymax=-0.5] {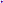};

\nextgroupplot[
axis background/.style={fill=white!89.80392156862746!black},
axis line style={white},
tick align=outside,
x grid style={white},
xmajorticks=false,
xmin=-0.5, xmax=2.5,
xtick style={color=white!33.33333333333333!black},
y grid style={white},
ymajorticks=false,
ymin=-0.5, ymax=2.5,
ytick style={color=white!33.33333333333333!black}
]
\addplot graphics [includegraphics cmd=\pgfimage,xmin=-0.5, xmax=2.5, ymin=2.5, ymax=-0.5] {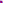};

\nextgroupplot[
axis background/.style={fill=white!89.80392156862746!black},
axis line style={white},
tick align=outside,
x grid style={white},
xmajorticks=false,
xmin=-0.5, xmax=2.5,
xtick style={color=white!33.33333333333333!black},
y grid style={white},
ymajorticks=false,
ymin=-0.5, ymax=2.5,
ytick style={color=white!33.33333333333333!black}
]
\addplot graphics [includegraphics cmd=\pgfimage,xmin=-0.5, xmax=2.5, ymin=2.5, ymax=-0.5] {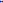};

\nextgroupplot[
axis background/.style={fill=white!89.80392156862746!black},
axis line style={white},
tick align=outside,
x grid style={white},
xmajorticks=false,
xmin=-0.5, xmax=2.5,
xtick style={color=white!33.33333333333333!black},
y grid style={white},
ymajorticks=false,
ymin=-0.5, ymax=2.5,
ytick style={color=white!33.33333333333333!black}
]
\addplot graphics [includegraphics cmd=\pgfimage,xmin=-0.5, xmax=2.5, ymin=2.5, ymax=-0.5] {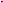};

\nextgroupplot[
axis background/.style={fill=white!89.80392156862746!black},
axis line style={white},
tick align=outside,
x grid style={white},
xmajorticks=false,
xmin=-0.5, xmax=2.5,
xtick style={color=white!33.33333333333333!black},
y grid style={white},
ymajorticks=false,
ymin=-0.5, ymax=2.5,
ytick style={color=white!33.33333333333333!black}
]
\addplot graphics [includegraphics cmd=\pgfimage,xmin=-0.5, xmax=2.5, ymin=2.5, ymax=-0.5] {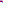};

\nextgroupplot[
axis background/.style={fill=white!89.80392156862746!black},
axis line style={white},
tick align=outside,
x grid style={white},
xmajorticks=false,
xmin=-0.5, xmax=2.5,
xtick style={color=white!33.33333333333333!black},
y grid style={white},
ymajorticks=false,
ymin=-0.5, ymax=2.5,
ytick style={color=white!33.33333333333333!black}
]
\addplot graphics [includegraphics cmd=\pgfimage,xmin=-0.5, xmax=2.5, ymin=2.5, ymax=-0.5] {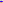};
\end{groupplot}

\end{tikzpicture}
		}
	}
	\caption{Visualization of the learned filter kernels for the real-valued CNN model. Filter kernels are normalized and the color map ranges from minus to plus one, where red is positive, blue is negative and white indicates a value of zero.}
	\label{fig:visualization_filter_kernels}
\end{figure}

While there is no simple approach of interpreting what the CNN model has learned, we can gain more insights by inspecting visualizations of learned filter kernels and activations (i.e. feature maps) for a specific input. Figure~\ref{fig:visualization_activations} shows RD map patches of model inputs, outputs and activations after each layer for the real-valued CNN. The same RD measurement is used, where the left side input includes interference and the right side input does not. We analyzed the activations corresponding to the input with interference and the ones without interference and made the following observations:
\begin{itemize}
	\item Inputs with interference have similar patterns in real and imaginary parts, including positive and negative values.
	\item All activations after the 1\textsuperscript{st} layer contain high magnitude values (positive and negative) at object positions.
	\item Some activations after the 1\textsuperscript{st} layer contain much more interference than others (compare patches with green borders) and seem to concentrate on different aspects of interference patterns. E.g. the activations marked with an orange border contain strong diagonal patterns while the activations marked with a red border contain more unstructured noise.
	\item Activations after the 2\textsuperscript{nd} layer are either all positive or all negative for inputs without interference. However, for inputs with interference, there are activations (compare patches with blue borders) that separate a patch into objects and interference patterns by means of positive and negative values.
	\item The activations at the outputs resemble the inputs at object positions while possible interference and noise are removed. Note, that the output from the sample with interference contains even less noise than the output from the sample without interference which stays almost unchanged.
\end{itemize}
Figure~\ref{fig:visualization_filter_kernels} shows the learned filter kernels of the real-valued CNN model for the first and last layers. The first and last layer filter kernels can be interpreted, because we have an intuition about the input and output of the model. In contrast, an interpretation of the second layer filter kernels is difficult, since it operates on activations from hidden layers. The learned filter kernels are diverse and resemble well known spatial filter types:
\begin{description}
	\item[Gaussian] filters smooth the input. These filters have all positive or all negative values, where the intensity is higher in the middle and lower at the corners. E.g. \includegraphics[height=\fontcharht\font`\B]{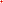}%
	,~\includegraphics[height=\fontcharht\font`\B]{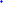}%
	\item[Gradient] filters can detect edges along different dimensions, i.e. horizontal, vertical and diagonal. They show first order spatial derivatives of the input and they themselves have edges of positive or negative values along some dimension. E.g. \includegraphics[height=\fontcharht\font`\B]{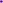}%
	,~\includegraphics[height=\fontcharht\font`\B]{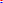}%
	,~\includegraphics[height=\fontcharht\font`\B]{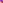}%
	\item[Laplacian] filters highlight regions of rapid intensity changes in the input. They show second order spatial derivatives of the input and have a positive or negative peak in the middle, that is surrounded by lower intensity values of the opposite sign. E.g. \includegraphics[height=\fontcharht\font`\B]{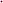}%
	,~\includegraphics[height=\fontcharht\font`\B]{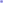}%
\end{description}

The first layer contains diverse variations of Gaussian and gradient filters. They are directly applied to the input and used for low level feature detection. The third layer contains Laplacian filters, which are used to reassemble real and imaginary parts from the second layer outputs. Note, that also in classical computer vision applications Laplacian filters are often used in combination with other filters (e.g. Gaussians) that are previously applied to the inputs. This visualization gives a valuable insight into learned filter kernels and how the model detects and removes the interference. The output is then reconstructed using the objects without including that interference.

\section{Conclusion}
In this paper, we investigate the capability to quantize CNN-based models for denoising and interference mitigation of radar signals. Our experiments emphasize the importance of small real-valued base models in order to obtain memory efficient models after quantization. We conclude, that small architectures are not suitable for binarization in the context of the considered regression task and instead multiple bits are required to retain high performance. For the considered task and selected base model, the quantization of activations has a substantially higher impact on the overall memory than the quantization of weights.
The bit-width can be learned in addition to the model weights resulting in a memory reduction of up to 80\% for the selected base model. However, for simplicity and practical reasons, we advocate 8-bit quantization of all weights and activations yielding a memory footprint reduction of approximately 75\% compared to the real-valued model without any noteworthy performance degradation.
Furthermore, we analyze the effects of training distributions over discrete weights in contrast to quantization aware training with the STE and find that competitive results can be achieved including additional information about uncertainty estimates in interference mitigated RD maps. As future work we plan to conduct quantitative evaluations of real-world interference signals. Further interesting directions are model architectures that exploit sequential and multichannel information.

\section*{Acknowledgments}
This work was supported by the Austrian Research Promotion Agency (FFG) under the project SAHaRA (17774193) and NVIDIA by providing GPUs.

\ifCLASSOPTIONcaptionsoff
  \newpage
\fi

\bibliographystyle{ieeetr}
\bibliography{mybibliography}
\vspace{-9mm}
\begin{IEEEbiography}
[{\includegraphics[width=1in,height=1.25in,clip,keepaspectratio]{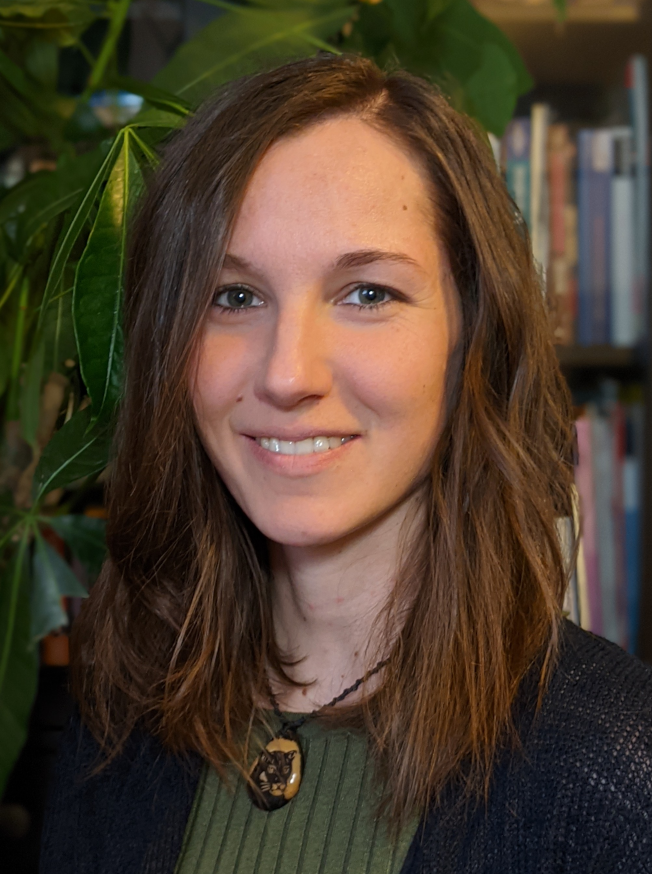}}]%
{Johanna Rock}
received her MSc degree in computer science from Graz University of Technology, Austria, in 2018. Currently she is pursuing towards her PhD as a research associate at the Signal Processing and Speech Communication Laboratory at Graz University of Technology. Her main research interests are in the area of machine learning and pattern recognition with a focus on real-world signals, automotive-radar, interference mitigation, target detection, resource-efficient deep learning and uncertainty for robust neural networks.
\end{IEEEbiography}
\vspace{-9mm}
\begin{IEEEbiography}
[{\includegraphics[width=1in,height=1.25in,clip,keepaspectratio]{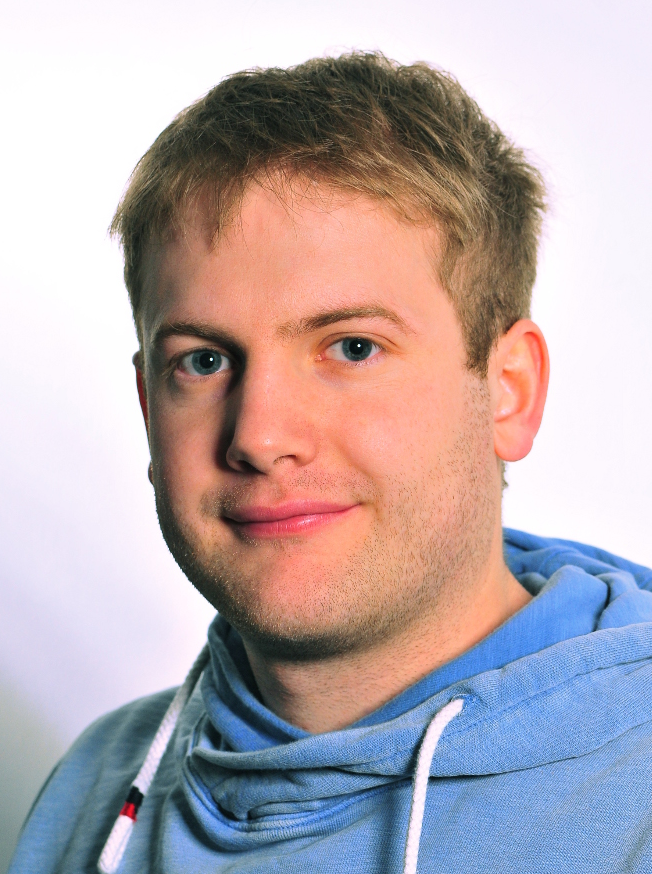}}]%
{Wolfgang Roth}
received his Msc degree in computer science from Graz University of Technology, Austria, in 2015. He is currently a PhD student at the Signal Processing and Speech Communication Laboratory at Graz University of Technology. His research interests include Bayesian inference, deep learning, and resource-efficient models.
\end{IEEEbiography}
\vspace{-9mm}
\begin{IEEEbiography}
[{\includegraphics[width=1in,height=1.25in,clip,keepaspectratio]{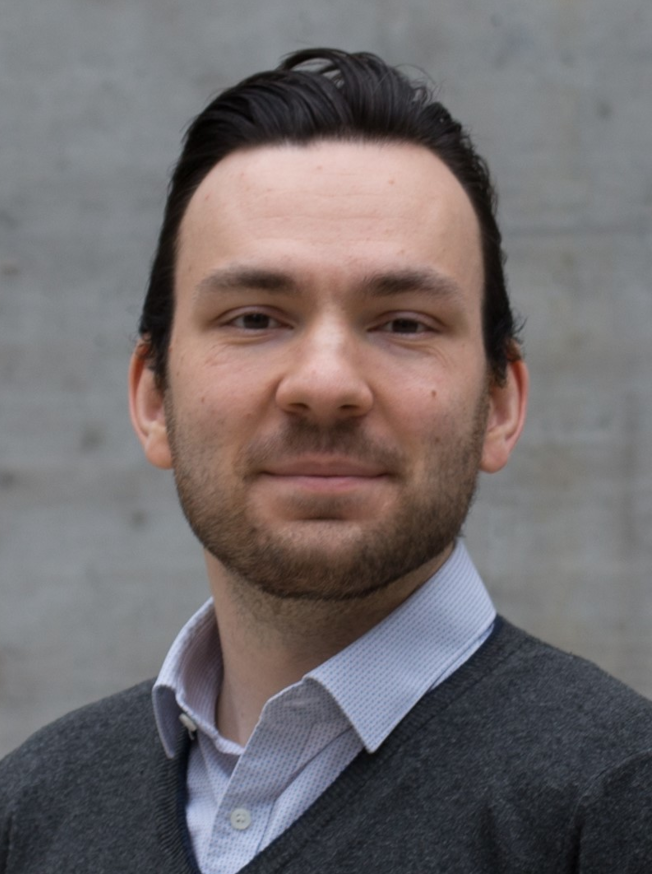}}]%
{Mate Toth}
received his MSc degree in electrical engineering from Graz University of Technology, Austria, in 2018. He is currently pursuing a PhD at Infineon Technologies Austria in cooperation with Graz University of Technology. His research interests include robust and efficient signal processing for sensors and communication, with a current focus on signal denoising and parameter estimation in automotive radar.
\end{IEEEbiography}
\vspace{-9mm}
\begin{IEEEbiography}
[{\includegraphics[width=1in,height=1.25in,clip,keepaspectratio]{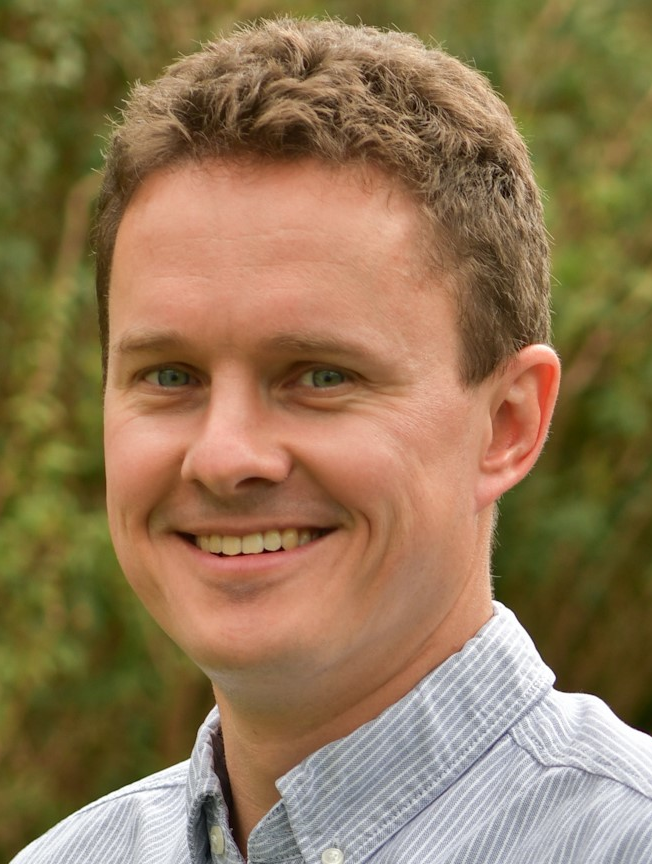}}]%
{Paul Meissner}
received his MSc degree in information and communications engineering in 2009 and the PhD degree in electrical engineering in 2014 from Graz University of Technology, respectively. He is currently a concept engineer for automotive radar at Infineon Technologies Austria, focusing on receiver architectures for radar MMICs. His research interests are statistical signal processing, system modeling for complex sensor systems, and data processing algorithms for radar sensors.
\end{IEEEbiography}
\vspace{-9mm}
\begin{IEEEbiography}
[{\includegraphics[width=1in,height=1.25in,clip,keepaspectratio]{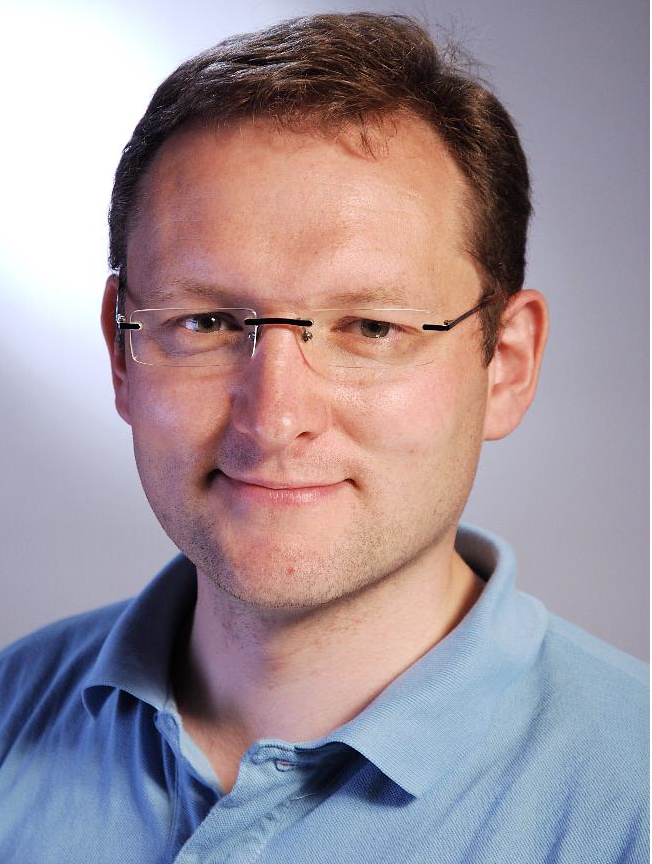}}]%
{Franz Pernkopf}
received his PhD degree from the University of Leoben in 2002. He was awarded the Erwin Schr\"{o}dinger Fellowship and was a research associate at the Department of Electrical Engineering at the University of Washington, Seattle, from 2004 to 2006. Since 2010 (Associate) and 2019 (Full) he is a Professor for Intelligent Systems at the Signal Processing and Speech Communication Laboratory at Graz University of Technology, Austria. His research is focused on machine learning and data analysis with a wide range of applications including signal and speech processing. He is particularly interested in probabilistic graphical models for reasoning under uncertainty, discriminative and hybrid learning paradigms, deep neural networks and sequence modeling.
\end{IEEEbiography}




\vfill


\end{document}